\documentclass[onecolumn,10pt,letterpaper]{elsarticle}

\usepackage{epsf,amsmath,amsfonts}

\usepackage{graphics,psfrag}
\usepackage{graphicx,psfrag}
\usepackage[export]{adjustbox}

\usepackage{wrapfig}
\usepackage{color}

\usepackage{verbatim}

\usepackage{xcolor}

\usepackage{lineno,hyperref}
\modulolinenumbers[5]
\hypersetup{
    colorlinks,
    linkcolor={blue!80!black},
    citecolor={blue!50!black},
    urlcolor={blue!80!black}
}

\usepackage{etoolbox,calc}

\makeatletter
\def\appendixname{Appendix}
\appto\appendix{%
  \addtocontents{toc}{\patch@l@section}
  \appto\appendixname{ }
}
\protected\def\patch@l@section{%
  \patchcmd{\l@section}{1.5em}{\widthof{\appendixname\space}+2.5em}{}{}%
}
\makeatother

\newcommand{\bea}{\begin{eqnarray}}
\newcommand{\eea}{\end{eqnarray}}
\newcommand{\beast}{\begin{eqnarray*}}
\newcommand{\eeast}{\end{eqnarray*}}

\newcommand{\beq}{\begin{equation}}
\newcommand{\eeq}{\end{equation}}
\newcommand{\half}{\frac{1}{2}}
\newcommand{\intdr}{{\int}d^3r\,}
\newcommand{\figref}[1]{Fig.~\ref{#1}}
\newcommand{\figtworef}[2]{Figures~\ref{#1} and \ref{#2}}
\newcommand{\tabref}[1]{Table~\ref{#1}}

\newcommand{\secref}[1]{Sec.~(\ref{#1})}
\newcommand{\appref}[1]{App.~(\ref{#1})}
\newcommand{\eqnref}[1]{Eq.~(\ref{#1})}
\newcommand{\eqntworef}[2]{Eqs.~(\ref{#1}) and (\ref{#2})}

\newcommand{\bra}[1]{ \langle {#1} |}
\newcommand{\ket}[1]{| {#1} \rangle }
\newcommand{\spr}[2]{ \langle {#1} | {#2} \rangle }
\renewcommand{\v}[1]{\mathbf{#1}}
\newcommand{\refsec}[1]{Sec.~\ref{#1}}
\newcommand{\refapp}[1]{~\ref{#1}}

\begin{document}

\begin{frontmatter}
\title{Kubo-Greenwood Electrical Conductivity Formulation 
 and Implementation for Projector Augmented Wave Datasets}

\author[ufl]{ L.~Calder\'in \fnref{fntlc} \corref{cor1}}
\ead{lcalderin@email.arizona.edu}

\author[ufl]{V.V.~Karasiev}
\author[ufl]{S.B.~Trickey}

\address[ufl]{QTP, Depts.\ of Physics and Chemistry, Univ.\ of Florida, 
Gainesville, FL 32611-8440, USA }

\fntext[fntlc]{Present address: Dept. of Materials Science and Engineering,
University of Arizona, Tucson, AZ 85721-0012, USA}
\cortext[cor1]{Corresponding author}

\begin{abstract}
As the foundation for a new computational implementation, we survey the calculation of the complex electrical 
conductivity tensor based on the Kubo-Greenwood (KG) formalism
(J.\ Phys.\ Soc.\ Jpn. \textbf{12}, 570 (1957); Proc.\ Phys.\ Soc.\ \textbf{71}, 585 (1958)), 
with emphasis on derivations and technical aspects pertinent 
to use of 
projector augmented wave datasets with plane wave basis sets (Phys.\ Rev.\ B \textbf{50}, 17953 (1994)). 
New analytical results and a full implementation of the KG approach in 
an open-source Fortran 90 post-processing code for use with Quantum Espresso 
(J.\ Phys.\ Cond.\ Matt.\ \textbf{21}, 395502 (2009)) are presented.
Named KGEC ([K]ubo [G]reenwood [E]lectronic [C]onductivity), the code 
 calculates the full complex conductivity tensor (not just the average trace).
 It supports use of either the original KG formula or the popular one approximated  
 in terms of a Dirac delta function. It provides both Gaussian and 
Lorentzian representations of the Dirac delta function (though
the Lorentzian is preferable on basic grounds).  
KGEC provides decomposition of the conductivity into intra- and 
inter-band contributions as well as degenerate state contributions. 
It calculates the dc conductivity tensor directly. 
It is MPI parallelized over k-points, bands, and plane waves, with 
an option to recover the plane wave processes for their use in band  
parallelization as well. 
It is designed to provide rapid convergence with respect to 
$\mathbf k$-point density. Examples of its use are given.
\end{abstract}


\end{frontmatter}

\newpage
 \tableofcontents
\newpage

\section{Introduction}
Calculation of transport properties of matter is a venerable but still very 
active research area in part because of the physical significance of 
transport coefficients and in part because of the major theoretical 
and computational challenges involved.  The 
computational goal of the present work is
to design algorithms for the calculation of the Kubo-Greenwood (KG) 
electrical conductivity \cite{KUBO1957,GREENWOOD1958} and implement 
them as a post-processing tool for
the widely used Quantum Espresso  \cite{QE2009} (QE) code. We begin by 
reviewing the state
of the art of KG electrical conductivity calculations, with 
emphasis upon derivations and their technical implications.  The
computational context of the formulation is projector augmented wave (PAW)
datasets used with plane wave (PW) basis sets \cite{PAW} for the
solution of the Kohn-Sham (KS) equations \cite{PhysRev.140.A1133}.  The 
resultant new program is named  
KGEC, from the initial letters of Kubo-Greenwood Electrical Conductivity.


Though the primary goal was computational, 
that reconsideration of the underlying analysis also has proved 
fruitful, as will become apparent, for example, in the treatment
of contributions of intra-band and degenerate band transitions 
to the conductivity.  Beyond the obvious 
goal of providing new capability for users of QE, the project also
was motivated by the opportunity to include finite-temperature
effects via free energy density functionals 
\cite{PhysRevLett.112.076403,PhysRevE.93.063207} and 
to provide benefits from orbital-free density functional theory (DFT) 
molecular dynamics via the Profess@QE package \cite{KARASIEV20143240}. The coupling of KGEC with 
these developments opens a wide range of possibilities 
for simulations of systems over a wide range of state conditions,
e.g.\ warm dense matter.

Starting with the KG general formula in the next section
(\refsec{sec:conductivity}) we derive in detail all of the 
mathematical expressions necessary for a full KG implementation. 
In \refsec{sec:paw} we provide the essential ingredients of the PAW
method, followed by derivation of the expression for the matrix
elements of the gradient operator 
(\refsec{sec:grad-matrix-elements}). Next, 
\refsec{sec:qe-implementation} provides an overview of the work flow
in KGEC, its installation, execution, input, output and MPI parallelization. 
We also
present results from various tests in
\secref{sec:kgec-tests}, including a comparison with similar Abinit
calculations \cite{Gonze2002478}. Underlying difficulties 
including numerical problems are
discussed in \secref{sec:difficulties},
while remarks
and comments about future work 
are in \secref{sec:remarks}.

\section{\label{sec:conductivity} The Kubo-Greenwood electrical conductivity formula}
\subsection{\label{sec:kg-general}General expression}
The KG expression \cite{KUBO1957,GREENWOOD1958} for the frequency-dependent 
complex electrical conductivity tensor is 
\begin{equation}
 \sigma(\omega)=i\frac{2 e^2 \hbar^3}{m_e^2 V}
 \sum_m  
 \sum_{m'} 
 \frac{(f(\epsilon_{m'})-f(\epsilon_m))}{(\epsilon_m-\epsilon_{m'})} 
\frac{\bra{m}\nabla\ket{m'}\bra{m'}\nabla\ket{m}}{\epsilon_m-\epsilon_{m'}-\hbar\omega+i\delta/2}
\label{KGdefnA}
\end{equation}
or in more compact form
\begin{equation}
 \sigma(\omega)=i\frac{2 e^2 \hbar^3}{m_e^2 V}
 \sum_m  
 \sum_{m'} \frac{\Delta f_{m'm}}{\Delta\epsilon_{mm'}} 
\frac{\bra{m}\nabla\ket{m'}\bra{m'}\nabla\ket{m}}{(\Delta\epsilon_{mm'}-\hbar\omega+i\delta/2)} \; .
\label{KGdefnB}
\end{equation}

Before proceeding, note an unconventional aspect compared to the usual 
KG presentation. In 
both equations (\ref{KGdefnA}) and (\ref{KGdefnB}), the 
expression $\bra{m}\nabla\ket{m'}\bra{m'}\nabla\ket{m}$
is a dyadic in the coordinate indices of the gradients.  For 
didactic clarity, in a Cartesian system, Eq.\ (\ref{KGdefnB}) becomes 
\begin{equation}
 \sigma_{x,z}(\omega)=i\frac{2 e^2 \hbar^3}{m_e^2 V}
 \sum_m
 \sum_{m'}  
\frac{\Delta f_{m'm}}{\Delta\epsilon_{mm'}} 
\frac{\bra{m}{\frac{\partial}{\partial x}}\ket{m'}\bra{m'}{\frac{\partial}{\partial z}}\ket{m}}{(\Delta\epsilon_{mm'}-\hbar\omega+i\delta/2)} 
\label{KGinCartesian}
\end{equation}
for the $x$-$z$ element of the conductivity tensor.  The more 
familiar version comes from taking the trace.  

In these expressions $m$, $m^\prime$ label non-spin-polarized 
single-particle states with $\epsilon_{m}$, $\epsilon_{m^\prime}$ the corresponding eigenvalues and associated Fermi-Dirac 
occupation numbers $f(\epsilon_{m})$, $f(\epsilon_{m^\prime})$. (For
simplicity of notation, the temperature is suppressed for now.)  In practice
and in our implementation, the 
states and occupations are from a KS DFT calculation, though the 
analysis presented in this
section and the next one does
not depend upon that particular choice of mean-field Hamiltonian. (Note
that because of the spin-unpolarized formulation,
the net occupation of each KS orbital is $2f(\epsilon_{m})$.)  Then 
$\Delta\epsilon_{mm'}=\epsilon_m-\epsilon_{m'}$ and 
$\Delta f_{m'm}= f(\epsilon_{m'})-f(\epsilon_m)$.  The constants $e$,
$\hbar$, $m_e$ and $V$ are the electron charge, Planck's constant, 
electron mass, and system volume, respectively. The $i\delta/2$ is an
imaginary factor related to damping or relaxation effects. In the
Drude model for the electrical conductivity, it is identified 
with the inverse of the average inter-collision 
time.  

If the matrix element dyadic 
product $\bra{m}\nabla\ket{m'}\bra{m'}\nabla\ket{m}$ 
is real, the real and imaginary parts of $\sigma(\omega)$ can
be separated by multiplying and dividing 
by $(\Delta\epsilon_{mm'}-\hbar\omega-i\delta/2)$, leading to 
\begin{equation}
\sigma(\omega)=\sigma_1(\omega)+i\sigma_2(\omega),
 \label{eq:sigmaReImag}
\end{equation} 
with
\begin{equation}
 \sigma_1(\omega)=\frac{2 e^2 \hbar^3}{m_e^2 V}
 \sum_m
 \sum_{m'}  
 \frac{\Delta f_{m'm}}{\Delta\epsilon_{mm'}}
 \bra{m}\nabla\ket{m'}\bra{m'}\nabla\ket{m}
 \frac{\delta/2}{(\Delta\epsilon_{mm'}-\hbar\omega)^2+\delta^2/4}
 \label{eq:sigma1-general}
\end{equation}
and 
\begin{equation}
 \sigma_2(\omega)=\frac{2 e^2 \hbar^3}{m_e^2 V}
 \sum_m
 \sum_{m'}  
 \frac{\Delta f_{m'm}}{\Delta\epsilon_{mm'}}
 \bra{m}\nabla\ket{m'}\bra{m'}\nabla\ket{m}
 \frac{(\Delta\epsilon_{mm'}-\hbar\omega)}{(\Delta\epsilon_{mm'}-\hbar\omega)^2+\delta^2/4}.
 \label{eq:sigma2-general}
\end{equation}
Again be reminded that both $\sigma_1$ and $\sigma_2$ are tensors, not
scalars.  

Commonly it is argued that for small $\delta$, the Lorentzian 
in $\sigma_1(\omega)$ behaves like a Dirac delta function, that is
\begin{equation}
 \frac{\delta/2}{(\Delta\epsilon_{mm'}-\hbar\omega)^2+\delta^2/4}\approx\pi \delta(\Delta\epsilon_{mm'}-\hbar\omega), 
\label{Lorentzian-Delta}
\end{equation}
which allows $\sigma_1(\omega)$ to be written as 
\begin{equation}
 \sigma_1(\omega)=\frac{2 \pi e^2 \hbar^3}{m_e^2 V}
 \sum_m
 \sum_{m'}
 \frac{\Delta f_{m'm}}{\Delta\epsilon_{mm'}}
 \bra{m}\nabla\ket{m'}\bra{m'}\nabla\ket{m}
 \delta(\Delta\epsilon_{mm'}-\hbar\omega),
 \label{eq:sigma1-general-2a}
\end{equation}
or
\begin{equation}
 \sigma_1(\omega)=\frac{2 \pi e^2 \hbar^2}{m_e^2 V\omega}
 \sum_m
 \sum_{m'}  
 \Delta f_{m'm}
 \bra{m}\nabla\ket{m'}\bra{m'}\nabla\ket{m}
 \delta(\Delta\epsilon_{mm'}-\hbar\omega).
 \label{eq:sigma1-general-2b}
\end{equation}
Both forms commonly are encountered.
We will label \eqnref{eq:sigma1-general-2a} ``the Dirac-delta form''
(notation ``D-d'') or ``the exact form or expression''.  
Note that if one starts with it and represents the
Dirac delta function by a Lorentzian, the original Kubo-Greenwood expression 
is recovered.
Similarly \eqnref{eq:sigma1-general-2b} will be labeled ``the approximated formula or expression''  
because one cannot recover the exact Kubo-Greenwood formula from it 
by simple substitution for the delta function. 

In computation, the Dirac 
delta function in $\sigma_1$ often is 
represented by a Gaussian, even though 
its natural representation is a Lorentzian. Distinctions among these
representations should disappear as $\delta \rightarrow 0$, but in 
practice they are manifest even for a small, non-zero $\delta$.  We return to
that in the discussion of numerical tests in \secref{sec:kgec-tests}. 
 Notice also that, because $\omega > 0$ 
the Dirac delta function in \eqnref{eq:sigma1-general-2b}  selects only 
states with positive energy differences, but the original expression included
contributions from states with negative energy differences. That 
discrepancy can be resolved
by introduction of the $\delta(\Delta\epsilon_{mm'}+\hbar \omega)$ term as well.
Another problem is that only non-degenerate inter-band contributions 
are included in the approximated formula.  We return
to that below as well.

\subsection{\label{sec:kg-solid}KG formula in the Bloch picture }
We focus on periodic systems, so the state indices $m$ and $m'$ 
become band index and Brillouin zone wave vector pairs
$n,\v{k}$ and $n',\v{k'}$ for Bloch states. Because the gradient  
matrix elements between 
$\v{k}$ and $\v{k'}$ states are zero if $\v{k}\neq\v{k'}$, the KG formulae,  
(\eqntworef{eq:sigma1-general}{eq:sigma2-general}), become
\begin{equation}
\tilde{\sigma}_1(\omega)=\frac{2 e^2 \hbar^3}{m_e^2 \Omega}
 \sum_\v{k} w_{\v{k}}
 \sum_{nn'}  
 \frac{\Delta f_{n'\v{k},n\v{k}}}{\Delta\epsilon_{n\v{k},n'\v{k}}}
 \bra{\Psi_{n\v{k}}}\nabla\ket{\Psi_{n'\v{k}}}\bra{\Psi_{n'\v{k}}}\nabla\ket{\Psi_{n\v{k}}}
 \frac{\delta/2}{(\Delta\epsilon_{n\v{k},n'\v{k}}-\hbar\omega)^2+\delta^2/4}
 \end{equation}
and
\begin{equation}
 \tilde{\sigma}_2(\omega)=\frac{2 e^2 \hbar^3}{m_e^2 \Omega}
 \sum_\v{k} w_{\v{k}}
 \sum_{nn'}  
 \frac{\Delta f_{n'\v{k},n\v{k}}}{\Delta\epsilon_{n\v{k},n'\v{k}}}
 \bra{\Psi_{n\v{k}}}\nabla\ket{\Psi_{n'\v{k}}}\bra{\Psi_{n'\v{k}}}\nabla\ket{\Psi_{n\v{k}}}
 \frac{(\Delta\epsilon_{n\v{k},n'\v{k}}-\hbar\omega)}{(\Delta\epsilon_{n\v{k},n'\v{k}}-\hbar\omega)^2+\delta^2/4} \; .
\end{equation}
Here $\Omega$ is the unit cell volume and $w_{\v{k}}$ are the $\mathbf k$-point 
integration weights. 
We have also used a tilde {\~{  } }  atop the $\sigma$s to 
highlight that they both become complex because the 
matrix element tensor product no longer is necessarily real (since 
the Bloch wave functions are, in the most general case, complex).  

Both $\sigma_1$ and $\sigma_2$ can be recovered by means of the 
elementary relations
${\sigma}_1(\omega)  =\Re(\tilde{\sigma_1}+i\tilde{\sigma_2})$, 
$ {\sigma}_2(\omega)  =\Im(\tilde{\sigma_1}+i\tilde{\sigma_2})$ 
and use of the fact that the real part of $\sigma$ must be even 
and the imaginary part odd with respect to $\omega$. 
It follows that 
\begin{align}
{\sigma}_1(\omega)
&=\Re(\tilde{\sigma}_1(\omega))-\Im(\tilde{\sigma}_2(\omega)) = \Re(\tilde{\sigma}_1(\omega))
\\
&=\frac{2 e^2 \hbar^3}{m_e^2 \Omega}
 \sum_\v{k} w_{\v{k}}
 \sum_{nn'}  
\frac{\Delta f_{n'\v{k},n\v{k}}}{\Delta\epsilon_{n\v{k},n'\v{k}}}
 \Re{(\bra{\Psi_{n\v{k}}}\nabla\ket{\Psi_{n'\v{k}}}\bra{\Psi_{n'\v{k}}}\nabla\ket{\Psi_{n\v{k}}})}
 \frac{\delta/2}{(\Delta\epsilon_{n\v{k},n'\v{k}}-\hbar\omega)^2+\delta^2/4}
 \label{eq:sigma1-solid}
\end{align}
and
\begin{align}
{\sigma}_2(\omega)
&=\Im(\tilde{\sigma}_1(\omega))+\Re(\tilde{\sigma}_2(\omega)) = \Re(\tilde{\sigma}_2(\omega))
\\
&=\frac{2 e^2 \hbar^3}{m_e^2 \Omega}
 \sum_\v{k} w_{\v{k}}
 \sum_{nn'}  
 \frac{\Delta f_{n'\v{k},n\v{k}}}{\Delta\epsilon_{n\v{k},n'\v{k}}}
 \Re{(\bra{\Psi_{n\v{k}}}\nabla\ket{\Psi_{n'\v{k}}}\bra{\Psi_{n'\v{k}}}\nabla\ket{\Psi_{n\v{k}}})}
 \frac{(\Delta\epsilon_{n\v{k},n'\v{k}}-\hbar\omega)}{(\Delta\epsilon_{n\v{k},n'\v{k}}-\hbar\omega)^2+\delta^2/4}  \; .
\end{align}
Sum rules also emerge, to wit
\begin{equation}
 \sum_\v{k} w_{\v{k}}
 \sum_{nn'}  
\frac{\Delta f_{n\v{k},n'\v{k}}}{\Delta\epsilon_{n\v{k},n'\v{k}}}
 \Im{(\bra{\Psi_{n\v{k}}}\nabla\ket{\Psi_{n'\v{k}}}\bra{\Psi_{n'\v{k}}}\nabla\ket{\Psi_{n\v{k}}})}
 \frac{\delta/2}{(\Delta\epsilon_{n\v{k},n'\v{k}}-\hbar\omega)^2+\delta^2/4}
 =0
\end{equation}
and
\begin{equation}
 \sum_\v{k} w_{\v{k}}
 \sum_{nn'}  
 \frac{\Delta f_{n\v{k},n'\v{k}}}{\Delta\epsilon_{n\v{k},n'\v{k}}}
 \Im{(\bra{\Psi_{n\v{k}}}\nabla\ket{\Psi_{n'\v{k}}}\bra{\Psi_{n'\v{k}}}\nabla\ket{\Psi_{n\v{k}}})}
 \frac{(\Delta\epsilon_{n\v{k},n'\v{k}}-\hbar\omega)}{(\Delta\epsilon_{n\v{k},n'\v{k}}-\hbar\omega)^2+\delta^2/4}=0.
\end{equation}
We return to them below.

In correspondence with the general KG formulae of the 
preceding section, for the solid we have the D-d form 
\begin{align}
{\sigma}_1^{D-d}(\omega)
=\frac{2\pi e^2 \hbar^3}{m_e^2 \Omega}
 \sum_\v{k} w_{\v{k}}
 \sum_{nn'}  
\frac{\Delta f_{n'\v{k},n\v{k}}}{\Delta\epsilon_{n\v{k},n'\v{k}}}
 \,\Re{(\bra{\Psi_{n\v{k}}}\nabla\ket{\Psi_{n'\v{k}}}\bra{\Psi_{n'\v{k}}}\nabla\ket{\Psi_{n\v{k}}})}
\,\delta( \Delta\epsilon_{n\v{k},n'\v{k}} - \hbar\omega)
 \label{eq:sigma1-solid-Dd}
\end{align}
and the approximated form 
\begin{align}
{\sigma}_1^{a}(\omega)
=\frac{2\pi e^2 \hbar^2}{m_e^2 \Omega \omega}
 \sum_\v{k} w_{\v{k}}
 \sum_{nn'}  
{\Delta f_{n'\v{k},n\v{k}}}
\,\Re{(\bra{\Psi_{n\v{k}}}\nabla\ket{\Psi_{n'\v{k}}}\bra{\Psi_{n'\v{k}}}\nabla\ket{\Psi_{n\v{k}}})}
\,\delta( \Delta\epsilon_{n\v{k},n'\v{k}} - \hbar\omega).
 \label{eq:sigma1-appr}
\end{align}

For calculations it may be numerically advantageous to enforce the even 
parity of $\sigma_1$ and use  
\begin{equation}
 \sigma_{1,calculated}(\omega) = \half\lbrack \sigma_1(\omega) + \sigma_1(-\omega)\rbrack \;.
\end{equation}

\subsubsection{\label{sec:sigma-contributions}Intra-band, degenerate state,
and inter-band contributions}
Practical use of the foregoing conductivity formulae requires  
resolution of the potential problems associated with 
$\Delta\epsilon_{n\v{k},n'\v{k}}$ going to zero. For that we return to 
Eq.\ (\ref{eq:sigma1-solid})
and separate the sums over band indices $n$ and $n'$ into one over $n=n'$, 
a second one for $n\neq n'$ and $\Delta\epsilon_{n\v{k},n'\v{k}}=0$, and a third 
sum for $n\neq n'$ and $\Delta\epsilon_{n\v{k},n'\v{k}} \ne 0$,
To treat the singularities in the first two sums, we add an
infinitesimal energy $\varepsilon$ and consider $\varepsilon \rightarrow 0$.
Details are
\begin{align}
 &\sum_{nn'}
 \frac{\Delta f_{n'\v{k},n\v{k}}}{\Delta\epsilon_{n\v{k},n'\v{k}}}
 \Re{(\bra{\Psi_{n\v{k}}}\nabla\ket{\Psi_{n'\v{k}}}\bra{\Psi_{n'\v{k}}}\nabla\ket{\Psi_{n\v{k}}})}
 \frac{\delta/2}{(\Delta\epsilon_{n\v{k},n'\v{k}}-\hbar\omega)^2+\delta^2/4}
\nonumber \\
= &  \lim_{\varepsilon\rightarrow 0}\sum_{n}
 \frac{f(\epsilon_{n\v{k}})-f(\epsilon_{n\v{k}}+\varepsilon)}{\varepsilon}
 {\Re(\bra{\Psi_{n\v{k}}}\nabla\ket{\Psi_{n\v{k}}}\bra{\Psi_{n\v{k}}}\nabla\ket{\Psi_{n\v{k}}})}
 \frac{\delta/2}{(\varepsilon-\hbar\omega)^2+\delta^2/4}
\nonumber\\
+
 &\lim_{\varepsilon\rightarrow 0}
 \sum_{\substack{ n\neq n' \\ \epsilon_{n\v{k}} = \epsilon_{n'\v{k}}}}
 \frac{f(\epsilon_{n\v{k}})-f(\epsilon_{n\v{k}}+\varepsilon)}{\varepsilon}
 {\Re(\bra{\Psi_{n\v{k}}}\nabla\ket{\Psi_{n'\v{k}}}\bra{\Psi_{n'\v{k}}}\nabla\ket{\Psi_{n\v{k}}})}
 \frac{\delta/2}{(\varepsilon-\hbar\omega)^2+\delta^2/4}
\nonumber\\
+
 &
\sum_{\substack{ n\neq n' \\ \epsilon_{n\v{k}}\neq \epsilon_{n'\v{k}}}}
 \frac{\Delta f_{n'\v{k},n\v{k}}}{\Delta\epsilon_{n\v{k},n'\v{k}}}
 {\Re(\bra{\Psi_{n\v{k}}}\nabla\ket{\Psi_{n'\v{k}}}\bra{\Psi_{n'\v{k}}}\nabla\ket{\Psi_{n\v{k}}})}
 \frac{\delta/2}{(\Delta\epsilon_{n\v{k},n'\v{k}}-\hbar\omega)^2+\delta^2/4} \;.
\end{align}
Taking the limits reduces the expression to 
\begin{align}
 &\sum_{nn'}
 \frac{\Delta f_{n'\v{k},n\v{k}}}{\Delta\epsilon_{n\v{k},n'\v{k}}}
 {\Re(\bra{\Psi_{n\v{k}}}\nabla\ket{\Psi_{n'\v{k}}}\bra{\Psi_{n'\v{k}}}\nabla\ket{\Psi_{n\v{k}}})}
 \frac{\delta/2}{(\Delta\epsilon_{n\v{k},n'\v{k}}-\hbar\omega)^2+\delta^2/4}
\nonumber \\
= &-\sum_{n}
 \frac{\partial f(\epsilon_{n\v{k}})}{\partial\epsilon_{n\v{k}}}
 {\Re(\bra{\Psi_{n\v{k}}}\nabla\ket{\Psi_{n\v{k}}}\bra{\Psi_{n\v{k}}}\nabla\ket{\Psi_{n\v{k}}})}
 \frac{\delta/2}{(\hbar\omega)^2+\delta^2/4}
\nonumber\\
- &\sum_{\substack{n \neq n'\\ \epsilon_{n\v{k}} = \epsilon_{n'\v{k}} }}
 \frac{ \partial f(\epsilon_{n\v{k}})}{\partial \epsilon_{n\v{k}}}
 {\Re(\bra{\Psi_{n\v{k}}}\nabla\ket{\Psi_{n'\v{k}}}\bra{\Psi_{n'\v{k}}}\nabla\ket{\Psi_{n\v{k}}})}
 \frac{\delta/2}{(\hbar\omega)^2+\delta^2/4}
\nonumber\\
+ &\sum_{\substack{ n\neq n' \\ \epsilon_{n\v{k}}\neq \epsilon_{n'\v{k}}}}
 \frac{\Delta f_{n'\v{k},n\v{k}}}{\Delta\epsilon_{n\v{k},n'\v{k}}}
 {\Re(\bra{\Psi_{n\v{k}}}\nabla\ket{\Psi_{n'\v{k}}}\bra{\Psi_{n'\v{k}}}\nabla\ket{\Psi_{n\v{k}}})}
 \frac{\delta/2}{(\Delta\epsilon_{n\v{k},n'\v{k}}-\hbar\omega)^2+\delta^2/4} \; .
 \end{align}
The result is 
\begin{align}
{\sigma}_1(\omega)=-\frac{2 e^2 \hbar^3}{m_e^2 \Omega}
 \sum_\v{k} w_{\v{k}}
\Big[
&\sum_{n}
 \frac{\partial f(\epsilon_{n\v{k}})}{\partial\epsilon_{n\v{k}}}
 {\Re(\bra{\Psi_{n\v{k}}}\nabla\ket{\Psi_{n\v{k}}}\bra{\Psi_{n\v{k}}}\nabla\ket{\Psi_{n\v{k}}})}
 \frac{\delta/2}{(\hbar\omega)^2+\delta^2/4}
\nonumber \\
 &+\sum_{\substack{n \neq n'\\ \epsilon_{n\v{k}} = \epsilon_{n'\v{k}} }}
 \frac{ \partial f(\epsilon_{n\v{k}})}{\partial \epsilon_{n\v{k}}}
 {\Re(\bra{\Psi_{n\v{k}}}\nabla\ket{\Psi_{n'\v{k}}}\bra{\Psi_{n'\v{k}}}\nabla\ket{\Psi_{n\v{k}}})}
 \frac{\delta/2}{(\hbar\omega)^2+\delta^2/4}
\nonumber\\
-&  \sum_{\substack{ n\neq n' \\ \epsilon_{n\v{k}}\neq \epsilon_{n'\v{k}}}}
 \frac{\Delta f_{n'\v{k},n\v{k}}}{\Delta\epsilon_{n\v{k},n'\v{k}}}
 {\Re(\bra{\Psi_{n\v{k}}}\nabla\ket{\Psi_{n'\v{k}}}\bra{\Psi_{n'\v{k}}}\nabla\ket{\Psi_{n\v{k}}})}
 \frac{\delta/2}{(\Delta\epsilon_{n\v{k},n'\v{k}}-\hbar\omega)^2+\delta^2/4}
\Big].
\label{eq:sigma1-full}
\end{align}
Similarly for $\sigma_2$ we have
\begin{align}
{\sigma}_2(\omega)=-\frac{2 e^2 \hbar^3}{m_e^2 \Omega}
 \sum_\v{k} w_{\v{k}}
\Big[ 
&\sum_{n}
 \frac{\partial f(\epsilon_{n\v{k}})}{\partial\epsilon_{n\v{k}}}
 {\Re(\bra{\Psi_{n\v{k}}}\nabla\ket{\Psi_{n\v{k}}}\bra{\Psi_{n\v{k}}}\nabla\ket{\Psi_{n\v{k}}})}
 \frac{\hbar\omega}{(\hbar\omega)^2+\delta^2/4}
\nonumber \\
+ &\sum_{\substack{n \neq n'\\ \epsilon_{n\v{k}} = \epsilon_{n'\v{k}} }}
 \frac{ \partial f(\epsilon_{n\v{k}})}{\partial \epsilon_{n\v{k}}}
 {\Re(\bra{\Psi_{n\v{k}}}\nabla\ket{\Psi_{n'\v{k}}}\bra{\Psi_{n'\v{k}}}\nabla\ket{\Psi_{n\v{k}}})}
 \frac{\hbar\omega}{(\hbar\omega)^2+\delta^2/4}
\nonumber \\
-&
 \sum_{\substack{ n\neq n' \\ \epsilon_{n\v{k}}\neq \epsilon_{n'\v{k}}}}
 \frac{\Delta f_{n'\v{k},n\v{k}}}{\Delta\epsilon_{n\v{k},n'\v{k}}}
 {\Re(\bra{\Psi_{n\v{k}}}\nabla\ket{\Psi_{n'\v{k}}}\bra{\Psi_{n'\v{k}}}\nabla\ket{\Psi_{n\v{k}}})}
 \frac{(\Delta\epsilon_{n\v{k},n'\v{k}}-\hbar\omega)}{(\Delta\epsilon_{n\v{k},n'\v{k}}-\hbar\omega)^2+\delta^2/4}
\Big].
\label{eq:sigma2-full}
\end{align}

The occupation number derivatives $\partial
f(\epsilon_{n\v{k}})/\partial \epsilon_{n\v{k}}$ have been discussed
in the closely related setting of density functional perturbation
theory \cite{RevModPhys.73.515} and in 
 consideration of intra-band
contributions in the KG context \cite{Allen2006}.
So far as we can 
tell, a full treatment for the KG formalism leading to
the appearance of such derivatives from both intra-band transitions
and from degeneracies 
has not been presented. Note that there has been work on
deriving the intra-band contributions using a 
band dispersion linearization technique \cite{PhysRevB.82.035104}.

\subsubsection{Drude and dc components}
A brief detour is useful.  If the inter-band, non-degenerate contribution 
is negligible for small $\omega$, then only the first two sums in 
\eqnref{eq:sigma1-full} contribute to the total and therefore we can write
\begin{align}
\sigma_{1}^D(\omega)&=
-\frac{2 e^2 \hbar^3}{m_e^2 \Omega}
 \frac{\delta/2}{(\hbar\omega)^2+\delta^2/4}  \nonumber \\
&\times
 \sum_\v{k} w_{\v{k}} \Big[ \sum_{n}
 \frac{\partial f(\epsilon_{n\v{k}})}{\partial\epsilon_{n\v{k}}}
 \Re{(\bra{\Psi_{n\v{k}}}\nabla\ket{\Psi_{n\v{k}}}\bra{\Psi_{n\v{k}}}\nabla\ket{\Psi_{n\v{k}}})}
\nonumber\\
+ &\sum_{n \neq n'}  \delta_{\epsilon_{n\v{k}}  \epsilon_{n'\v{k}}}
 \frac{ \partial f(\epsilon_{n\v{k}})}{\partial \epsilon_{n\v{k}}}
 \Re{(\bra{\Psi_{n\v{k}}}\nabla\ket{\Psi_{n'\v{k}}}\bra{\Psi_{n'\v{k}}}\nabla\ket{\Psi_{n\v{k}}})}
\Big] \; .
\label{DrudeEquiv}
\end{align}
If we identify the average inter-collision time as 
\beq
\tau=2\hbar/\delta
\eeq
and the effective charge-to-mass ratio as 
\begin{align}
\left(\frac{n_e}{m_e}\right)_{eff} &= %
-\frac{2 e^2 \hbar^2}{m_e^2 \Omega} %
 \sum_\v{k} w_{\v{k}} \left\lbrack \sum_{n} \frac{\partial f(\epsilon_{n\v{k}})}{\partial\epsilon_{n\v{k}}} %
 \Re{(\bra{\Psi_{n\v{k}}}\nabla\ket{\Psi_{n\v{k}}}\bra{\Psi_{n\v{k}}}\nabla\ket{\Psi_{n\v{k}}})} \right. \nonumber \\
 &+ \left. \sum_{yn \neq n'}  \delta_{\epsilon_{n\v{k}}  \epsilon_{n'\v{k}}} %
 \frac{ \partial f(\epsilon_{n\v{k}})}{\partial \epsilon_{n\v{k}}} %
 \Re{(\bra{\Psi_{n\v{k}}}\nabla\ket{\Psi_{n'\v{k}}}\bra{\Psi_{n'\v{k}}}\nabla\ket{\Psi_{n\v{k}}})} \right\rbrack \;,
\end{align}
then Eq.\ (\ref{DrudeEquiv}) becomes the Drude expression 
\cite{ANDP:ANDP19003060312,Allen2006}
\beq
 \sigma_{1}^D(\omega)=
 \frac{\left(\frac{n_e}{m_e}\right)_{eff}\tau}{1+(\omega\tau)^2}.
\eeq

The limit $\omega\rightarrow 0$ yields the direct current (dc) 
conductivity tensor in the Drude approximation
\begin{align}
 \sigma_{dc}^D=
-\frac{2 e^2 \hbar^2 \tau }{m_e^2 \Omega}
 \sum_\v{k} w_{\v{k}}
\Big[&\sum_{n}
 \frac{\partial f(\epsilon_{n\v{k}})}{\partial\epsilon_{n\v{k}}}
 \Re{(\bra{\Psi_{n\v{k}}}\nabla\ket{\Psi_{n\v{k}}}\bra{\Psi_{n\v{k}}}\nabla\ket{\Psi_{n\v{k}}})}
\nonumber \\
+ &\sum_{n \neq n'}  \delta_{\epsilon_{n\v{k}}  \epsilon_{n'\v{k}}}
 \frac{ \partial f(\epsilon_{n\v{k}})}{\partial \epsilon_{n\v{k}}}
 \Re{(\bra{\Psi_{n\v{k}}}\nabla\ket{\Psi_{n'\v{k}}}\bra{\Psi_{n'\v{k}}}\nabla\ket{\Psi_{n\v{k}}})}
\Big].
\label{eq:sigmaddc}
\end{align}

\subsubsection{Exact dc component} 
Without invoking the Drude approximation, simply taking the limit 
$\omega \rightarrow 0$ in \eqnref{eq:sigma1-full} gives 
\begin{align}
{\sigma}_{dc}=-\frac{ 2 e^2 \hbar^3}{m_e^2 \Omega} %
 & \sum_\v{k} w_{\v{k}} \Big[ %
 \frac{2}{\delta}  \sum_{n} %
 \frac{\partial f(\epsilon_{n\v{k}})}{\partial\epsilon_{n\v{k}}} %
 \Re{(\bra{\Psi_{n\v{k}}}\nabla\ket{\Psi_{n\v{k}}}\bra{\Psi_{n\v{k}}}\nabla\ket{\Psi_{n\v{k}}})}
\nonumber \\
 +& 
 \frac{2}{\delta}
 \sum_{\substack{n \neq n'\\ \epsilon_{n\v{k}} = \epsilon_{n'\v{k}} }}
 \frac{ \partial f(\epsilon_{n\v{k}})}{\partial \epsilon_{n\v{k}}}
 \Re{(\bra{\Psi_{n\v{k}}}\nabla\ket{\Psi_{n'\v{k}}}\bra{\Psi_{n'\v{k}}}\nabla\ket{\Psi_{n\v{k}}})}
\nonumber \\
-&
 \sum_{\substack{ n\neq n' \\ \epsilon_{n\v{k}}\neq \epsilon_{n'\v{k}}}}
 \frac{\Delta f_{n'\v{k},n\v{k}}}{\Delta\epsilon_{n\v{k},n'\v{k}}}
 \Re{(\bra{\Psi_{n\v{k}}}\nabla\ket{\Psi_{n'\v{k}}}\bra{\Psi_{n'\v{k}}}\nabla\ket{\Psi_{n\v{k}}})}
 \frac{\delta/2}{(\Delta\epsilon_{n\v{k},n'\v{k}})^2+\delta^2/4}
\Big] \; .
\label{eq:sigma_dc}
\end{align}
This expression includes all possible contributions to the dc conductivity, 
in contrast with \eqnref{eq:sigmaddc}, which
omits the non-degenerate inter-band contributions. 

\subsection{Sum rules}
Clearly a key ingredient in the KG conductivity is the set of
gradient operator matrix elements.  Computing them is a 
seemingly simple task that can be complicated by procedures 
(e.g. PAWs; see below) used in the underlying KS calculations.
Knowledge of the exact behavior of matrix  
element sums therefore has been used to test 
both implementations and calculations. 
Such sum rules are developed in this section 
and discussed in terms of 
their use as possible quality measures of an implementation or 
accuracy measures of results. 
  
\subsubsection{\label{sec:sum-rule-r}Sum rule in terms of $\hat{\v{r}}$ }

A seemingly round-about but fruitful way to begin is to use the    
commutator relation for the Cartesian component $\alpha$ of the position 
operator with the Hamiltonian $\hat H$
\begin{equation}
 [\hat{{r}}_\alpha, \hat H ]=i\frac{\hbar}{m_e} \hat{{p}}_\alpha. 
\label{eq:rh-commutator}
 \end{equation}
Then for the double commutator 
we have 
\begin{equation}
 [\hat{{r}}_\alpha, [\hat{{r}}_\alpha, \hat H ]]=i\frac{\hbar}{m_e} [\hat{{r}}_\alpha,\hat{{p}}_\alpha]=-\frac{\hbar^2}{m_e}.
\label{eq:rh-double-commutator}
\end{equation}

Formation of matrix elements of \eqnref{eq:rh-double-commutator} 
taken with $\bra{m}$ from 
the left and $\ket{n}$ from the right and use of the completeness 
relation $\sum_{m'}\ket{m'}\bra{m'}=\hat{I}$ gives 
\begin{equation}
\sum_{m'} \left(\bra{m} \hat{{r}}_\alpha \ket{m'}\bra{m'}[\hat{{r}}_\alpha, \hat H ]\ket{n}-\bra{m} [\hat{{r}}_\alpha, \hat H ] \ket{m'}\bra{m'} \hat{{r}}_\alpha \ket{n}\right)=-\frac{\hbar^2}{m_e}\spr{m}{n}\; .
\end{equation}
This reduces to
the general sum rule for each Cartesian 
component of $\hat{\v{r}}$
\begin{equation}
\sum_{m'} (  2\epsilon_{m'} -\epsilon_n - \epsilon_m  ) \bra{m} \hat{{r}}_\alpha \ket{m'} \bra{m'}\hat{{r}}_\alpha\ket{n}=\frac{\hbar^2}{m_e}\delta_{mn}.
\label{eq:x-sumrule-general}
\end{equation}
In particular, for $m=n$ we have the sum rule
\begin{equation}
2 \sum_{m'} ( \epsilon_{m'} -\epsilon_m ) |\bra{m'}\hat{{r}}_\alpha\ket{m}|^2=\frac{\hbar^2}{m_e}.
\label{eq:x-sumrule-nn1}
\end{equation}
or
\begin{equation}
2 \sum_{m'(\ne m)} ( \epsilon_{m'} -\epsilon_m ) |\bra{m'}\hat{{r}}_\alpha\ket{m}|^2=\frac{\hbar^2}{m_e} \; .
\label{eq:x-sumrule-nn2}
\end{equation}
Notice that there is no contribution in 
\eqnref{eq:x-sumrule-general} from states with
$\epsilon_{m'}=\epsilon_{m}$ or $\epsilon_{m'}=\epsilon_n$. That is, there
are neither self-contributions nor degenerate-state contributions.

\subsubsection{Sum rule in terms of $\hat{\v{p}}$}

Formation of matrix elements of \eqnref{eq:rh-commutator} from the left with 
$\bra{n}$ and $\ket{m}$ from the right yields
\begin{equation}
\bra{n}\hat{{r}}_\alpha\ket{m}=-i\frac{\hbar}{m_e}\frac{\bra{n}\hat{{p}}_\alpha\ket{m}}{(\epsilon_n-\epsilon_m)} 
\end{equation}
for 
the Cartesian components of $\hat{\v{p}}$ in the case $n \neq\ m$ and $\epsilon_{m} \neq \epsilon_{n}$. Since the result in 
\eqnref{eq:x-sumrule-nn2} is commensurate with that 
exclusion, simple  substitution yields 
\begin{equation}
\frac{2}{m_e} \sum_{\substack{m'(\neq n) \\ \epsilon_{m'} \neq \epsilon_n}}  
\frac{|\bra{m'}\hat{{p}}_\alpha\ket{n}|^2}{(\epsilon_{m'}-\epsilon_{n})}=1 \; .
\label{eq:p-sumrule-nn}
\end{equation}

\subsubsection{Sum rule involving occupation numbers}
Multiplication of \eqnref{eq:p-sumrule-nn} by the net occupation 
number of state $n$ and 
summation  over all states gives 
\begin{equation}
\frac{4}{m_e} \sum_{m=1}^\infty  \sum_{\substack{ n=1\\ (n\neq m)\\\epsilon_{m}\neq \epsilon_{n}}}^\infty f(\epsilon_n)
\frac{|\bra{m}\hat{{p}}_\alpha\ket{n}|^2}{(\epsilon_{m}-\epsilon_{n})}=2\sum_{n=1}^\infty f(\epsilon_n)=N_e,
\label{eq:p-sumrule-on}
\end{equation}
where 
$N_e$ is the total number of electrons. The left-hand side  can be written as 
the sum of two terms that are identical 
except for exchange of the summation indices in one of them:
\begin{equation}
2 \sum_{m=1}^\infty  \sum_{\substack{n=1 \\ (n\neq m)\\\epsilon_{m}\neq \epsilon_{n}}}^\infty f(\epsilon_n)
\frac{|\bra{m}\hat{{p}}_\alpha\ket{n}|^2}{(\epsilon_{m}-\epsilon_{n})}
+
2\sum_{n=1}^\infty  \sum_{\substack{m=1 \\ (m\neq n)\\\epsilon_{n}\neq \epsilon_{m}}}^\infty f(\epsilon_m)
\frac{|\bra{n}\hat{{p}}_\alpha\ket{m}|^2}{(\epsilon_{n}-\epsilon_{m})}
=m_e N_e \; .
\label{eq:p-sumrule-on-1}
\end{equation}
Thus one has the sum rule in terms of all the
occupation numbers and states,
\begin{equation}
S_f=\frac{2}{3 m_e N_e} \sum_{\alpha=1}^3 \sum_{m=1}^\infty  
\sum_{\substack{n=1 \\ (n\neq m) \\ \epsilon_{n}\neq\epsilon_{m}}}^\infty 
(f(\epsilon_n)-f(\epsilon_m))
\frac{|\bra{m}\hat{{p}}_\alpha\ket{n}|^2}{(\epsilon_{m}-\epsilon_{n})}
=1 \; .
\label{eq:sumrule-f}
\end{equation}

\subsubsection{Sum rule for the conductivity}
By introduction of a Dirac $\delta$-function, \eqnref{eq:sumrule-f} 
can be rewritten as 
\begin{equation}
S=
\frac{2}{3 m_e N_e} \int_{-\infty}^{\infty}d\omega
\sum_{\alpha=1}^3
\sum_{m=1}^\infty 
\sum_{\substack{n=1 \\ n\neq m\\\epsilon_{n}\neq\epsilon_{m}}}^\infty 
(f(\epsilon_n)-f(\epsilon_m))
{|\bra{m}\hat{{p}}_\alpha\ket{n}|^2}
\frac{\delta(\epsilon_{m}-\epsilon_{n}-\hbar\omega)}{\omega}
=1  \; .
\label{eq:p-sumrule-freq}
\end{equation}
This is the frequency sum rule. In terms of 
the trace of the conductivity tensor 
(\eqnref{eq:sigma1-general-2b}), it translates to
\begin{equation}
S_\omega=
\frac{ 2 m_e V}{3 \pi e^2 N_e}
\int_{0}^{\infty}d\omega \,Tr(\sigma_1(\omega))=1,
\label{eq:p-sumrule-cond-2}
\end{equation}
after taking into account that $\sigma_1$ is even.

However there is a problem.  The derivation of  
\eqnref{eq:p-sumrule-freq} specifically excludes 
contributions from states with 
the same labels and from degenerate states
(\secref{sec:sum-rule-r}). But we have also shown that
$\sigma_1(\omega)$ has both intra-band and degenerate-state 
contributions
(\secref{sec:sigma-contributions}).  Therefore,
\eqnref{eq:p-sumrule-cond-2} is valid only if the intra-band and
degenerate-state contributions are negligible. 
If they are not, then they always give a 
positive contribution to 
the integral in \eqnref{eq:p-sumrule-cond-2}. Therefore the general condition
in the limit $\delta\rightarrow 0$ is 
\begin{equation}
S_\omega=
\frac{ 2 m_e V}{3 \pi e^2 N_e}
\int_{0}^{\infty}d\omega \,Tr(\sigma_1(\omega))\ge 1 \;. 
\label{eq:p-sumrule-cond-3}
\end{equation}
The larger the difference of $S_\omega$ from one, the larger 
will be the intra-band and degenerate-state contributions to the conductivity.

Finally, to get the sum rules for solids, do all the following in the sum rule
of interest: replace  $\sum_{mm'}$ by
$\sum_{\v{k}}w_{k}\sum_{nn'}$, replace the spatial volume $V$ by
the unit cell volume  $\Omega$, and take $N_e$ to be the number of
electrons per unit cell.

\subsection{\label{sec:sum-rules-states}Sum rules for finite number of states}

The assumption of a complete set of states 
was instrumental to the
sum rule derivations.  For a finite set of states 
those rules break down, as can be seen just by
evaluating the left-hand side of \eqnref{eq:p-sumrule-nn} at the 
highest energy state in a finite set.  The resulting 
sum is strictly negative, hence cannot be equal to unity.  

The problem appears as an incomplete sum for \eqnref{eq:sumrule-f}.
To assist in the analysis, introduce the dimensionless variable
$
 x\equiv \beta (\epsilon - \epsilon_F) 
$
with $\epsilon_F$ as the Fermi energy,
and make the corresponding F-D occupation definition
\beq
f(\epsilon_m;\beta) = 1/\lbrack(\exp\beta(\epsilon_m-\epsilon_F) + 1\rbrack %
\rightarrow {\tilde f}(x) := 1/\lbrack(\exp(x) + 1\rbrack \label{eq:ftildedefn} \; .
 \eeq
Then the relevant ratio becomes 
\begin{equation}
\frac{f(\epsilon_n;\beta)-f(\epsilon_m;\beta)}{(\epsilon_{m}-\epsilon_{n})} %
= \beta \frac{{\tilde f}(x_n)-{\tilde f}(x_m)}{(x_{m}-x_{n})} = %
\beta\frac{\Delta {\tilde f}}{\Delta x}(x_m,x_n) \; .
\label{eq:DfDx}
\end{equation}
and 
$S_f$ in terms of dimensionless variables  
is
\begin{equation}
S_f=\frac{2\beta}{3 m_e N_e} \sum_{\alpha=1}^3 \sum_{m=1}^\infty  
\sum_{\substack{n=1 \\ (n\neq m) \\ x_{n}\neq x_{m}}}^\infty 
\frac{\Delta {\tilde f}}{\Delta x}(x_m,x_n)
{|\bra{m}\hat{{p}}_\alpha\ket{n}|^2}
=1 \; .
\label{eq:sumrule-f-DfDx}
\end{equation}

\begin{figure}
\includegraphics[width=\textwidth,viewport=1 1 710 495,clip]{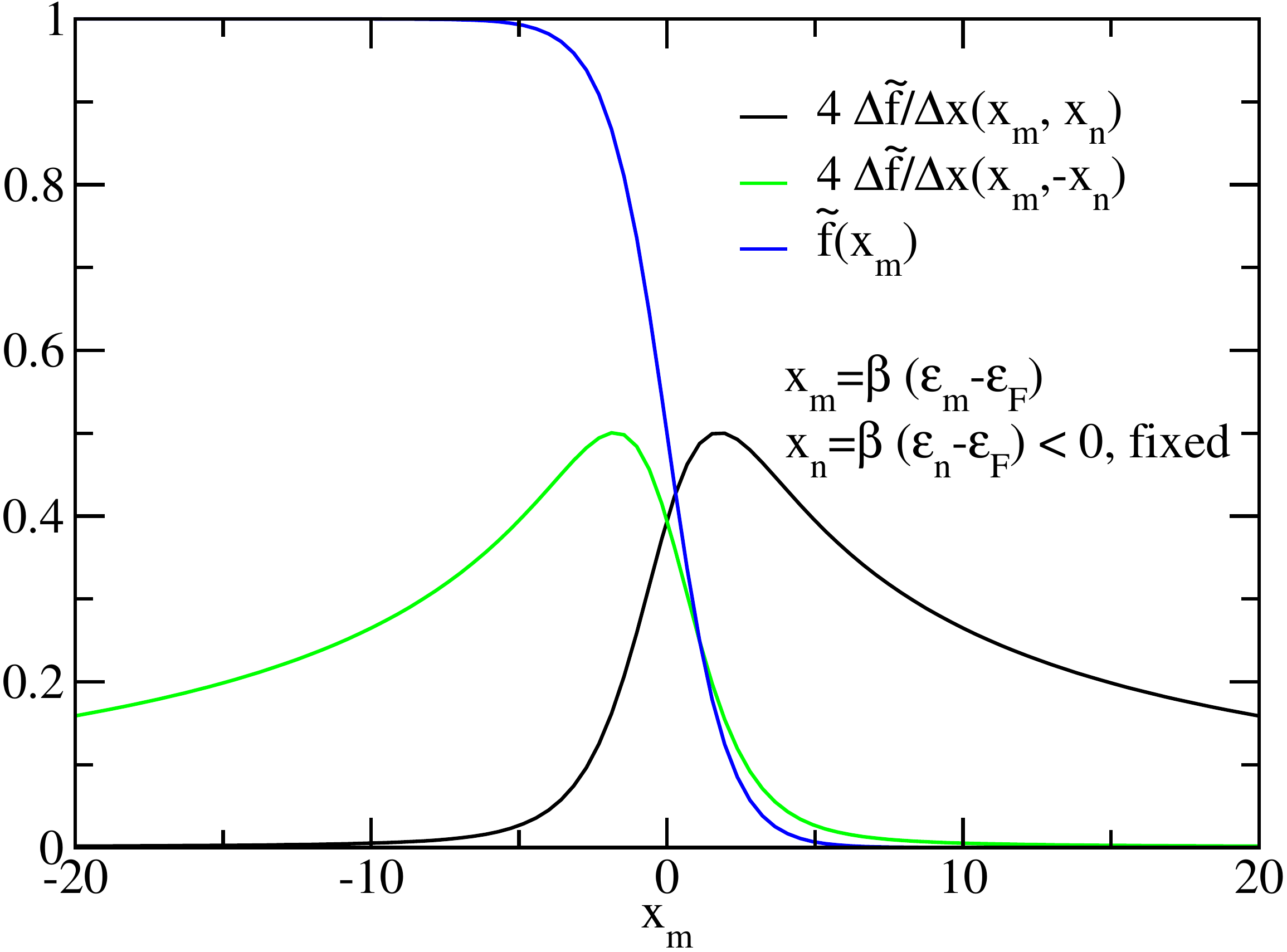}
\caption{\label{fig:dfdx} Behavior of \eqnref{eq:DfDx} as a function of 
$x_m$ for a fixed negative value of 
$x_n$, ($x_n = -5$) and for $-x_n$. Note that those two plots are 
magnified by a factor of 4 for clarity. The Fermi-Dirac distribution 
$f (x_m )$ as a function of the scaled variable $x_m$ also is shown.}
\end{figure}

\figref{fig:dfdx} shows the behavior of \eqnref{eq:DfDx}, divided by $\beta$, as a function 
of $x_m$ for a fixed negative value of  $x_n$ and for the symmetric case 
$-x_n$. We use $x_n = -5$. (Note the magnification in the figure.)  Observe 
that negative (positive) $x_m$ represent states with energies below (above)
$\epsilon_F$. The graph also depicts the  
Fermi-Dirac distribution as a function of the scaled variable $x_m$.  From it 
one sees that $x_m = 10$ ($\tilde f(10) =4.54 \times 10^{-5}$) 
is a reasonable maximum value for 
purposes of analysis. But, as also shown in 
\figref{fig:dfdx}, 
\eqnref{eq:DfDx} evaluated at negative $x_n$ has a significant contribution 
to the sums in \eqref{eq:sumrule-f-DfDx} for $x_m>10$.  Therefore 
if the sums were to be truncated at $x_m= x_n=10$,
$S_f$ would be incomplete and consequently less than unity.

In addition, the contributions of intra-band
transitions and degenerate states make  $S_w$ differ from unity. 
Only in the limits of large numbers of k-points, bands and a large
frequency interval will $S_w \rightarrow 1$, if there are only
non-degenerate inter-band contributions. If there are also 
intra-band or degenerate contributions, it will go to some value
greater than one.  However, the conductivity may reach convergence
over the entire frequency interval of interest long before $S_w$ reaches
convergence.  Conversely, the value of $S_w$ could be around one or
greater, depending on the afore-mentioned contributions, for a 
particular  set of
k-points and number of bands, but that does not mean that $S_w$ is
converged and therefore that the conductivity is as well.

In consequence, convergence analysis with respect to the number of 
k-points and bands of the calculated 
conductivity itself over the frequency interval of interest is unavoidable.

\section{\label{sec:paw}Projector augmented wave  method}
Ordinarily the KS equations are solved by expanding the KS orbitals
in a basis.  A PW basis commonly is used both because 
the orbitals of simple metals resemble PWs and, more critically, 
because they are not centered on nuclear sites. Site-independence
simplifies the use of KS DFT to drive ab initio molecular dynamics \cite{MarxHutter2000,Tse2002,MarxHutter2009}.

However, reproduction of the rapid oscillation of the 
KS orbitals near a nucleus would require an 
impracticably large PW basis. Conventionally that difficulty
was alleviated by use of pseudo-potentials, 
but it was really solved, at least in principle, by the 
introduction of the PAW method \cite{PAW}. Another significant advantage 
is that, distinct from pseudopotentials, the PAW approach allows for a 
significant simplification 
of the matrix elements of the operators while retaining 
the effect of core electrons.

The PAW method is based on the construction of a linear transformation 
which connects each KS orbital $\ket{\Psi}$ with a corresponding,
much smoother pseudo-orbital $\ket{\tilde\Psi}$, that is
\begin{equation}
 \ket{\Psi}=\ket{\tilde{\Psi}}+
 \sum_i \left[ \ket{\phi_i}
-\ket{\tilde{\phi_i}}\right] \spr{{\tilde{p_i}}}{{\tilde{\Psi}}} \; . 
\end{equation}
The set $\{\ket{\phi}\}$ is an orthonormal basis, 
while the sets $\{\ket{\tilde\phi}\}$ and $\{\ket{\tilde p}\}$ form 
a dual basis.
That is, besides the orthonormality and completeness conditions for the 
set $\ket{\phi}$s, one also has the duality conditions of completeness
\begin{equation}
 \sum_i \ket{\tilde{\phi}_i}\bra{\tilde{p}_i} =1
\end{equation}
and orthonormality
\begin{equation}
 \spr{\tilde{p}_i}{\tilde{\phi}_j}=\delta_{ij},
 \label{eq:dual-complete}
\end{equation}
between the other two sets.  Physically, the set  $\{\ket{\tilde\phi}\}$ is
to be smoothed relative to the set  $\{\ket{\phi}\}$, hence amenable to
efficient plane-wave expansion.  

The transformation connecting $\ket{\Psi}$ and $\ket{\tilde\Psi}$ is unitary and therefore any operator $A$ can be transformed to its smoothed 
version $\tilde A$ according to
\begin{align}
\tilde{A}=
A
+ \sum_{ij} \ket{\tilde{p_i}} ( \bra{\phi_i} A \ket{\phi_j}
-\bra{\tilde{\phi}_i} A \ket{\tilde{\phi}_j}   ) \bra{\tilde{p_j}}.
\label{eq:blochtransform}
\end{align}

In practice the $\ket{\phi_i}$s are taken as ground state atomic
orbitals of a chemical element augmented with other eigenfunctions 
of the same Hamiltonian operator.  The $\ket{\tilde\phi_i}$s
are pseudized forms of the corresponding $\ket{\phi_i}$s.   
The $\ket{\tilde p_i}$s are defined as zero outside a sphere centered 
at the atom (augmentation sphere) and constructed to be the dual basis 
to the pseudized set inside the augmentation sphere.  On the assumption
that there is no overlap between augmentation spheres, the sum in 
\eqnref{eq:blochtransform} reduces from pairwise to a single atom.
That is the so-called one-center
approximation.  It requires computational care to ensure negligible overlap
of augmentation spheres in practice.

\subsection{\label{sec:grad-matrix-elements}The $\bra{\Psi_{n\v{k}}} \nabla \ket{\Psi_{n'\v{k}}}$ matrix elements }
Matrix elements of the velocity operator in the PAW representation follow
from \eqnref{eq:blochtransform} as
\begin{align}
\bra{\Psi_{n\v{k}}} \nabla \ket{\Psi_{n^\prime \v{k}}}
=&
\bra{\tilde\Psi_{n\v{k}}} \nabla \ket{\tilde\Psi_{n'\v{k}}} + 
\nonumber\\
&
+ \sum_{i}\sum_{ \ell m } \sum_{ \ell^\prime m^\prime } 
\spr{\tilde\Psi_{n\v{k}}}{\tilde{p}_{i \ell m }} 
\left[ 
\bra{\varphi_{i \ell m}} \nabla \ket{\varphi_{i \ell^\prime m^\prime}}
-\bra{\tilde{\varphi}_{i\ell m}} \nabla \ket{\tilde{\varphi}_{i\ell^\prime m^\prime}} 
\right ]
\spr{\tilde{p}_{i\ell^\prime m^\prime}}{\tilde\Psi_{n^\prime \v{k}}},
\label{eq:blochtransform2}
\end{align}
with the atomic orbitals $\ket{\varphi_{i\ell m}}$, pseudo-orbitals 
$\ket{\tilde{\varphi}_{i \ell m}}$, and projectors $\ket{\tilde{p}_{i \ell m}}$ 
of atom $i$ (and associated augmentation region).  Those are  
defined in terms of products of 
radial functions and 
spherical harmonics  $Y_{\ell m}(\theta,\phi)$ (see \refapp{app:csh}) as
\begin{align}
 \varphi_{i\ell m}(\v{r}-\v{R}_i)= &R_{i\ell}(|\v{r}-\v{R}_i|) Y_{\ell m}(\theta,\phi),
\\
 \tilde{\varphi}_{ilm}(\v{r}-\v{R}_i)= &\tilde{R}_{i\ell}(|\v{r}-\v{R}_i|) Y_{\ell m}(\theta,\phi), \\
 \tilde{p}_{i\ell m}(\v{r}-\v{R}_i)&= \tilde{p}_{i\ell}(|\v{r}-\v{R}_i|) Y_{\ell m}(\theta,\phi) \; .
  \end{align}
The one oddity (anticipating the practice in Quantum Espresso \cite{QE2009}) is that the principal quantum number is suppressed.  One may think of the
atom index $i$ as being a compound of site and principal quantum number.  
In compressed notation  
\begin{align}
\tilde{\nabla}_{nn'}^{\v{k}}\equiv& \bra{\tilde\Psi_{n\v{k}}} \nabla \ket{\tilde\Psi_{n'\v{k}}}\; , \\
\gamma_{i l m n \v{k}}^\dag \equiv&\spr{\tilde\Psi_{n\v{k}}}{\tilde{p}_{ilm}} \; , \\
\nabla_{i l m l' m'} \equiv & \bra{\varphi_{ilm}} \nabla \ket{\varphi_{i l' m'}}\; , \\
\tilde{\nabla}_{i l m l' m'} \equiv & \bra{\tilde{\varphi}_{ilm}} \nabla \ket{\tilde{\varphi}_{il'm'}}\; ,
\end{align}
\eqnref{eq:blochtransform2} becomes 
\begin{align}
\nabla_{nn'}^{\v{k}}\equiv
\bra{\Psi_{n\v{k}}} \nabla \ket{\Psi_{n'\v{k}}}
=
\tilde{\nabla}_{nn'}^{\v{k}}
+ \sum_{i}\sum_{lm}\sum_{l'm'} 
\gamma_{i l m n\v{k}}^\dag
\left[ 
\nabla_{i l m l' m'}
-\tilde{\nabla}_{i l m l' m'} 
\right ]
\gamma_{i l' m' n' \v{k}}  \;. 
\label{eq:blochtransform-2}
\end{align}
The task is to find expressions for all the foregoing matrix elements.

It is straightforward to prove that
\begin{align}
\tilde{\nabla}_{nn'}^{\v{k}}= i \sum_{\v{G}} C_{n\v{k}\v{G}}^*\, C_{n'\v{k}\v{G}} \, (\v{k}+\v{G});
\end{align}
For $\nabla_{i\ell m \ell^\prime m^\prime}$ we have 
\begin{align}
\nabla_{ilm l'm'} 
=& \intdr \varphi^*_{ilm}(\v{r}-\v{R}_i) \nabla  \varphi_{il'm'}(\v{r}-\v{R}_{i})
\nonumber\\
=& \intdr \varphi^*_{ilm}(\v{r}) \nabla  \varphi_{il'm'}(\v{r})
\end{align}
where 
\begin{align}
 \nabla  \varphi_{i l'm'}(\v{r}) 
=&
\frac{d R_{l'}({r}) }{d r} Y_{l' m'}(\theta,\varphi )       
\hat{ \v{e} }_r(\theta,\varphi)  
+
\frac{R_{l'}({r})}{r} 
\left[ 
\frac{\partial Y_{l'm'}( \theta, \varphi )}{\partial \theta} 
\hat{\v{e}}_\theta(\theta,\varphi) \nonumber\right. \\
& \left. +
\frac{1}{\sin{\theta}}
\frac{\partial Y_{l'm'}( \theta, \varphi )}{\partial \varphi}  \
\hat{\v{e}}_\varphi(\theta,\varphi) 
\right] \;.
\end{align}
Therefore,
\begin{align}
\nabla_{ilm l'm'} 
=& 
\underbrace{\int_0^\infty r^2 dr R^*_{l}(r) \frac{d R_{l'}({r}) }{d r}}_{R^{(d)}_{ll'}} 
\underbrace{\int_0^{\pi} \sin(\theta) d\theta \int_0^{2\pi} d\varphi Y^*_{lm}(\theta,\varphi)  Y_{l' m'}(\theta,\varphi )
\hat{ \v{e} }_r(\theta,\varphi)
}_{=\v{I}^{(r)}_{lml'm'}}   
\nonumber\\
+&
\underbrace{\int_0^\infty r dr R^*_l(r)R_{l'}({r})}_{R_{ll'}} 
\big[ 
\underbrace{\int_0^{\pi} \sin(\theta) d\theta \int_0^{2\pi} d\varphi Y^*_{lm}(\theta,\varphi ) 
\frac{\partial Y_{l'm'}( \theta, \varphi )}{\partial \theta} 
\hat{\v{e}}_\theta(\theta,\varphi) 
}_{=\v{I}^{(\theta)}_{lml'm'}}
\nonumber\\
+&
\underbrace{\int_0^{\pi} d\theta \int_0^{2\pi} d\varphi Y^*_{lm}(\theta,\varphi ) \frac{\partial Y_{l'm'}( \theta, \varphi )}{\partial \varphi}
   \hat{\v{e}}_\varphi(\theta,\varphi) 
}
_{= \v{I}^{(\varphi)}_{lml'm}}
\big],
\end{align}
or
\begin{align}
\nabla_{ilm l'm'} 
=&
R^{(d)}_{ll'} \v{I}^{(r)}_{lml'm'} 
+ 
R_{ll'} \big[ \v{I}^{(\theta)}_{lml'm} + \v{I}^{(\varphi)}_{lml'm'})\big].
\end{align}

The matrices $R^{d}$ and $R$ are calculated numerically while the vector
matrices $\v{I}$ are reduced to analytical forms (\refapp{app:i-integrals-csh}):
\begin{align}
I^{(r)}_{lml'm',\hat x}=&
P^{(1)}_{lml'm'}A^{(c)}_{mm'},
\end{align}
\begin{align}
I^{(r)}_{lml'm',\hat y}=&
P^{(1)}_{lml'm'}A^{(s)}_{mm'},
\end{align}
\begin{align}
I^{(r)}_{lml'm',\hat z}=&
{P^{(2)}_{lml'm'}} 
\delta_{mm'},
\end{align}
\begin{align}
I^{(\theta)}_{lml'm',\hat x}=&
P^{(3)}_{lml'm'}
{A^{(c)}_{mm'}},
\end{align}
\begin{align}
I^{(\theta)}_{lml'm',\hat y}=&
P^{(3)}_{lml'm'}
{A^{(s)}_{mm'}},
\end{align}
\begin{align}
I^{(\theta)}_{lml'm',\hat z}=&
{P^{(4)}_{lml'm'}}\delta_{mm'},
\end{align}
\begin{align}
I^{(\varphi)}_{lml'm',\hat x}=&
-im'
{P^{(5)}_{lml'm'}}
A^{(s)}_{mm'},
\end{align}
\begin{align}
I^{(\varphi)}_{lml'm',\hat y}=&
im'
{P^{(5)}_{lml'm'}}
A^{(c)}_{mm'}
\end{align}
and
\begin{align}
I^{(\varphi)}_{lml'm',\hat z}=0 \;.
\end{align}
The matrices $P^{(i)} (i=1...5)$ are developed in 
\refapp{app:p-integrals}, while $A^{(c)}$ and $A^{(s)}$ are provided 
in \refapp{app:a-integrals}.

Similarly for $\tilde{\nabla}_{ilm l'm'} $ we have
\begin{align}
\tilde{\nabla}_{ilm l'm'} 
=& 
\underbrace{\int_0^\infty r^2 dr \tilde{R}^*_{l}(r) \frac{d \tilde{R}_{l'}({r}) }{d r}}_{\tilde{R}^{(d)}_{ll'}} 
\underbrace{\int_0^{\pi} \sin(\theta) d\theta \int_0^{2\pi} d\varphi Y^*_{lm}(\theta,\varphi)  Y_{l' m'}(\theta,\varphi )
\hat{ \v{e} }_r(\theta,\varphi)
}_{=\v{I}^{(r)}_{lml'm'}}   
\nonumber\\
+&
\underbrace{\int_0^\infty r dr \tilde{R}^*_l(r)\tilde{R}_{l'}({r})}_{\tilde{R}_{ll'}} 
\big[ 
\underbrace{\int_0^{\pi} \sin(\theta) d\theta \int_0^{2\pi} d\varphi Y^*_{lm}(\theta,\varphi ) 
\frac{\partial Y_{l'm'}( \theta, \varphi )}{\partial \theta} 
\hat{\v{e}}_\theta(\theta,\varphi) 
}_{={I}^{(\theta)}_{lml'm'}}
\nonumber\\
+&
\underbrace{\int_0^{\pi} d\theta \int_0^{2\pi} d\varphi Y^*_{lm}(\theta,\varphi ) \frac{\partial Y_{l'm'}( \theta, \varphi )}{\partial \varphi}
   \hat{\v{e}}_\varphi(\theta,\varphi) 
}
_{= {I}^{(\varphi)}_{lml'm'}}
\big],
\end{align}
or
\begin{align}
\tilde{\nabla}_{ilm l'm'} 
=&
\tilde{R}^{(d)}_{ll'} \v{I}^{(r)}_{lml'm'} 
+ 
\tilde{R}_{ll'} \big[ \v{I}^{(\theta)}_{lml'm'} + \v{I}^{(\varphi)}_{lml'm'}\big].
\end{align}

The formulae for the $\v{I}$ integrals in terms of real spherical 
harmonics (\refapp{app:csh}) are given in \refapp{app:i-integrals-rsh} (see also \refapp{app:a-integrals}). The $P$ matrices are the same as for the 
complex spherical harmonics.

\section{\label{sec:qe-implementation}The KGEC code}
\subsection{\label{sec:qe-overview}Overview}

On the foundations just laid, the 
KGEC code implements calculation of the full \emph{complex}
Kubo-Greenwood electrical 
conductivity \emph{tensor} using the KS orbitals calculated by QE with either PAW datasets or norm-conserving
pseudopotentials. (Note, however, that the latter case is {\it
  without} the non-local corrections.) KGEC is a post-processing tool
for QE programmed in Fortran 90.  It is modular and MPI-parallelized
over k-points, bands, and plane waves. Details of parallelization
are discussed below.

\begin{figure}
{\includegraphics[scale=1.0,width=\textwidth]{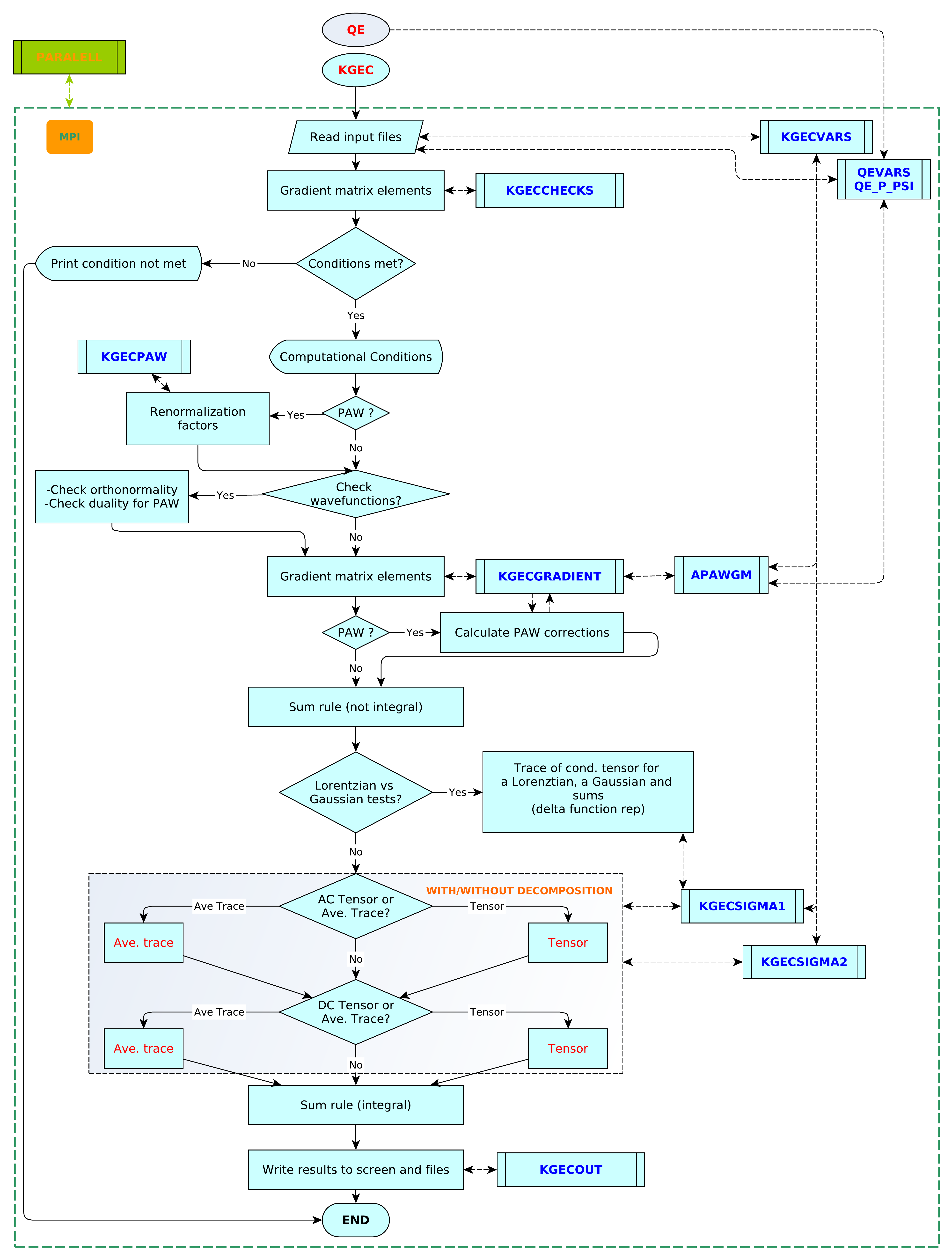}} 
\caption{\label{fig:workflow}KGEC general work flow.}
\end{figure}

KGEC work flow is presented in \figref{fig:workflow}.  It presumes 
an ordinary QE calculation has been done which provides the KS
orbitals, orbital energies, occupation numbers, temperature, 
and other relevant data via storage in
the usual outdir directory. All that data is made accessible to KGEC
by the QEVARS and QE\_P\_PSI modules.  

KGEC starts by reading an input file which provides the computational
conditions for the conductivity calculation and the location of
the QE data to use.  It then verifies that the conditions for which 
the code was designed are met.  If they are not, KGEC stops
with a message about the problem and possible solutions, if discernible.
Conversely, if the condition checks are satisfactory,the code  
proceeds to renormalize the PAW wave-functions (if PAWs are used) to avoid
small errors in the normalization introduced by the construction 
of the PAW orbitals.  
Subsequently, if requested by the user, the code checks orthonormality 
and duality conditions for 
the pseudo-atomic orbitals and projectors (for PAWs).

Next comes calculation of the gradient matrix elements, via the KGEGRADIENT 
module, for pseudo-orbitals
provided from QE and, if PAW datasets are utilized, a calculation
of the PAW corrections is done via the APAWGM module of KGEC.
Once the gradient matrix
elements are completed, the sum rule without a delta function or frequency 
dependence (\eqnref{eq:sumrule-f}) is calculated.

If selected by the user input, there follows the optional analysis of 
the effect of four different choices for numerical evaluation of the 
delta function (recall eq.\ \ref{Lorentzian-Delta}), namely calculation  
of the average trace of 
the conductivity tensor 
done for a Lorentzian, a Gaussian, the sum of two Lorentzians,
and the sum of two Gaussians (to eliminate problems at the origin).

Continuing, the code then proceeds to calculate either the full electrical 
conductivity tensor (including the average trace and the dc components), 
the average trace only (including the dc components), or the dc components 
only; all with or without decomposition.
Those implementations 
are contained in the KGECSIGMA1 and KGECSIGMA2
modules. The sum rule for integration of the conductivity over
frequencies (\eqnref{eq:p-sumrule-cond-2}) is calculated next.

Lastly, KGEC writes some additional information to the standard output 
and to the corresponding files.

\subsection{MPI parallelization}

In general KGEC is MPI-parallelized over k-points, bands, and plane
waves via its PARALLEL module. That hierarchical order is the 
same as in QE. Parallelization over plane waves primarily is useful for
gradient matrix element calculations. Once those matrix
elements are done, the plane-wave-based parallelization is
needless. However, the number of plane-waves greatly exceeds the
number of bands. Usually that disparity is reflected in a larger
number of processes used for plane waves than for bands. KGEC thus has
the capacity to recover the MPI processes used for plane wave
parallelization after gradient matrix element completion and add 
the recovered processes to the band parallelization processes.

\begin{figure}[h]
{\includegraphics[scale=1.0,width=\textwidth]{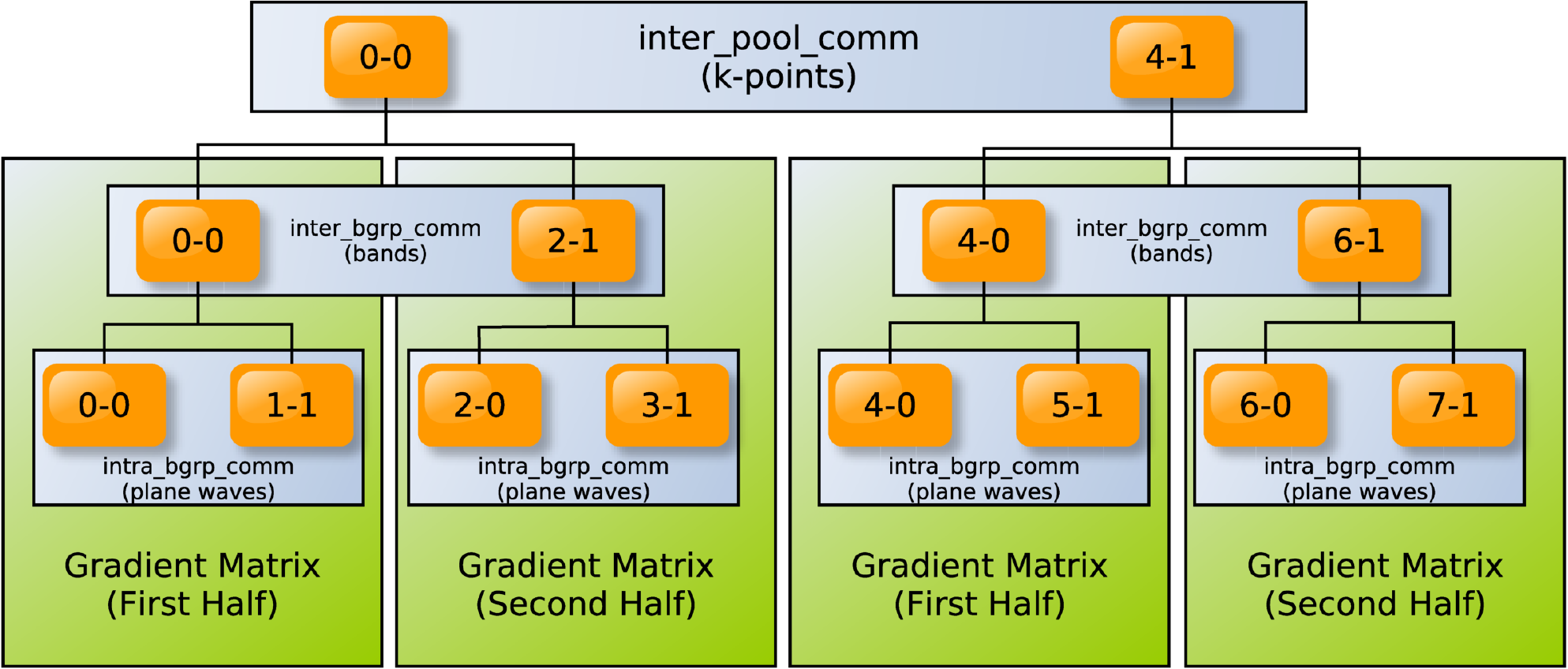}} 
\caption{\label{fig:mpi-processes-before-recovery}
Block diagram of 8 MPI processes with two of them dedicated to k-point 
parallelization, two dedicated to band parallelization for each k-point 
process, and the other two dedicated for plane wave 
parallelization for each band process. A process is represented by a yellow 
rectangle, a communicator by a light-blue rectangle, the black lines 
connect the related processes in  the parallel work flow, and the green 
rectangles represent the distribution of the gradient matrix.  The first 
number in each process is its rank in the world\_comm and the second is 
its rank in the communicator it belongs to or lies on.
}
\end{figure}

More specifically KGEC is parallelized over k-points using the QE MPI
communicator inter\_pool\_comm with $nk$ processes, over bands using
the inter\_bgrp\_comm with $nb$ processes and over plane-waves using
intra\_bgrp\_comm with $np$ processes. So, the total number of MPI
processes is $nk \times nb \times np$. An example for 8 MPI processes
is given in \tabref{tab:mpi-comms-with-nk-nb-npw-para}, with 2
processes dedicated to k-point parallelization, for each of them 2
dedicated to band parallelization, and for each of these 2 for
plane-wave parallelization.
\figref{fig:mpi-processes-before-recovery} shows all the MPI processes
and communicators in block form. With this scheme, the code can make
those processes in the same communicator, i.e. those lying on the same blue
rectangle in the figure, exchange information just by referencing
their communicator. That allows for  more efficient collective
operations (scatter, gather,reduce), as well as code simplicity.

\begin{center}
\begin{table}[h]
\begin{tabular}{l c c c c c c c c c c c}
\hline\hline
MPI Communicator     & \multicolumn{8}{c}{MPI Ranks}& Parallelization \\\hline
           world\_comm  & 0 & 1 & 2 & 3 & 4 & 5 & 6 & 7 & \\
inter\_pool\_comm   & 0 & 0 & 0 & 0 & 1 & 1 & 1 & 1 & over  k-points\\
inter\_bgrp\_comm     & 0 & 0 & 1 & 1 & 0 & 0 & 1 & 1  & over  bands   \\
intra\_bgrp\_comm     & 0 & 1 & 0 & 1 & 0 & 1 & 0 & 1  & over plane waves\\\hline\hline
\end{tabular}
\caption{\label{tab:mpi-comms-with-nk-nb-npw-para} 
MPI communicators and each process rank for a parallelization over 8 processes, two for k points, two for bands, and two for plane waves.}
\end{table}
\end{center}
\vspace*{-14pt}

Another key point is that the gradient matrix is distributed by the
number of bands processes.  Exact copies of those will end up stored
in all the plane wave processes associated with the same band
process. In other words, each process in an intra\_bgrp\_comm has
exactly the same copy of a fragment of the gradient matrix that
corresponds to the band process to which they are subordinated.  In the
specific case of \tabref{tab:mpi-comms-with-nk-nb-npw-para}, the 
gradient matrix is divided in halves, each of
them residing on processes belonging to an inter\_bgrp\_comm value and
replicated in the corresponding intra\_bgrp\_comm processes. That
means that for the 0-0 branch of the k-point parallelization the same
copy of the first half of the gradient matrix would be stored in the
0-0 and 1-1 processes, and the copy of the other half would be in the
processes 2-0 and 3-1. A similar situation holds for the branch 
4-1 of the k-point parallelization.

This structure is exploited for recovery of the plane wave processes that
otherwise would be idle, hence wasted, after the calculation of
gradient matrix elements, without any further communication.

\begin{figure}[h]
{\includegraphics[scale=1.0,width=\textwidth]{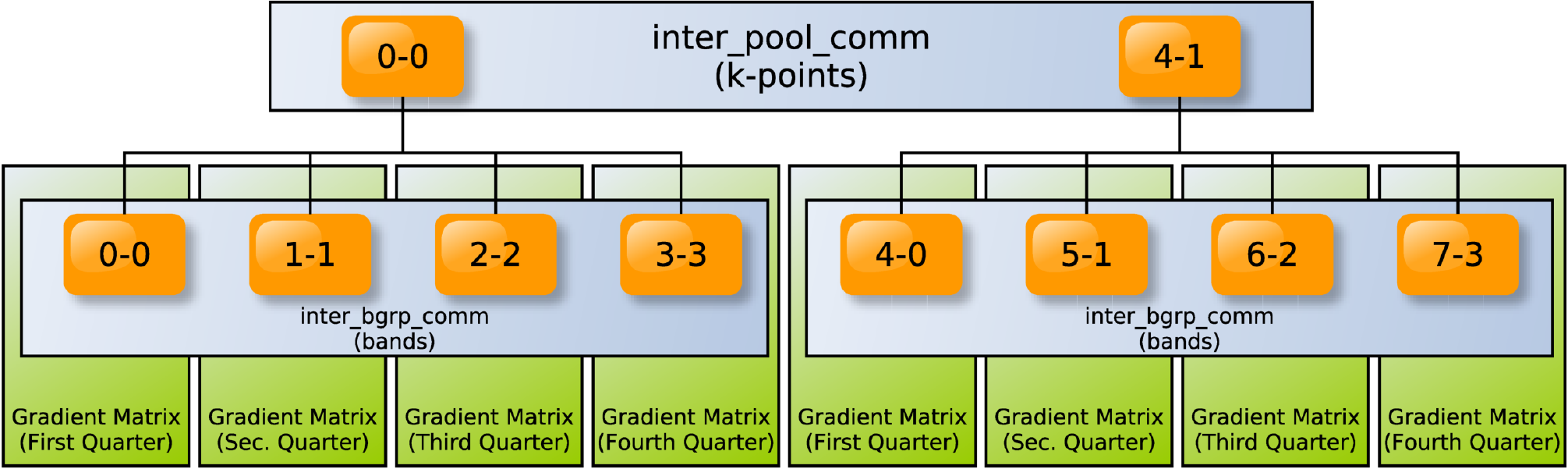}} 
\caption{\label{fig:mpi-processes-after-recovery}
Block diagram of 8 MPI processes after recovery of the plane waves 
processes to be used for bands parallelization. There still are two processes
 dedicated to k-point parallelization, but four (instead of two as before) 
dedicated  to band parallelization for each k-point process. A process 
is represented 
by a yellow rectangle, a communicator by a light-blue rectangle and the 
lines connect the related processes in the parallel work flow. The first 
number in each process is its rank in the world\_comm and the second is 
its rank in the communicator it belongs to or lies on.
}
\end{figure}

Therefore, if the option to recover the plane wave processes is set
to true (npwrecovery=.true.) then once the gradient matrix elements
have been calculated, KGEC redefines the MPI communicators to use the
plane wave processes for band parallelization. So, it goes from $nb$
band processes per k-point process to $nb \times np$ bands processes
per k-point process, the gradient matrix elements being redistributed
in place (without communication) between the $nb \times np$ bands
processes. Coming back to the example of 8 MPI processes, the
corresponding re-definition of the inter\_bgrp\_comm is given in
\tabref{tab:mpi-comms-with-nk-nb-para} and the block diagram in
\figref{fig:mpi-processes-after-recovery}.  One sees that the band 
processes have expanded from 0 to 1 for each k-point process to 
0,1,2,3. The half of the gradient matrix residing previously in each
intra\_bgrp\_comm processes is divided by 2 and each bands process
uses its own half from that point on. The rank 0 process in the
previous intra\_bgrp\_comm keeps the first half and destroys the
second one, while the rank 1 does the opposite. The final distribution
is then one-quarter of the total columns of the gradient matrix per each process in
the new inter\_bgrp\_comm.

\begin{center}
\begin{table}[h]
\begin{tabular}{l c c c c c c c c c c c}
\hline\hline
MPI Communicator     & \multicolumn{8}{c}{MPI Ranks}& Parallelization \\\hline
           world\_comm  & 0 & 1 & 2 & 3 & 4 & 5 & 6 & 7 & \\
inter\_pool\_comm   & 0 & 0 & 0 & 0 & 1 & 1 & 1 & 1 & over  k-points\\
inter\_bgrp\_comm     & 0 & 1 & 2 & 3 & 0 & 1 & 2 & 3  & over  bands   \\
intra\_bgrp\_comm     & - & - & - & - & - & - & - & -  & over plane waves\\\hline\hline
\end{tabular}
\caption{\label{tab:mpi-comms-with-nk-nb-para} 
MPI communicators and each process rank  for a parallelization over 8 processes after recovery of the plane waves processes.}
\end{table}
\end{center}

\subsection{Prerequisites}
The prerequisites for KGEC installation are:
\begin{itemize}
 \item MPI for parallel compilation
 \item Fortran 90 compiler (Makefiles for Intel Linux Fortran provided ).
 \item Quantum Espresso 5.1.2, 5.2.1, 5.4.0, 6.0 or 6.1  installed for either serial or mpi-parallel execution or QE 5.2.1 compiled for use with Profess@QE 
\cite{KARASIEV20143240}.
\end{itemize}
(Remark: all of our installations have been in Linux with the Bourne-again shell.) 

Both a README file and a more detailed User Guide 
are provided with the source code at download.  They 
give installation instructions, along with instructions on how to do a 
simple example calculation.  Input and reference output files
for that calculation are provided.  The example is fcc Aluminum 
with four atoms per unit cell at bulk density $\rho =2.7 $ g/cm$^3$
and temperature of 0.05 Rydberg (approximately 7,894 K).   
Note that if the example calculation (or any other for that matter) is
run on  more than one core, 
there will be differences with respect to the results from a serial 
calculation for the same input data.  Such differences are the 
consequence of 
floating point arithmetic differences. However, as the  number 
of k-points and bands are increased, the
serial and parallel results should converge to the same values.

\section{\label{sec:kgec-tests}KGEC tests}
\subsection{Comparison with Abinit}
\begin{figure}
\includegraphics[width=0.48\textwidth,viewport=1 1 710 495,clip]
{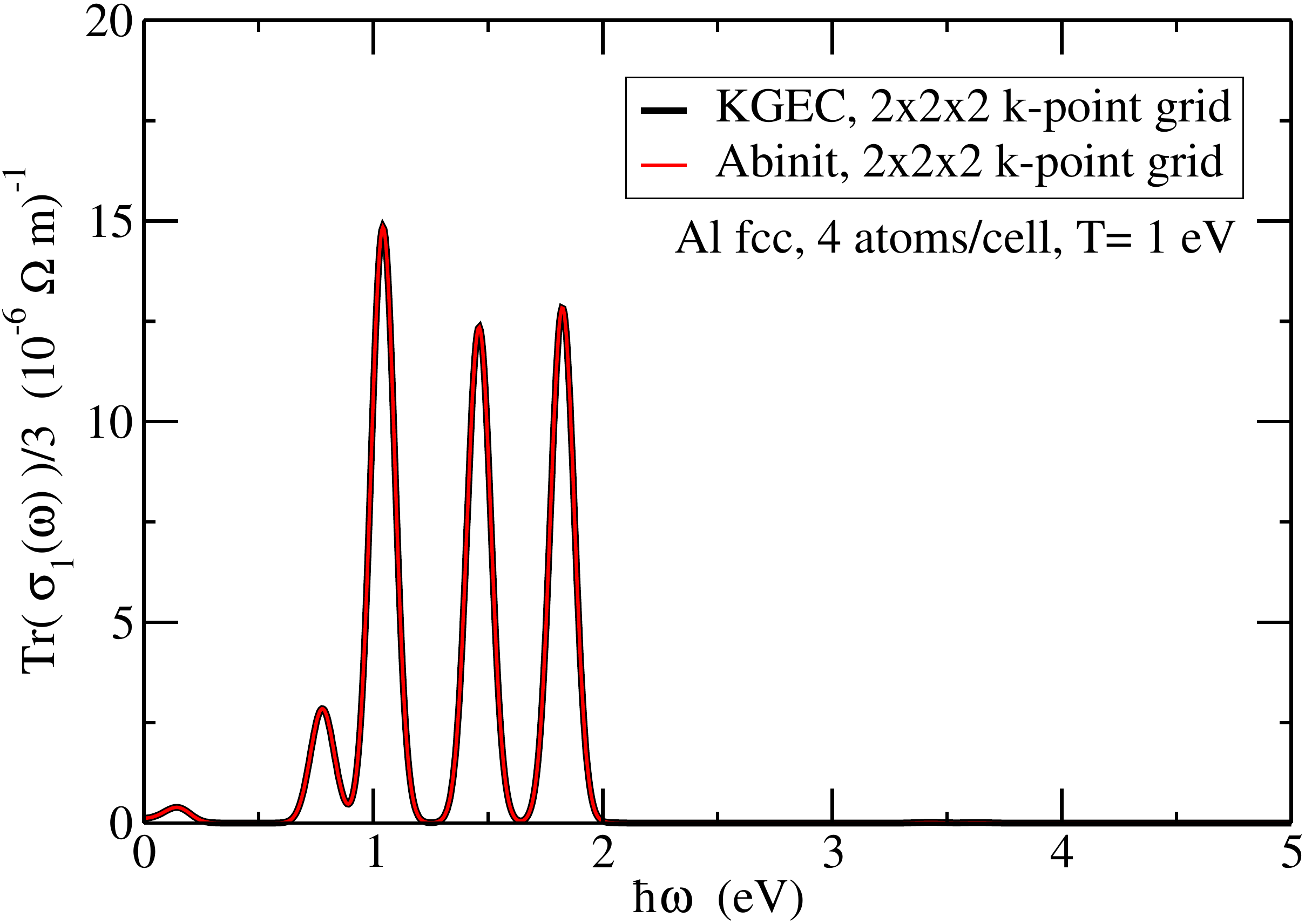}%
\includegraphics[width=0.48\textwidth,viewport=1 1 710 495,clip]
{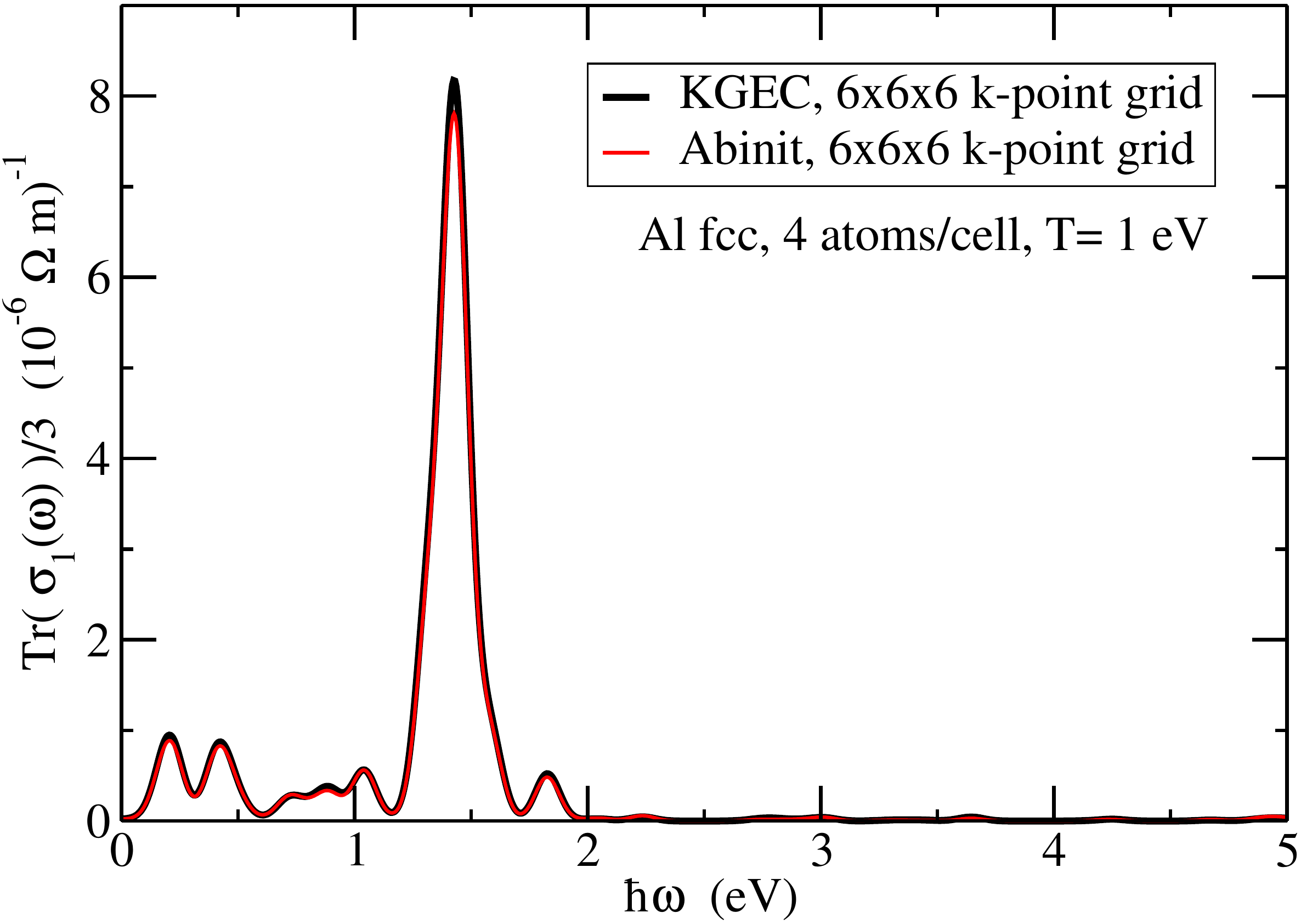}\\%
\includegraphics[width=0.48\textwidth,viewport=1 1 710 495,clip]
{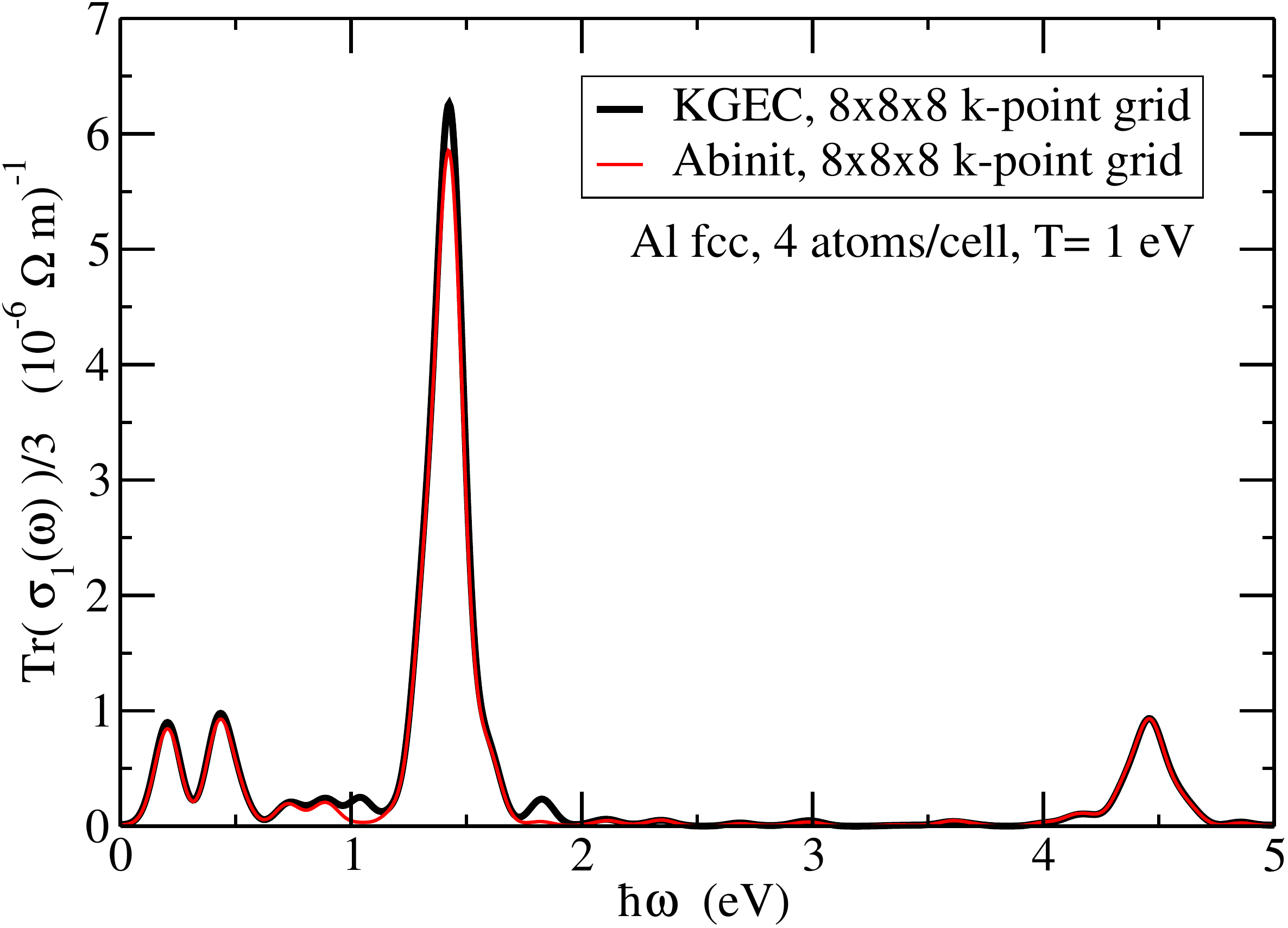}%
\includegraphics[width=0.48\textwidth,viewport=1 1 710 495,clip]
{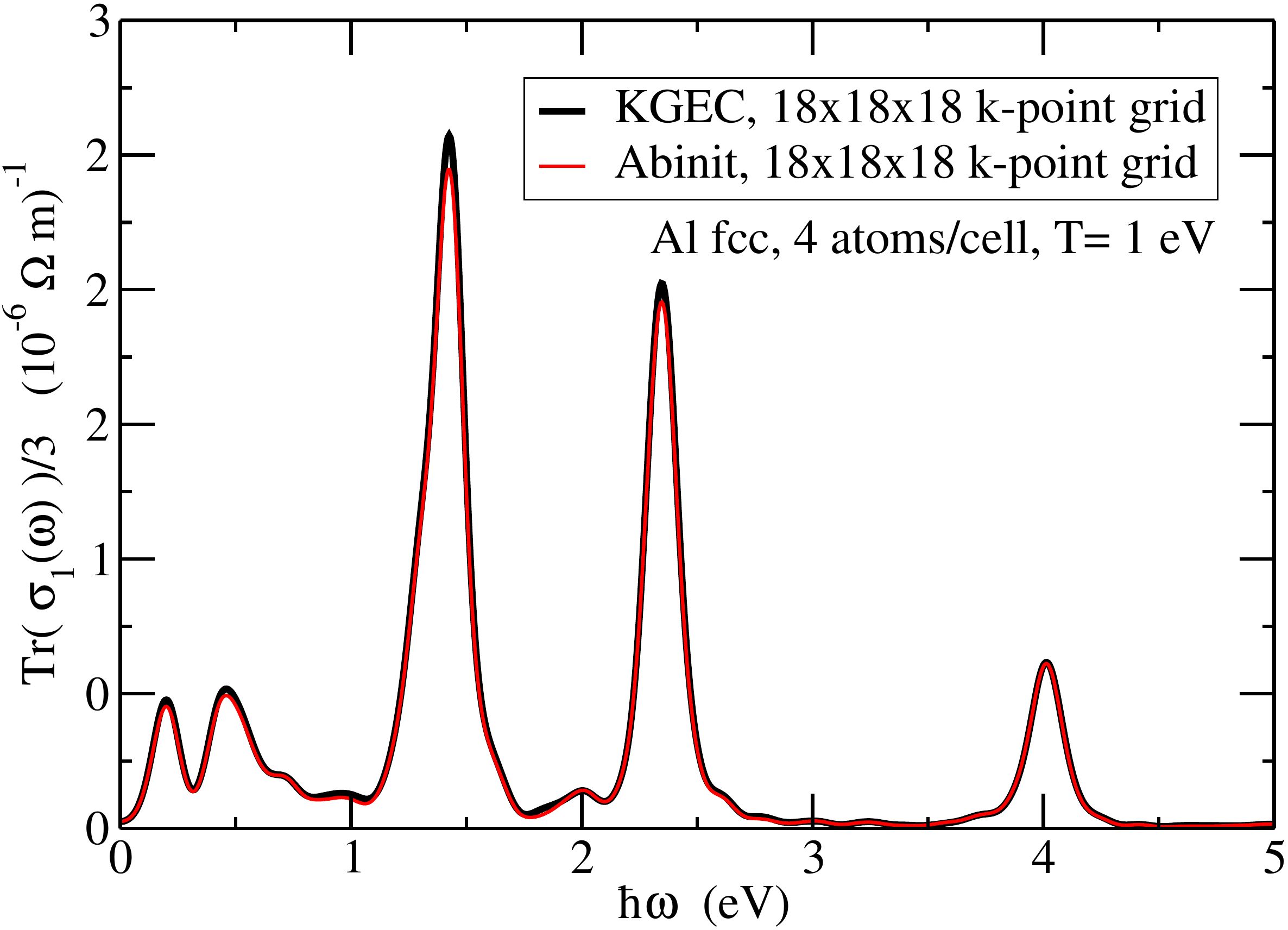}%
\caption{\label{fig:code-convergence} Comparison of KGEC and Abinit for different k-points in an ordered system; 4 atom/cell fcc Al at a density of 2.70 g/cm$^3$.}
\end{figure}

We have calculated the average trace of the electrical conductivity using 
the approximated formula with two 
Gaussians (enforcing even parity of the conductivity)
for Al fcc at bulk density $\rho = 2.7$ g/cm$^3$ and temperature $T =1$ eV 
for various numbers of k-points using KGEC and, for comparison, using   
Abinit. \cite{Gonze20092582,Gonze2002478,Torrent2008337}
The results are in very good agreement as \figref{fig:code-convergence} shows.

\begin{figure*}
\includegraphics[width=0.48\textwidth,viewport=1 1 710 495,clip]
{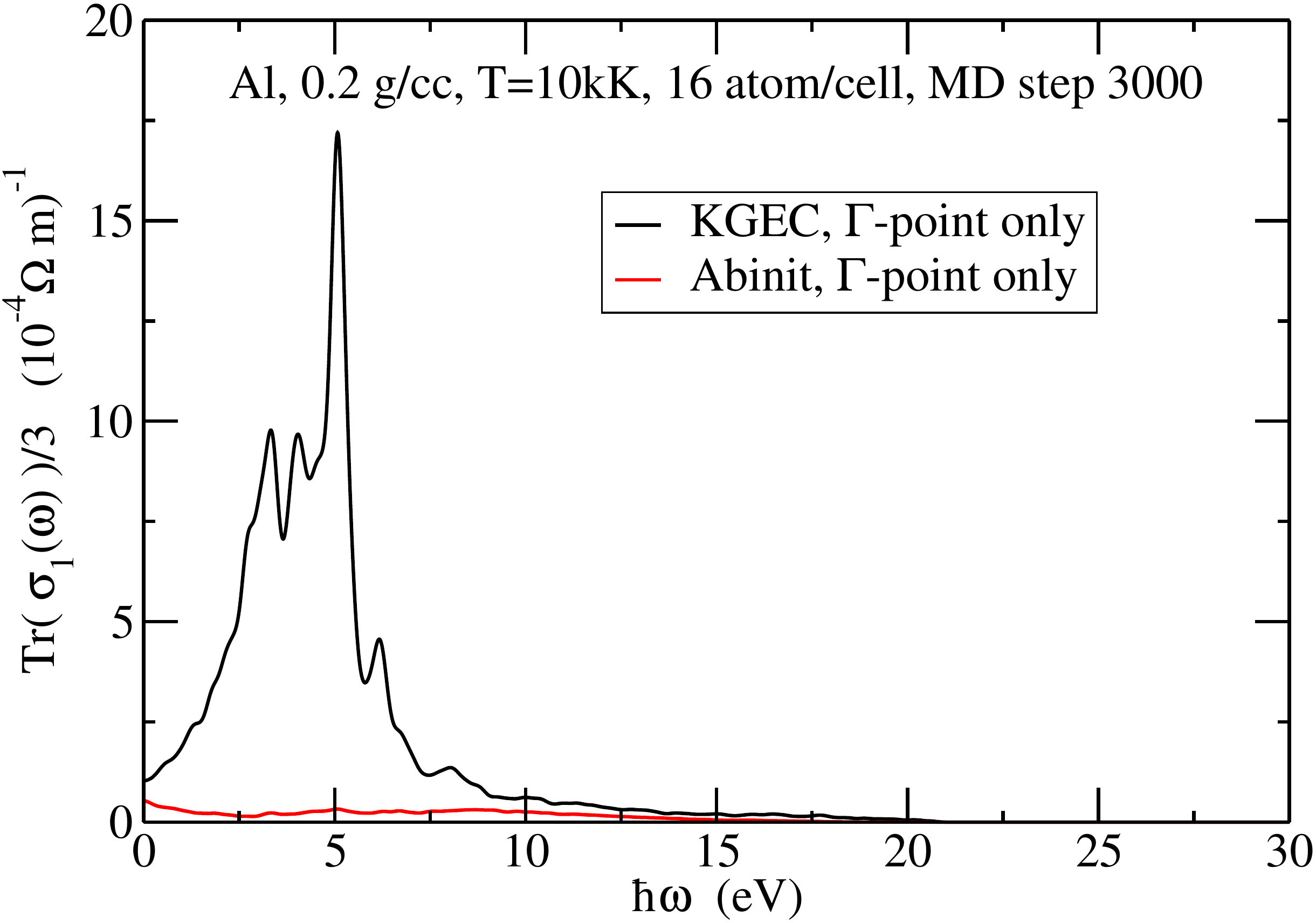}%
\includegraphics[width=0.48\textwidth,viewport=1 1 710 495,clip]
{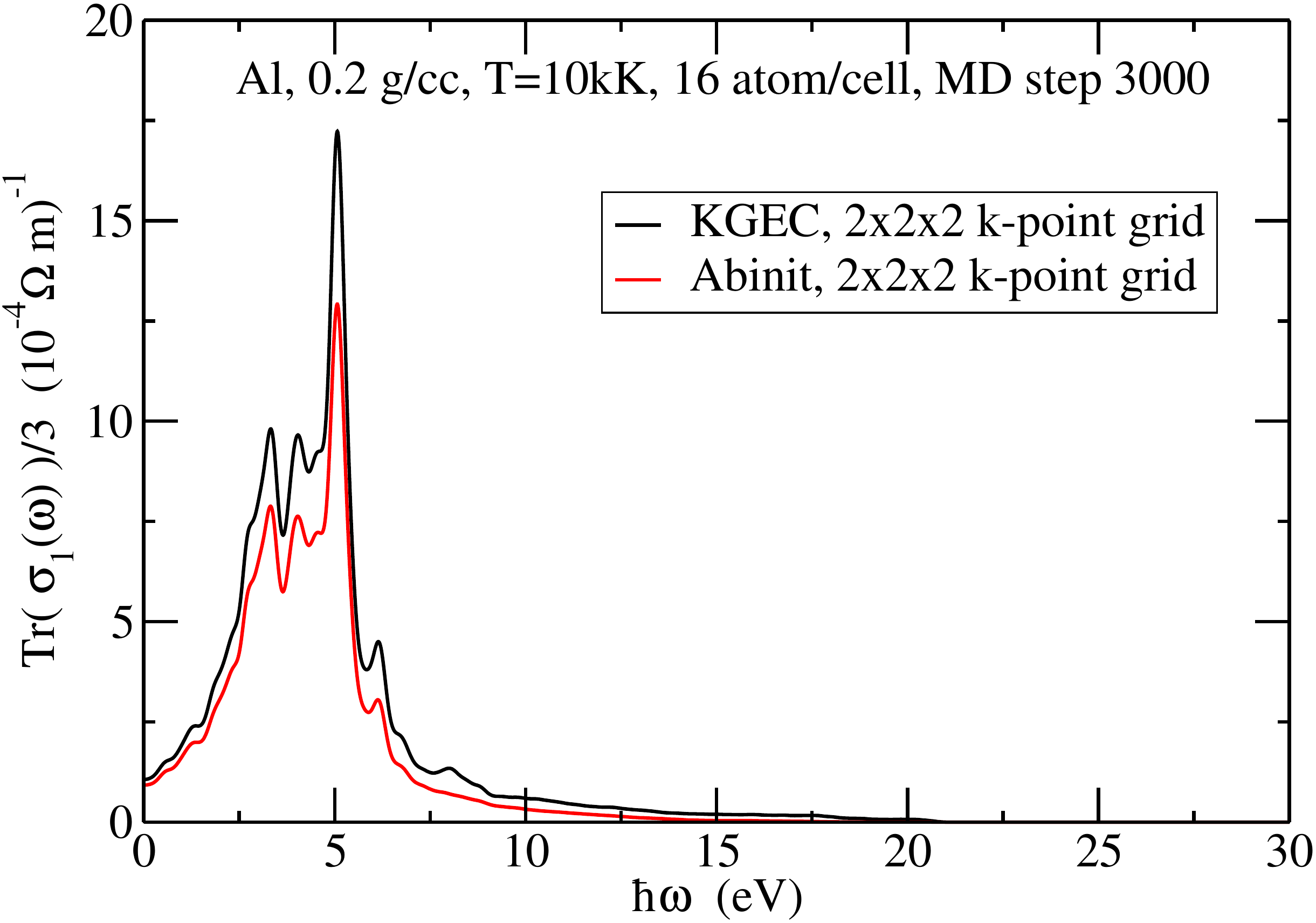}\\%
\includegraphics[width=0.48\textwidth,viewport=1 1 710 495,clip]
{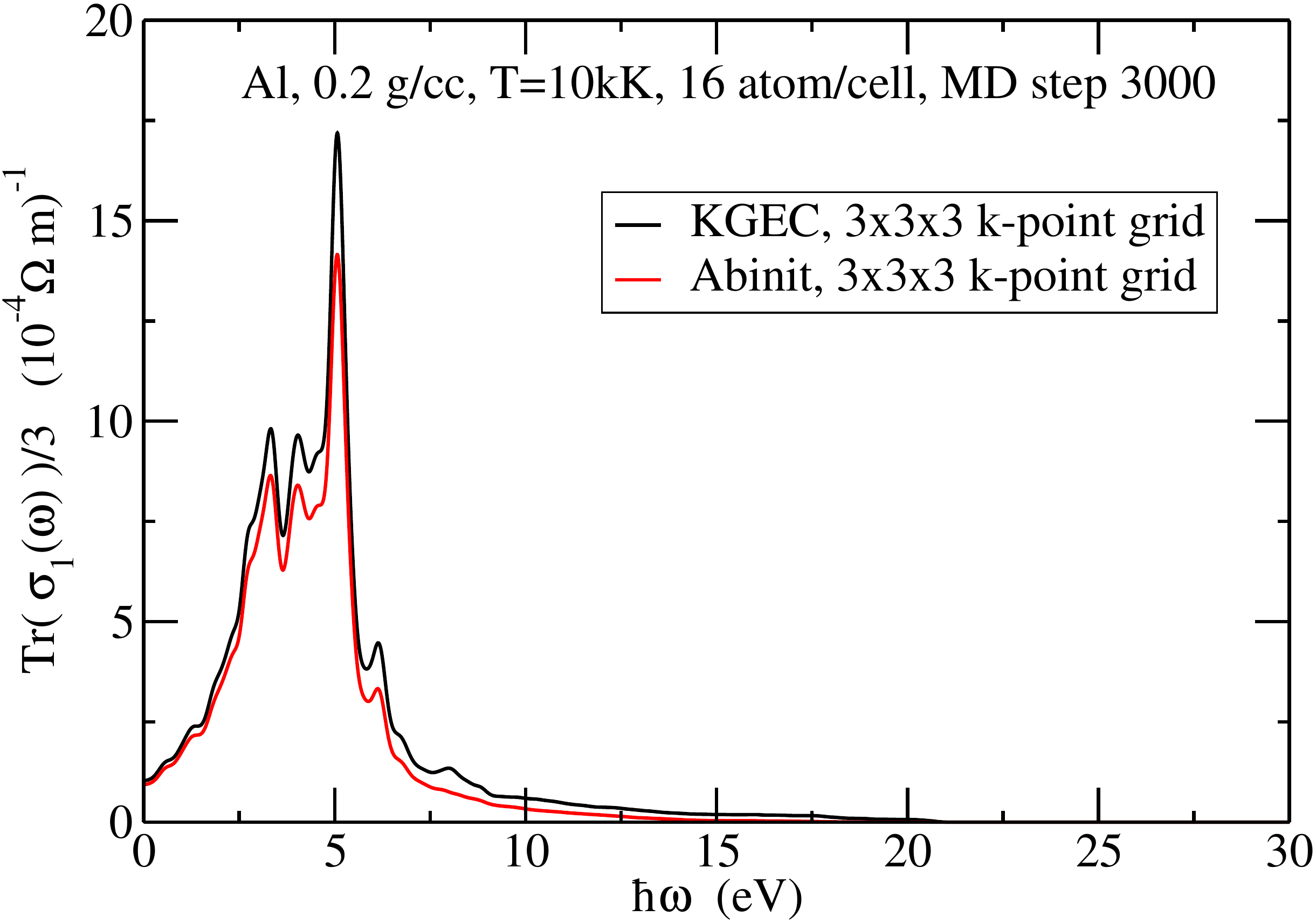}%
\includegraphics[width=0.48\textwidth,viewport=1 1 710 495,clip]
{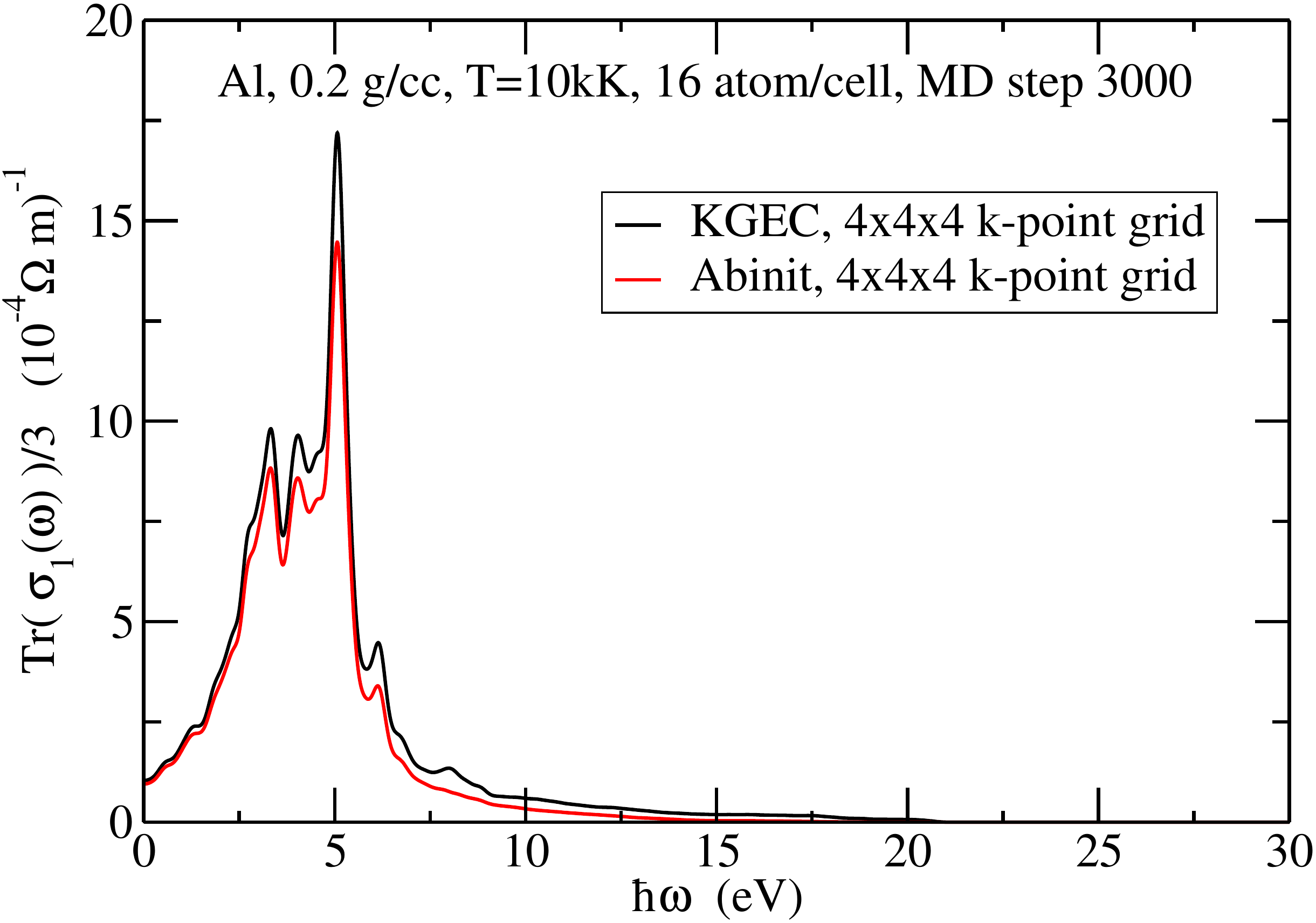}%
\caption{\label{fig:al16-md-3000-kgec-vs-abinit} Comparison of KGEC and Abinit for different k-point grid densities in a disordered system: 16 atom/cell Al at an arbitrarily chosen MD step.}
\end{figure*}
\begin{figure*}
\includegraphics[width=0.48\textwidth,viewport=1 1 710 495,clip]{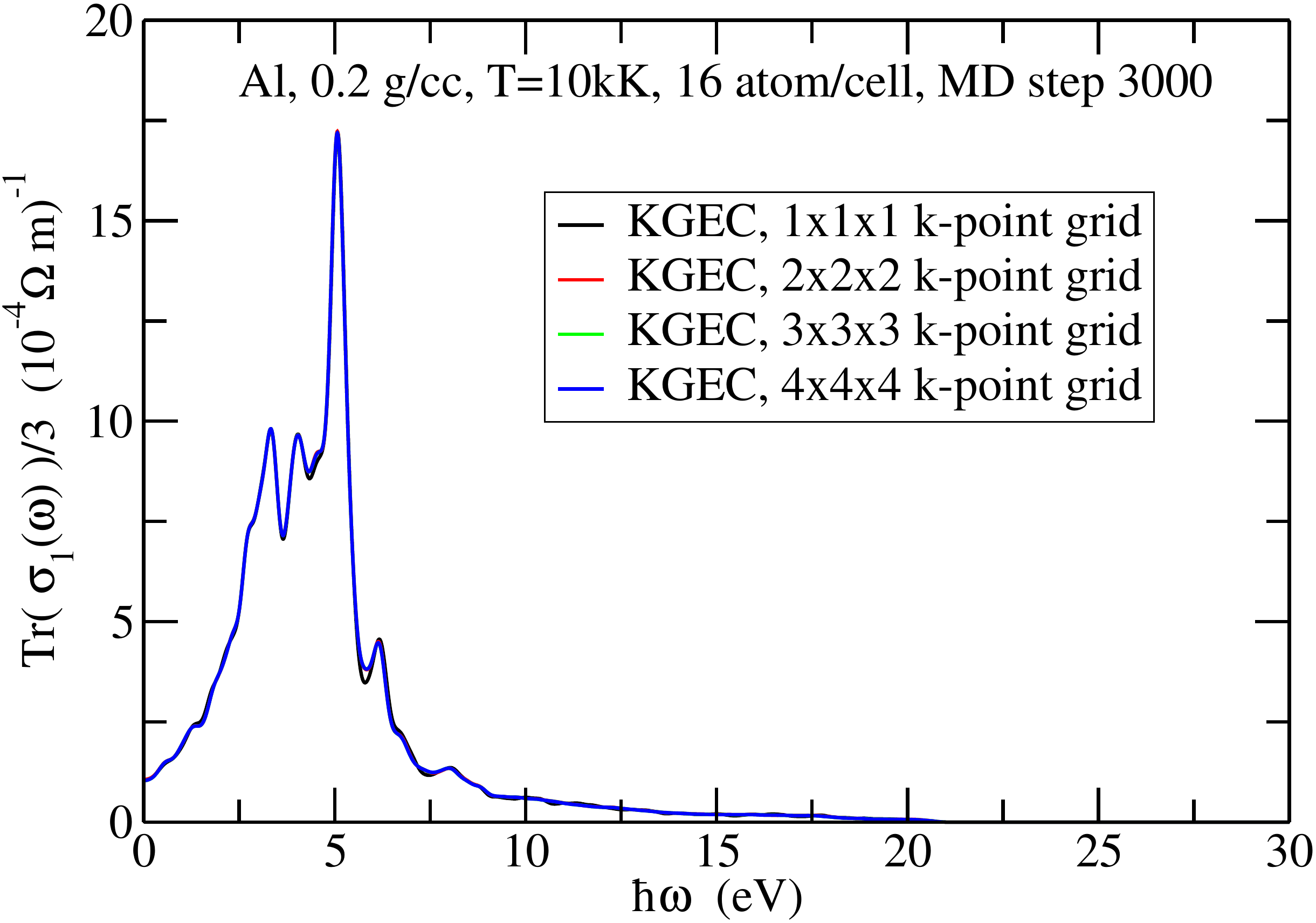}
\includegraphics[width=0.48\textwidth,viewport=1 1 710 495,clip]{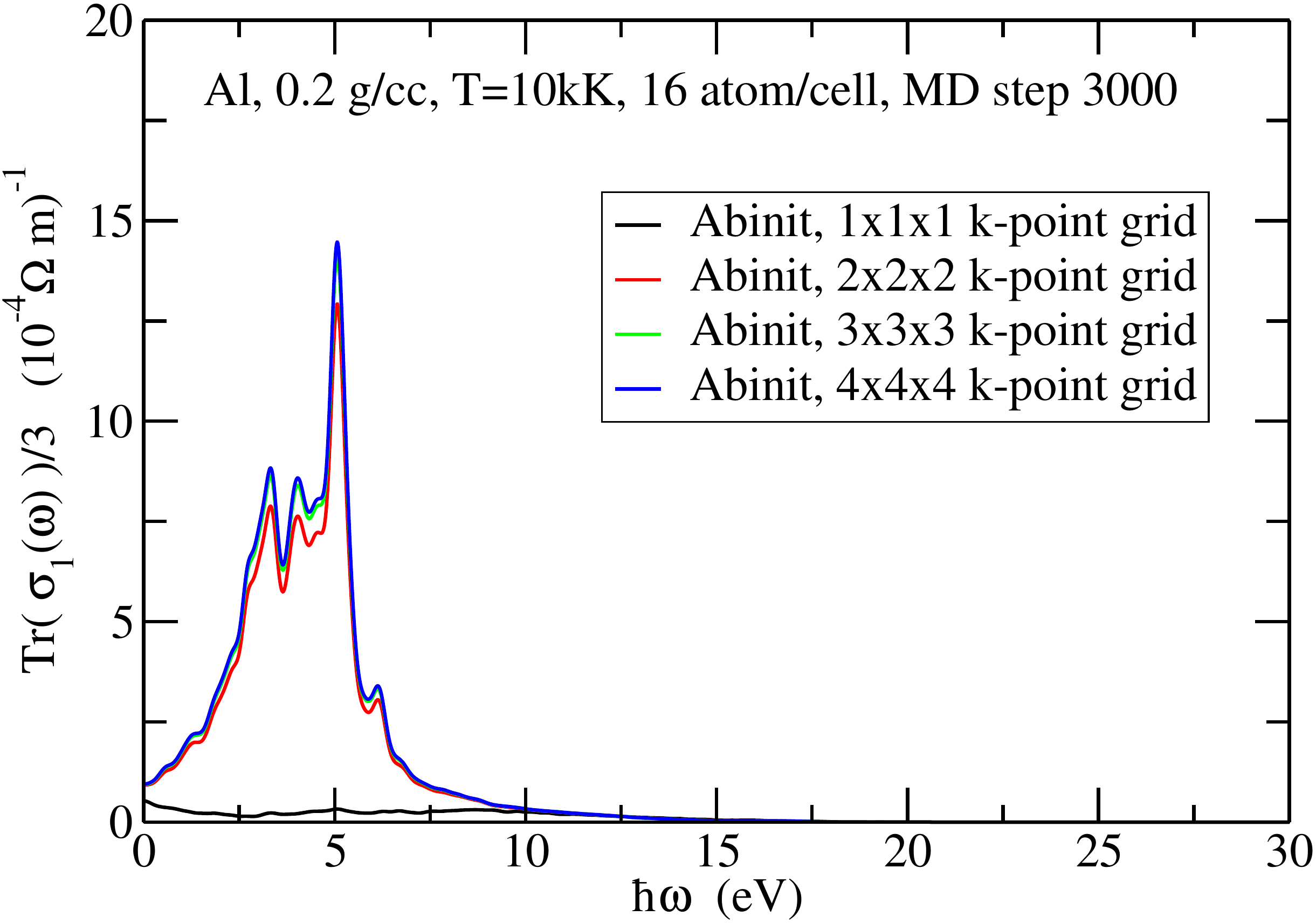}
\caption{\label{fig:code-convergence2} Convergence of KGEC (on the right) vs Abinit (on the left).}
\end{figure*}

However, for a more disordered system the results are sensitive to
the k-point grid density.  An example is for the ionic configuration  
from an arbitrarily selected  
molecular dynamics step of a 16 atom/cell Al system  at $\rho = 0.3$ g/cm$^3$ 
and $10$ kK (about 0.86 eV).  Results for the two codes differ for a $4\times 4
\times 4$ k-point grid; see (\figref{fig:al16-md-3000-kgec-vs-abinit}). But
comparison in \figref{fig:code-convergence2} shows that the KGEC results 
are already converged at that grid density while those from Abinit are not.

\subsection{\label{sec:kgec-tests-consistency}Consistency test}

\begin{figure*}[h]
\includegraphics[width=0.48\textwidth,viewport=1 1 710 495,clip]{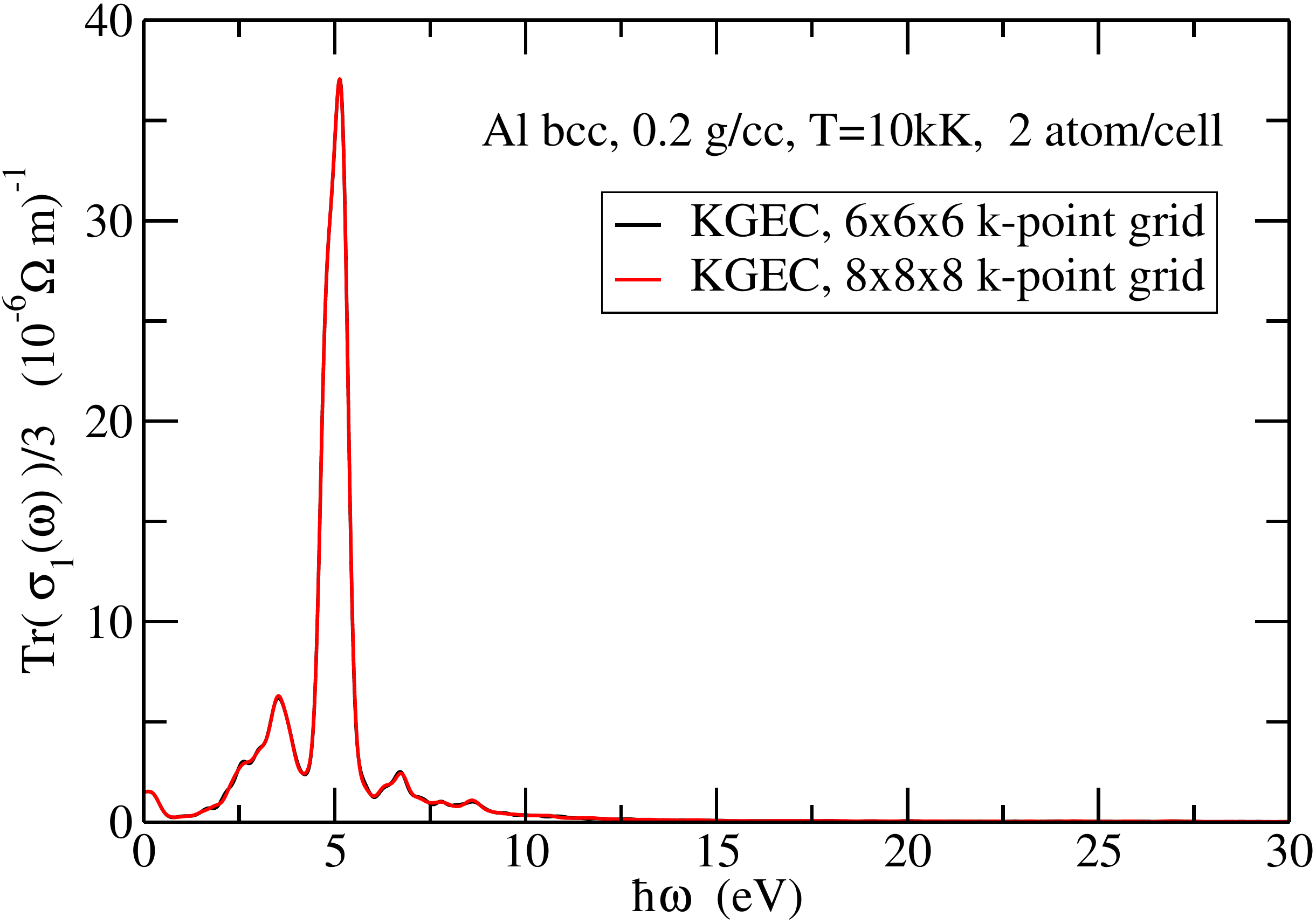}
\includegraphics[width=0.48\textwidth,viewport=1 1 710 495,clip]{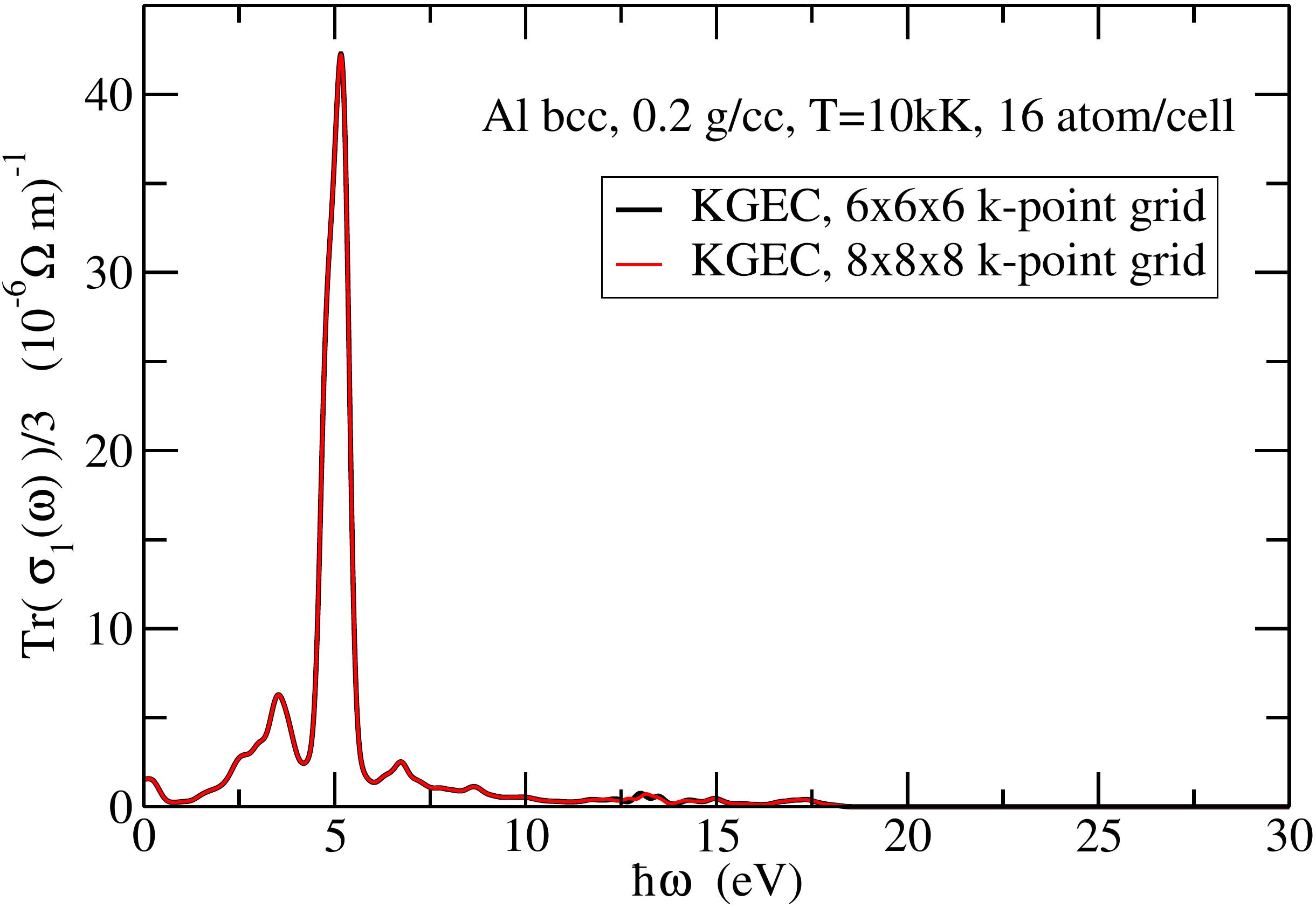}
\caption{\label{fig:code-consistency-1} Convergence of KGEC for bcc Al with 
$2$ and $16$ atoms per unit cell at $\rho = 0.2$ g/cm$^3$ and $T=10$ kK.}
\end{figure*}

A consistency test also was performed by calculating the average of
the conductivity for bcc Al with $2$ and $16$ atoms per unit cell at
$\rho=0.2 $ g/cm$^3$ and $T =10$ kK. This low-density regime is of 
intrinsic physical interest \cite{PhysRevE.93.063207}.  Convergence with 
k-point grid density was reached for both systems at the 
$8\times 8 \times 8$ grid, as can be seen in
\figref{fig:code-consistency-1}.

\begin{figure*}[h]
\begin{center}
\includegraphics[width=0.48\textwidth,viewport=1 1 710 495,clip]
{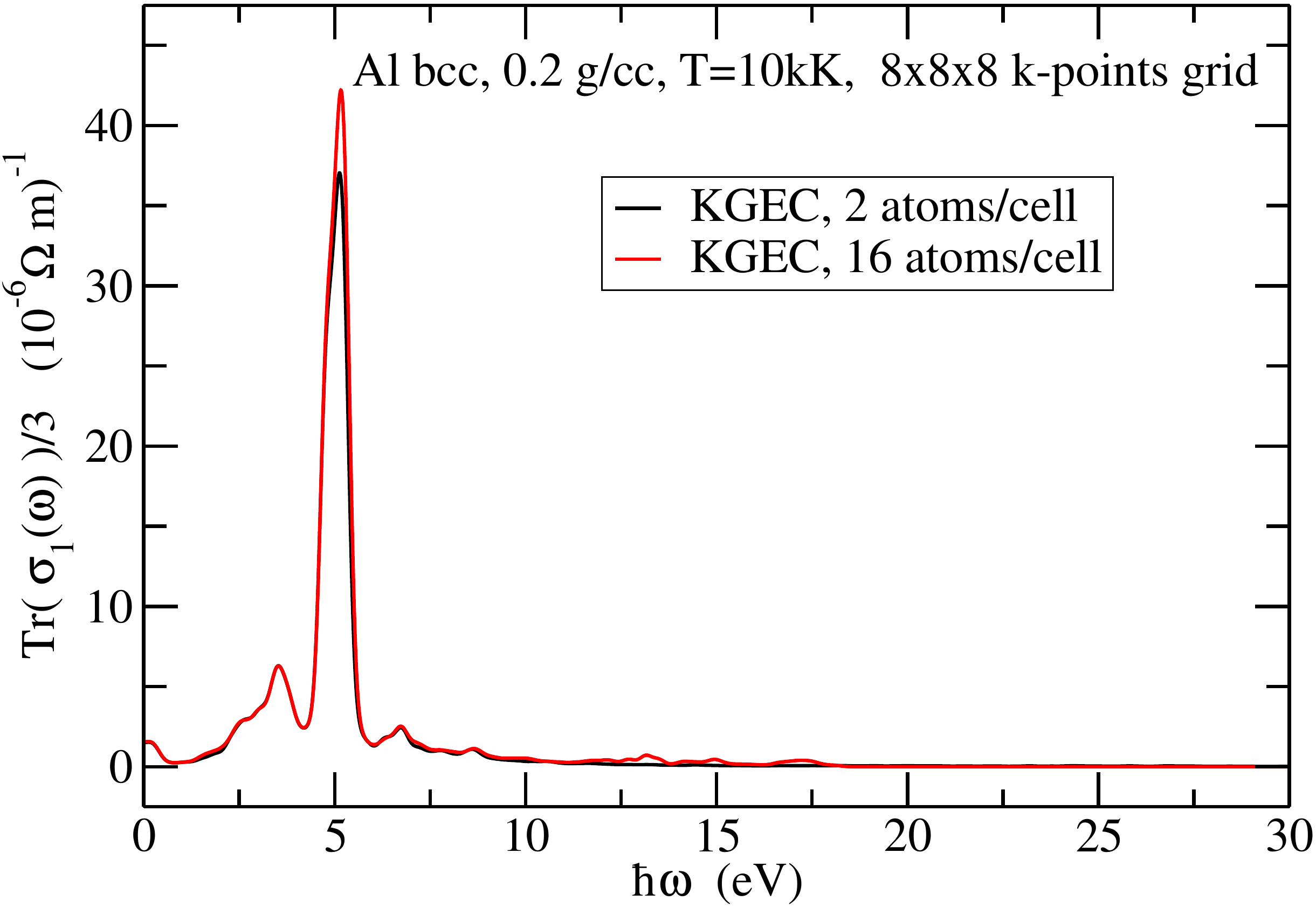}%
\caption{\label{fig:code-consistency-2} Comparison of the converged results for Al bcc $2$ and $16$ atoms per unit cell at $\rho = 0.2$ g/cm$^3$ and $T=10$ kK.}
\end{center}
\end{figure*}

However, comparison of the calculations for both systems performed with the 
$8 \times 8 \times 8$ mesh (\figref{fig:code-consistency-2}) reveals that there are some discrepancies in the intensities of the highest peak and 
in the smaller peaks around $15$ eV in frequency.
 The changes are related to temperature and unit cell size effects. Notice however the very good agreement at low frequencies.

\pagebreak
\section{\label{sec:difficulties}Difficulties}
\subsection{\label{sec:delta-function}Representation of the Dirac delta function}
It is frequent practice to use what we have called the ``approximated expression'',  \eqnref{eq:sigma1-general-2b}, with a Gaussian representation for the 
Dirac delta function. \eqnref{eq:sigma1-general-2b} also can be written as
\begin{equation}
 \sigma_1(\omega)=\frac{2 \pi e^2 \hbar^2}{m_e^2 V}\sum_m\sum_{m'}
 \frac{\Delta f_{m'm}}{\Delta \epsilon_{mm'}}
 \bra{m}\nabla\ket{m'}\bra{m'}\nabla\ket{m}
 \delta(\Delta\epsilon_{mm'}-\hbar\omega),
 \label{eq:sigma1-exact}
\end{equation}
denoted the as ``Dirac-delta form'' in the opening discussion.
Observe that the main distinction
between \eqnref{eq:sigma1-general-2b} and \eqnref{eq:sigma1-exact} is that
the $\Delta\epsilon_{mm'}$ in \eqnref{eq:sigma1-exact} is replaced by
$\omega$ in \eqnref{eq:sigma1-general-2b}.

We need to find the limits of \eqnref{eq:sigma1-exact} and \eqnref{eq:sigma1-general} for $\omega$ going to zero in the cases in which a Lorentzian 
or a Gaussian (\appref{app:kgec-l-g}) is used to represent the Dirac delta function. The issue reduces to evaluating four limits, to wit
\begin{align}
 \lim_{\omega\rightarrow 0}f^{(D-d)}_{L}(\omega)=%
\lim_{\omega\rightarrow 0}\frac{1}{\pi} \frac{\Delta f}{\Delta \epsilon} \frac{\delta/2}{(\Delta\epsilon-\hbar\omega)^2+\delta^2/4} = %
\frac{1}{\pi}\frac{\Delta f}{\Delta \epsilon} \frac{\delta/2}{(\Delta\epsilon)^2+\delta^2/4} 
\label{eq:feL-limit1}
 \end{align}

\begin{align}
  \lim_{\omega\rightarrow 0}f^{(D-d)}_{G}(\omega)=  \lim_{\omega\rightarrow 0} %
\frac{\Delta f}{\Delta \epsilon}  \frac{1}{\sigma_g \sqrt{\pi}}\; %
 \exp{ \left( -\frac{(\Delta\epsilon-\hbar\omega)^2}{\sigma^2_g}\right)} %
= \frac{\Delta f}{\Delta \epsilon} 
 \frac{1}{\sigma_g \sqrt{\pi}}\;
\exp{ \left( -\frac{(\Delta\epsilon)^2}{\sigma^2_g}\right)} 
\label{eq:feG-limit1}
 \end{align}

\begin{align}
  \lim_{\omega\rightarrow 0}f^{(a)}_{L}(\omega) =  \lim_{\omega\rightarrow 0}%
 \frac{\Delta f}{\pi\omega} \frac{\delta/2}{(\Delta\epsilon-\hbar\omega)^2+\delta^2/4}%
 = \pm \infty,
\label{eq:faL-limit1}
 \end{align}
and
\begin{align}
  \lim_{\omega\rightarrow 0}f^{(a)}_{G}(\omega)=  \lim_{\omega\rightarrow 0}%
\frac{\Delta f}{\omega}  \frac{1}{\sigma_g \sqrt{\pi}}\;
\exp{ \left( -\frac{(\Delta\epsilon-\hbar\omega)^2}{\sigma^2_g}\right)}%
= \begin{cases} 
      0 & \exp(-(\Delta\epsilon)^2)/\sigma_g^2)= 0 \\
      \pm\infty &  \exp(-b(\Delta\epsilon)^2/\sigma_g^2)\neq 0.
   \end{cases}
\label{eq:faG-limit1}
 \end{align}



First notice that the approximated expressions, $f^{(a)}_L(\omega)$ and
$f^{(a)}_G(\omega)$, do not have the same limit as the corresponding
D-d expressions, $f^{(D-d)}_L(\omega)$ and $f^{(D-d)}_G(\omega)$. Instead, 
the approximated expressions are singular at $\omega=0$.
Further, the two D-d versions $f^{(D-d)}_L(\omega)$ and
$f^{(D-d)}_G(\omega)$ do not have the same limit, though they should 
be the same in the limit of the delta-width of the
Lorentzian and the Gaussian going to zero.

The singularity of the approximated expressions can be lifted by using
the even parity of $\sigma_1(\omega)$. In that case the limits are
\begin{align}
\lim_{\omega\rightarrow 0}  f^{(a)}_{L}(\omega)
=& 
\lim_{\omega\rightarrow 0}
\frac{\Delta f}{2\pi\omega}
\left(
\frac{\delta/2}{(\Delta\epsilon-\hbar\omega)^2+\delta^2/4} -  \frac{\delta/2}{(\Delta\epsilon+\hbar\omega)^2+\delta^2/4}\right)
\nonumber\\
=&
\frac{\delta\Delta f}{4\pi}
\lim_{\omega\rightarrow 0}
\frac{1}{\omega}
\left(
\frac{1}{(\Delta\epsilon-\hbar\omega)^2+\delta^2/4} -  \frac{1}{(\Delta\epsilon+\hbar\omega)^2+\delta^2/4}\right)
\nonumber\\
=&
\frac{\delta\Delta f}{4\pi}
\lim_{\omega\rightarrow 0}
\frac{1}{\omega}
\left(
 \frac{4\Delta\epsilon  \hbar\omega}{((\Delta\epsilon-\hbar\omega)^2+\delta^2/4)((\Delta\epsilon+\hbar\omega)^2+\delta^2/4)}\right)
\nonumber\\
 =&
\frac{\hbar}{\pi}{\delta\Delta f \, \Delta\epsilon}
\lim_{\omega\rightarrow 0}
\left(
 \frac{1}{((\Delta\epsilon-\hbar\omega)^2+\delta^2/4)((\Delta\epsilon+\hbar\omega)^2+\delta^2/4)}\right)
\nonumber\\
 =&
 \frac{\hbar}{\pi}
\frac{\delta\Delta f \, \Delta\epsilon}{\lbrack(\Delta\epsilon)^2+\delta^2/4\rbrack^2},
\label{eq:faL-limit2}
\end{align}
and
\begin{align}
\lim_{\omega\rightarrow 0}
 f^{(a)}_{G}(\omega)
 =&
 \frac{\Delta f}{2\sigma_g \sqrt{\pi}}\;
 \lim_{\omega\rightarrow 0}
 \frac{1}{\omega} 
 \left[ \exp{\left(-\frac{(\Delta\epsilon-\hbar\omega)^2}{\sigma_g^2}\right)}
 - \exp{\left(-\frac{(\Delta\epsilon+\hbar\omega)^2}{\sigma_g^2}\right)}\right]
\nonumber\\
=&
\frac{2\hbar\Delta f\Delta\epsilon}{\sigma_g^3 \sqrt{\pi}}
\exp{\left(-\frac{(\Delta\epsilon)^2}{\sigma_g^2}\right)}\;.
\label{eq:faG-limit2}
\end{align}
But these limits are not the same as those in \eqnref{eq:feL-limit1} and \eqnref{eq:feG-limit1} either. 

Just for completeness let us calculate the limit of the D-d 
expressions for the $f$s also taking into account the even 
parity of $\sigma_1$. For those we have
\begin{align}
\lim_{\omega\rightarrow 0}
 f^{(D-d)}_{L}(\omega)
 =&
\frac{1}{2\pi}\frac{\Delta f}{\Delta \epsilon}
\lim_{\omega\rightarrow 0}
 \left(
 \frac{\delta/2}{(\Delta\epsilon-\hbar\omega)^2+\delta^2/4}
 +
  \frac{\delta/2}{(\Delta\epsilon+\hbar\omega)^2+\delta^2/4}
\right)
\nonumber\\
=&
\frac{1}{\pi}
\frac{\Delta f}{\Delta \epsilon}
 \frac{\delta/2}{(\Delta\epsilon)^2+\delta^2/4},
 \label{eq:feL-limit2}
\end{align}
and
\begin{align}
\lim_{\omega\rightarrow 0}
 f^{(e)}_{G}(\omega)
 =&
 \frac{\Delta f}{2\Delta \epsilon} 
 \frac{1}{\sigma_g \sqrt{\pi}}\;
 \lim_{\omega\rightarrow 0}
 \left[\exp{\left(-\frac{(\Delta\epsilon-\hbar\omega)^2}{\sigma_g^2}\right)}
 +\exp{\left(-\frac{(\Delta\epsilon+\hbar\omega)^2}{\sigma_g^2}\right)}\right]
\nonumber\\
=&
 \frac{\Delta f}{\Delta \epsilon} 
 \frac{1}{\sigma_g \sqrt{\pi}}\;
 \exp{\left(-\frac{(\Delta\epsilon)^2}{\sigma_g^2}\right)},
 \label{eq:feG-limit2}
\end{align}
results which are identical with \eqnref{eq:feL-limit1} 
and \eqnref{eq:feG-limit1}.

The dc expressions then are 
\begin{equation}
 \sigma_{dc}^{D-d,L}=\frac{2 e^2 \hbar^3}{m_e^2 V}
 \sum_{m}\sum_{m'}
 \frac{\Delta f_{m'm}}{\Delta\epsilon_{mm'}}
 \bra{m}\nabla\ket{m'}\bra{m'}\nabla\ket{m}
 \frac{\delta/2}{(\Delta\epsilon_{mm'})^2+\delta^2/4},
 \label{eq:sigma1-exact-lor-dc}
\end{equation}

\begin{equation}
 \sigma_{dc}^{D-d,G}=
 \frac{2 \sqrt{\pi} e^2 \hbar^3}{m_e^2 V\sigma_g}
 \sum_{m}\sum_{m'}
 \frac{\Delta f_{m'm}}{\Delta\epsilon_{mm'}}
 \bra{m}\nabla\ket{m'}\bra{m'}\nabla\ket{m}
 \; \exp{\left(-\frac{(\Delta\epsilon_{mm'})^2}{\sigma_g^2}\right)},
 \label{eq:sigma1-exact-ggau-dc}
\end{equation}

\begin{equation}
 \sigma_{dc}^{a,L}=
 \frac{4 e^2 \hbar^3}{m_e^2 V}
 \sum_{m}\sum_{m'}
 {\Delta f_{m'm}} \Delta\epsilon_{mm'}
 \bra{m}\nabla\ket{m'}\bra{m'}\nabla\ket{m}
 \frac{\delta/2}{((\Delta\epsilon_{mm'})^2+\delta^2/4)^2},
 \label{eq:sigma1-app-lor-dc}
\end{equation}
and
\begin{equation}
 \sigma_{dc}^{a,G}=
 \frac{4 \sqrt{\pi} e^2 \hbar^3}{m_e^2 V\sigma_g^3}
 \sum_{m}\sum_{m'}
 {\Delta f_{m'm}} \Delta\epsilon_{mm'}
 \bra{m}\nabla\ket{m'}\bra{m'}\nabla\ket{m}
\; \exp{\left(-\frac{(\Delta\epsilon_{mm'})^2}{\sigma_g^2}\right)}\;.
 \label{eq:sigma1-app-gau-dc}
\end{equation}

This simple analysis shows that in general the approximated 
$\sigma_1$ expressions \eqnref{eq:sigma1-general-2b}  do not have correct
low-frequency behavior, nor does the D-d form when evaluated with a
Gaussian.  Only the D-d expression with the Lorentzian recovers the
exact limit of $\sigma_1$ for any value of the delta-width. 

\begin{figure}
\includegraphics[width=0.48\textwidth,viewport=1 1 740 495,clip]
{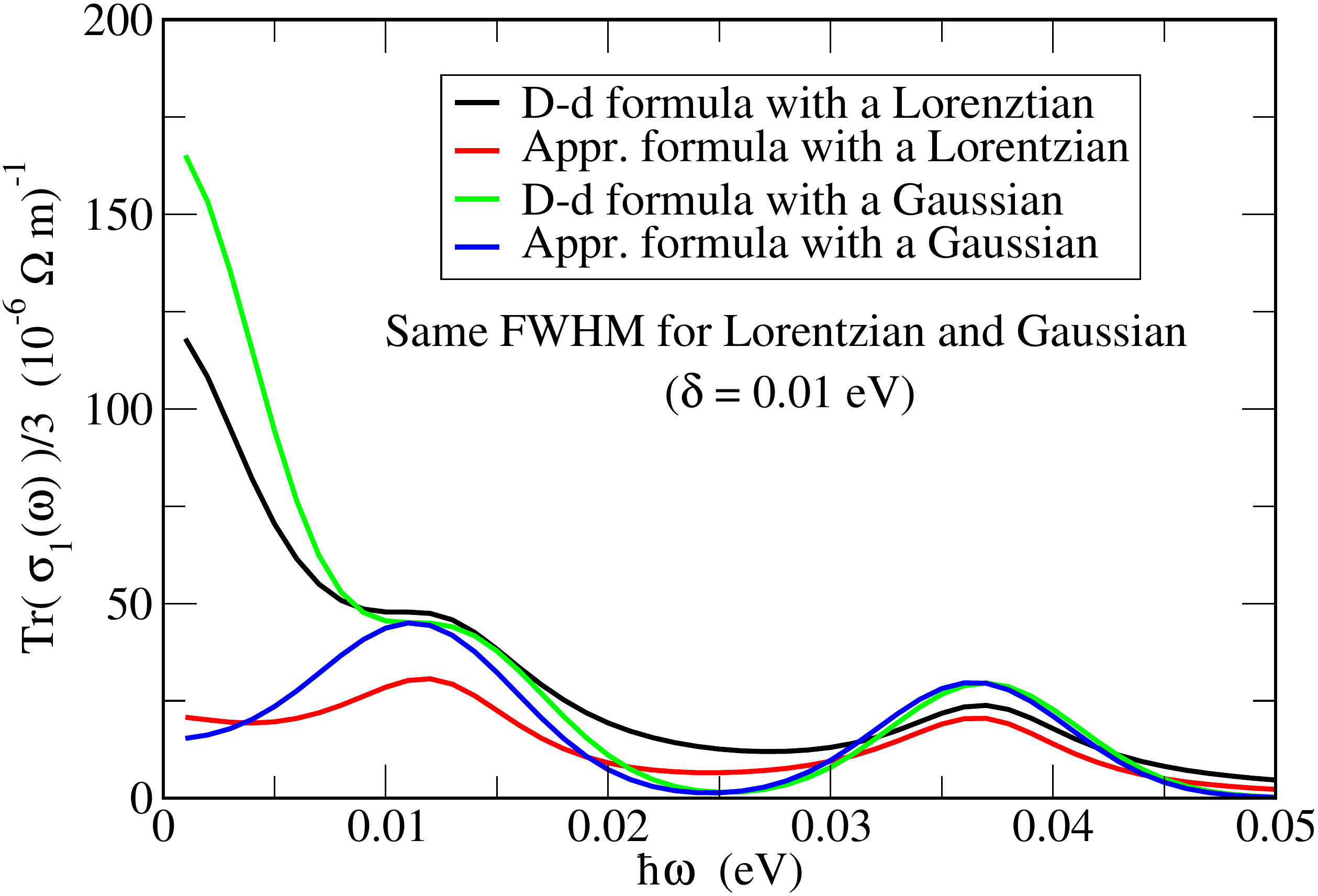}%
\includegraphics[width=0.48\textwidth,viewport=1 1 740 495,clip]
{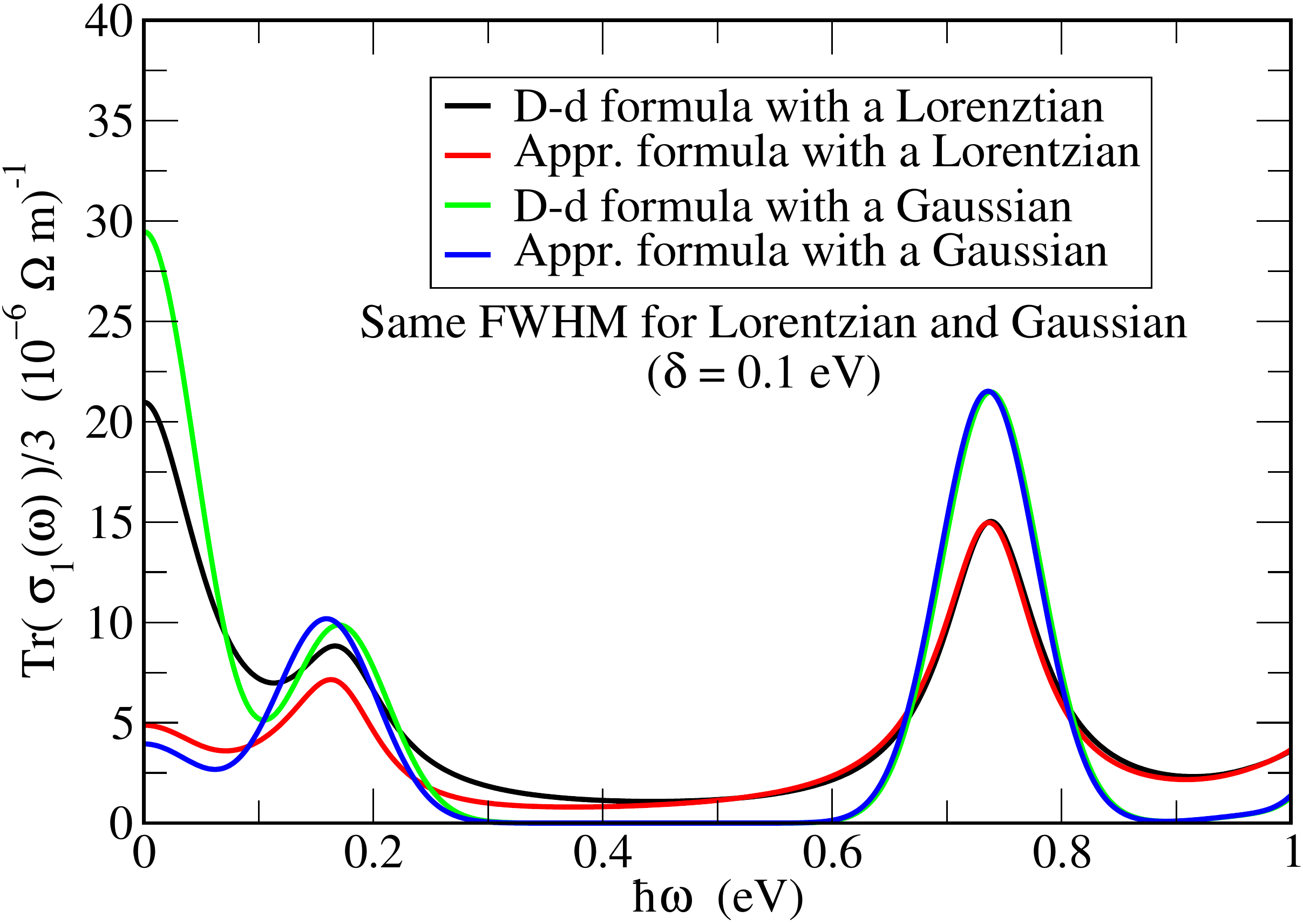}\\%
\includegraphics[width=0.48\textwidth,viewport=1 1 740 495,clip]
{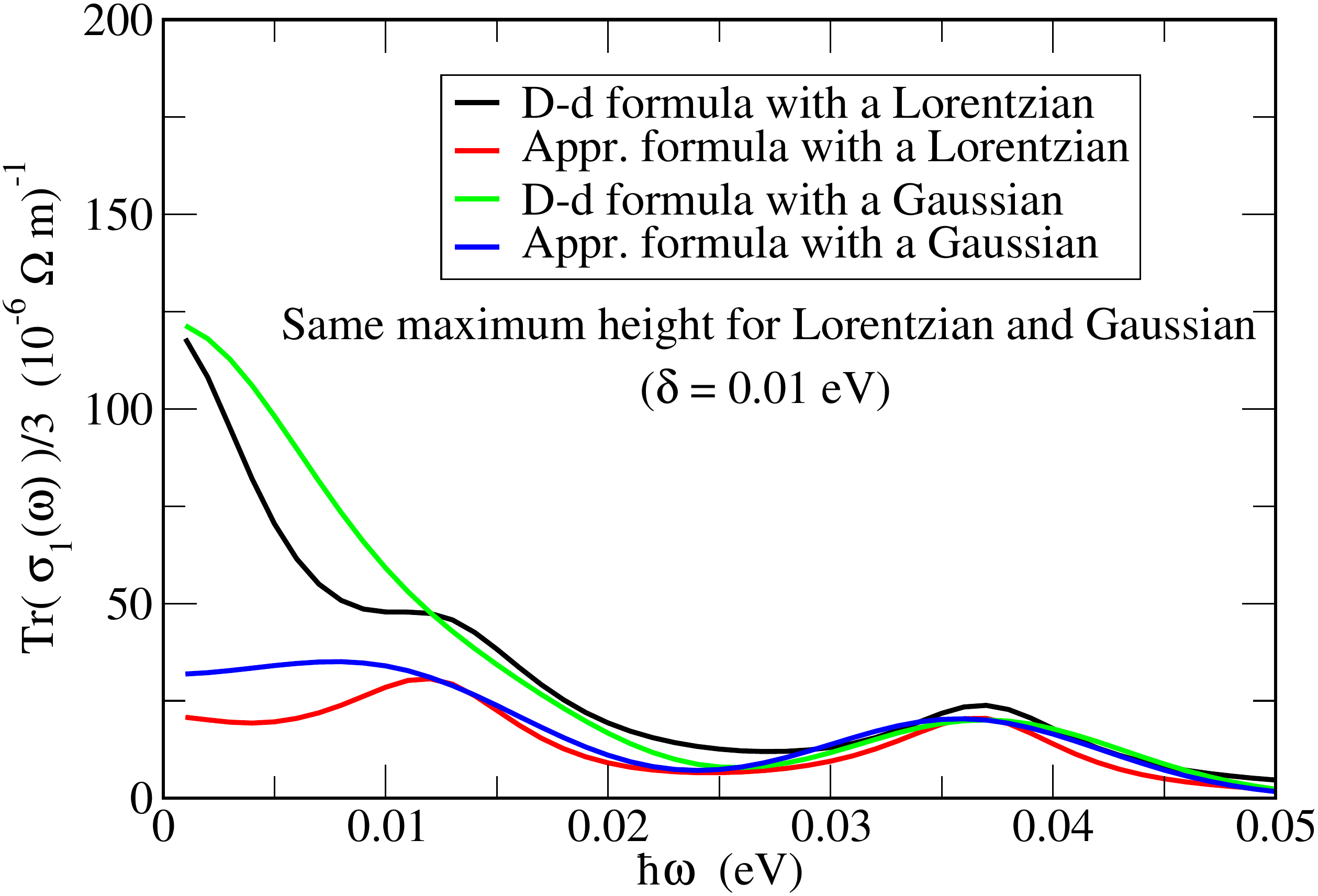}%
\includegraphics[width=0.48\textwidth,viewport=1 1 740 495,clip]
{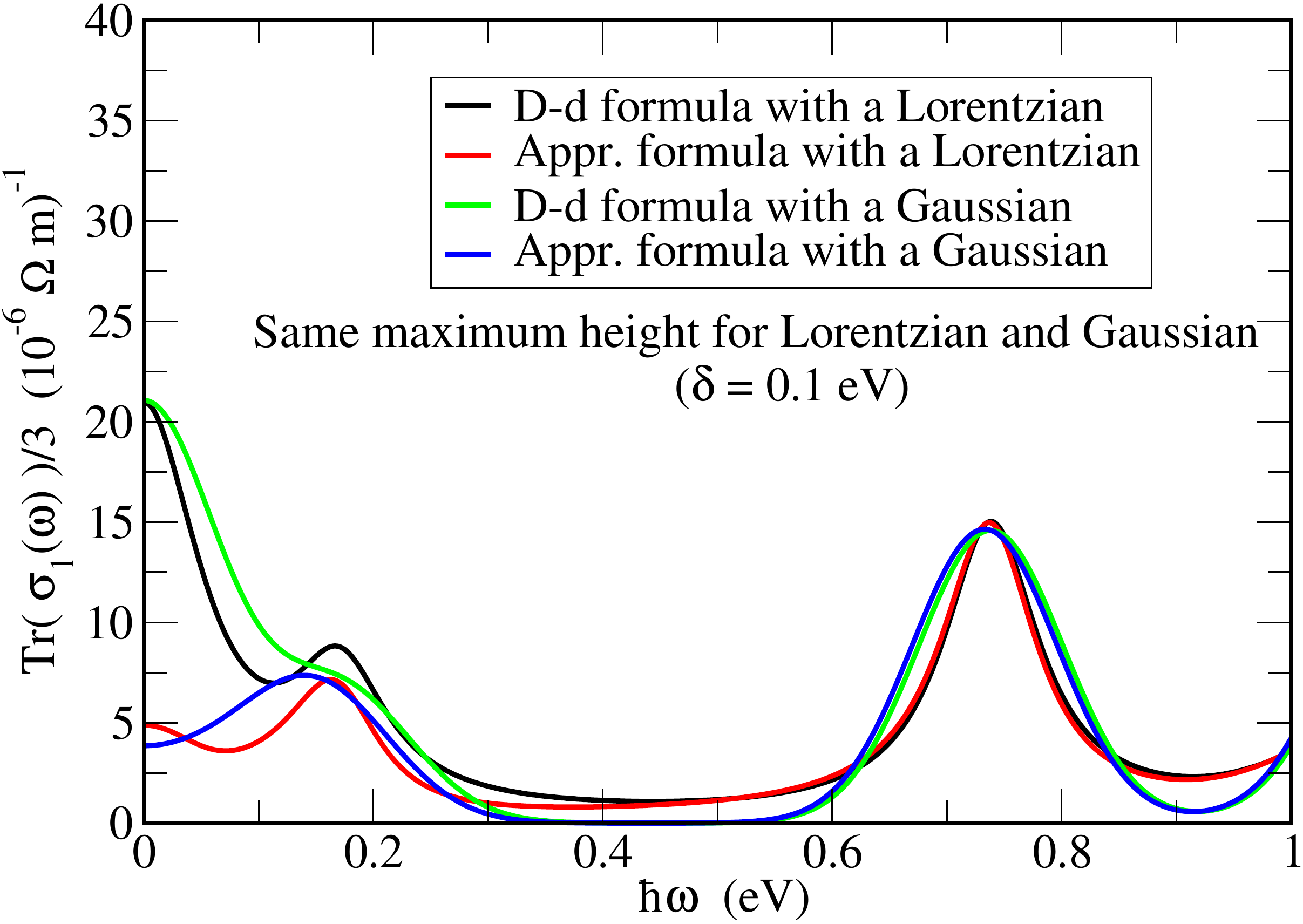}%
\caption{\label{fig:effect-of-delta-rep-al4}  
$Tr( \sigma_1)/3$ calculated with approximations for the Dirac delta function for Al fcc with four atoms per unit cell density $= 2.7$ g/cm$^3$ and  
temperature $= 31.6$ kK. 
The left column shows two figures for $\delta=0.01$ eV: the  upper one 
shows results for the same FWHM for the Lorentzian and Gaussian 
representations, while the lower one is for the same maximum heights. 
The right column shows the same comparison for the case $\delta=0.1$ eV.
In all panels the  curves labeled ``D-d formula ...'' show $Tr(\sigma_1)/3$ calculated with \eqnref{eq:sigma1-exact} and 
curves labeled  ``Appr.\ formula ...'' show $Tr(\sigma_1)/3$ calculated with \eqnref{eq:sigma1-general-2b}.
}
\end{figure}

Numerical examples of the behavior of the $\sigma_1$ conductivity
expressions are provided in two sets of figures. The first set
(\figref{fig:effect-of-delta-rep-al4}) shows results from calculations
performed for fcc Al with four atoms per unit cell at a density of
$2.7$ g/cm$^3$ and a temperature of $31.6$ kK. To compare the effect of
the Lorentzian versus Gaussian we did a set of calculations matching their
full width at half-maximum (FWHM), and another matching their maximum heights.

As anticipated analytically, in general the approximated formulae lead
to incorrect dc values, and also distort the spectra (peak shapes are
changed) at low frequencies. The D-d formula with matched maximum
heights for the Lorentzian and Gaussian leads to similar dc values,
albeit with more distortion of the peak shape introduced by the
Gaussian. Matching of the FWHM yields incorrect dc values but improves
the line shapes for the Gaussian.

\begin{figure}
\includegraphics[width=0.48\textwidth,viewport=1 1 740 495,clip]
{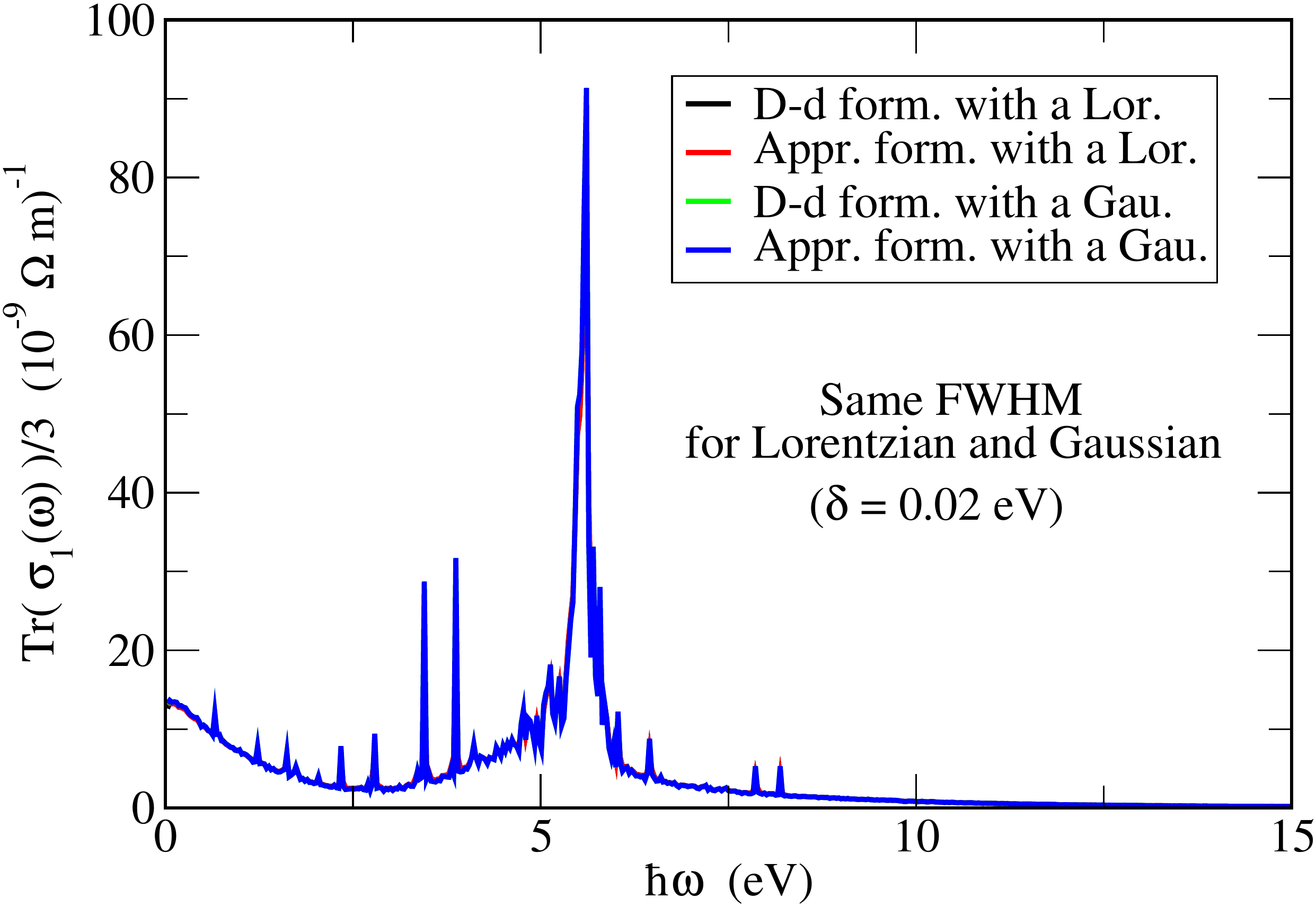}%
\includegraphics[width=0.48\textwidth,viewport=1 1 740 495,clip]
{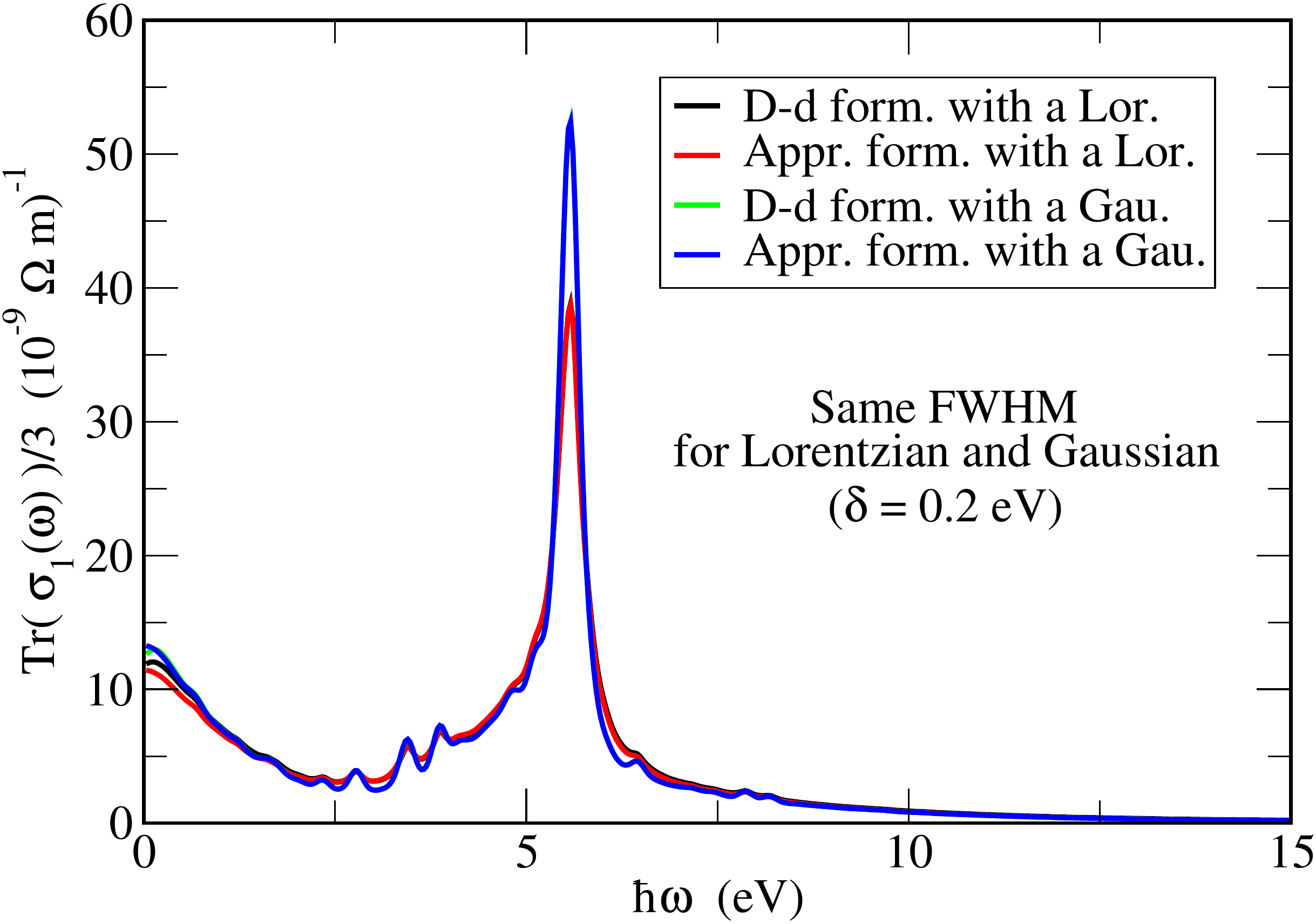}\\%
\includegraphics[width=0.48\textwidth,viewport=1 1 740 495,clip]
{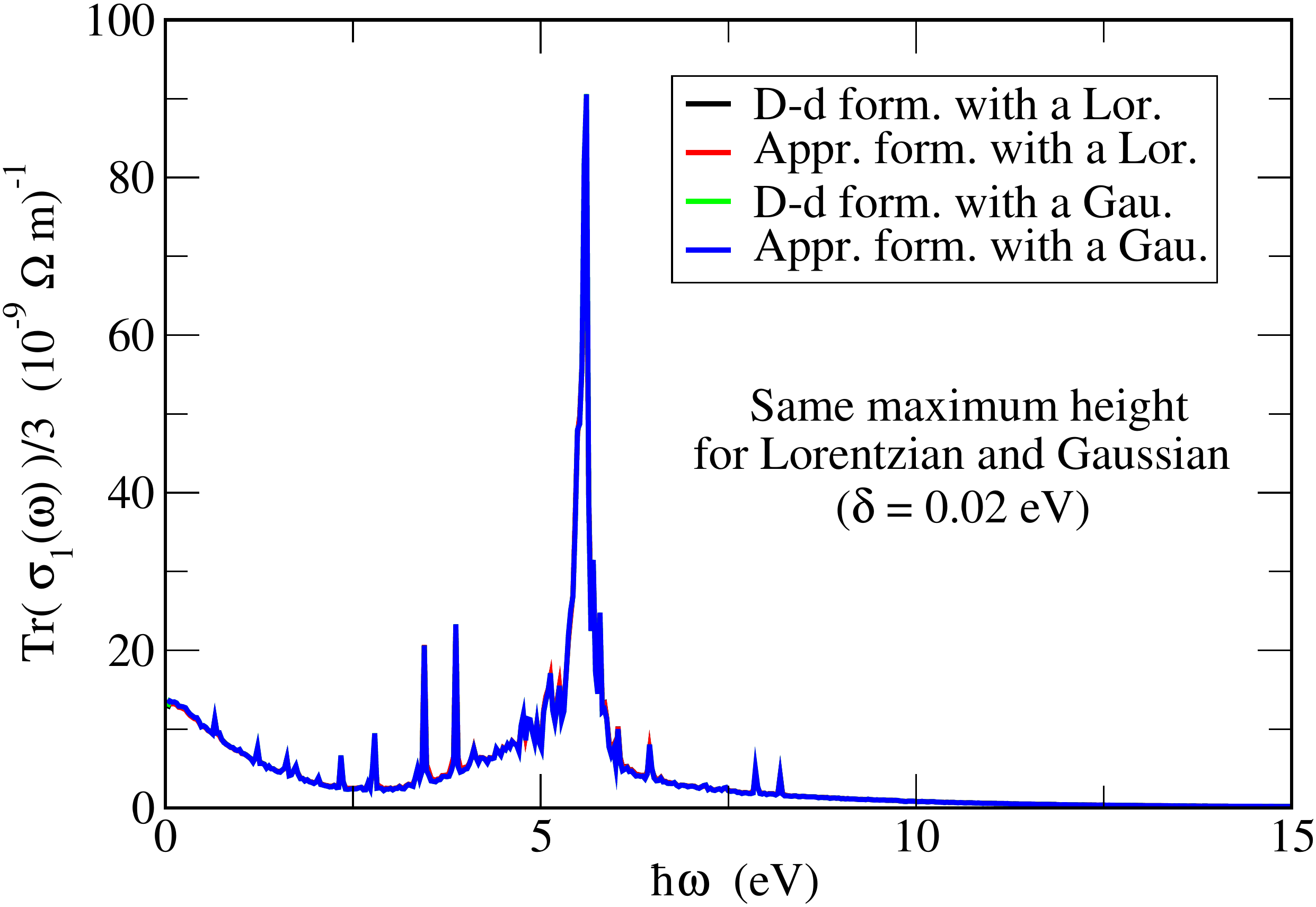}%
\includegraphics[width=0.48\textwidth,viewport=1 1 740 495,clip]
{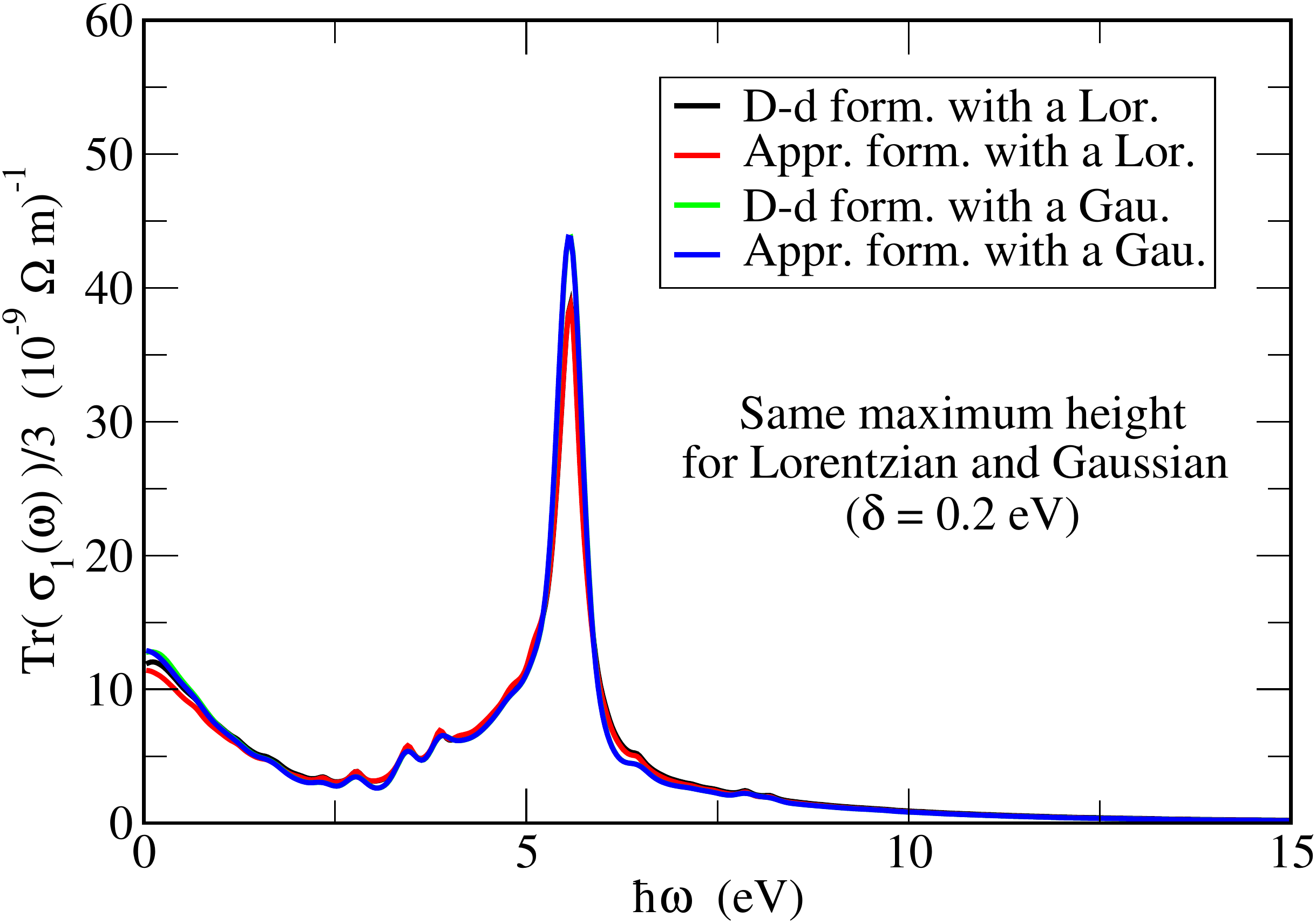}%
\caption{\label{fig:effect-of-delta-rep-al16}
$Tr( \sigma_1)/3$ calculated with different representations of the Dirac delta function for a molecular dynamics step of 
Al with $16$ atoms per unit cell at a density of $0.1$ g/cm$^3$ and a temperature of $30$ kK. Otherwise as in \figref{fig:effect-of-delta-rep-al4}.
}
\end{figure}

The second set of figures (\figref{fig:effect-of-delta-rep-al16})
shows calculations done for the ionic configuration from a molecular dynamics step of Al with 16 
atoms per unit cell at 0.1 g/cm$^3$ and a temperature of 30kK for 8 kps and 3096
bands.  The panels are ordered the same way as in the preceding figure. 
In contrast with
\figref{fig:effect-of-delta-rep-al4}, they show that there are cases
for which the delta function representations and width values are 
less important for the dc conductivity.

Another interesting aspect shown in \figref{fig:effect-of-delta-rep-al16} is that the smaller value of $\delta$
produces better convergence of all the delta function representations
in the case of the MD step even when the spectrum gets noisier
than the one calculated with a larger value of $\delta$.

Because of the relatively high temperature, in both cases the inter-band
contributions dominate the dc conductivity and therefore the
conductivity is not of Drude nature although the graphs look
Lorentzian-like close to zero.

To get an idea of why the results are so different, we compare the
cases of the D-d formula with the Lorentzian and the corresponding
Gaussians with matching FWHM and maximum height. For that it will
prove convenient to re-write the dc conductivity first in terms of one
sum over bands by reducing the pair of $mm^\prime$ labels to one $i$-label,
that is

\begin{equation}
 \sigma_{dc}=
 \frac{2 \pi e^2 \hbar^2}{3m_e^2 \Omega}
 \sum_{k} w_{\v{k}}
 \sum_i
 \frac{\Delta f_{i\v{k}}}{\Delta \epsilon_{i\v{k}}}
 \sum_\alpha
| \bra{\Psi_{m(i)\v{k}}}\nabla_\alpha\ket{\Psi_{m'(i)\v{k}}}|^2
 \delta(\Delta\epsilon_{i\v{k}}),
 \label{eq:sigma1-dc-one-band-index}
\end{equation}
and secondly by introducing $N(\Delta\epsilon_j)$ as the number of pairs 
of states with the same difference in energy
$\Delta \epsilon_j$ to get
\begin{equation}
 \sigma_{dc}=
 \frac{2 \pi e^2 \hbar^2}{3m_e^2 \Omega}
 \sum_j
N(\Delta\epsilon_j)
\left[
 \sum_{\v{k}} w_{\v{k}}
 \sum_i
 \delta_{\Delta \epsilon_{i\v{k}} \Delta \epsilon_j}
 \frac{\Delta f_{i\v{k}}}{\Delta \epsilon_{i\v{k}}}
 \sum_\alpha
| \bra{\Psi_{m(i)\v{k}}}\nabla_\alpha\ket{\Psi_{m'(i)\v{k}}}|^2
 \right]
 \delta(\Delta\epsilon_j).
 \label{eq:sigma1-dc-grouping-same-energy-difference}
\end{equation}

\begin{figure}
\begin{center}
\includegraphics[width=\textwidth,viewport=1 1 710 495,clip]{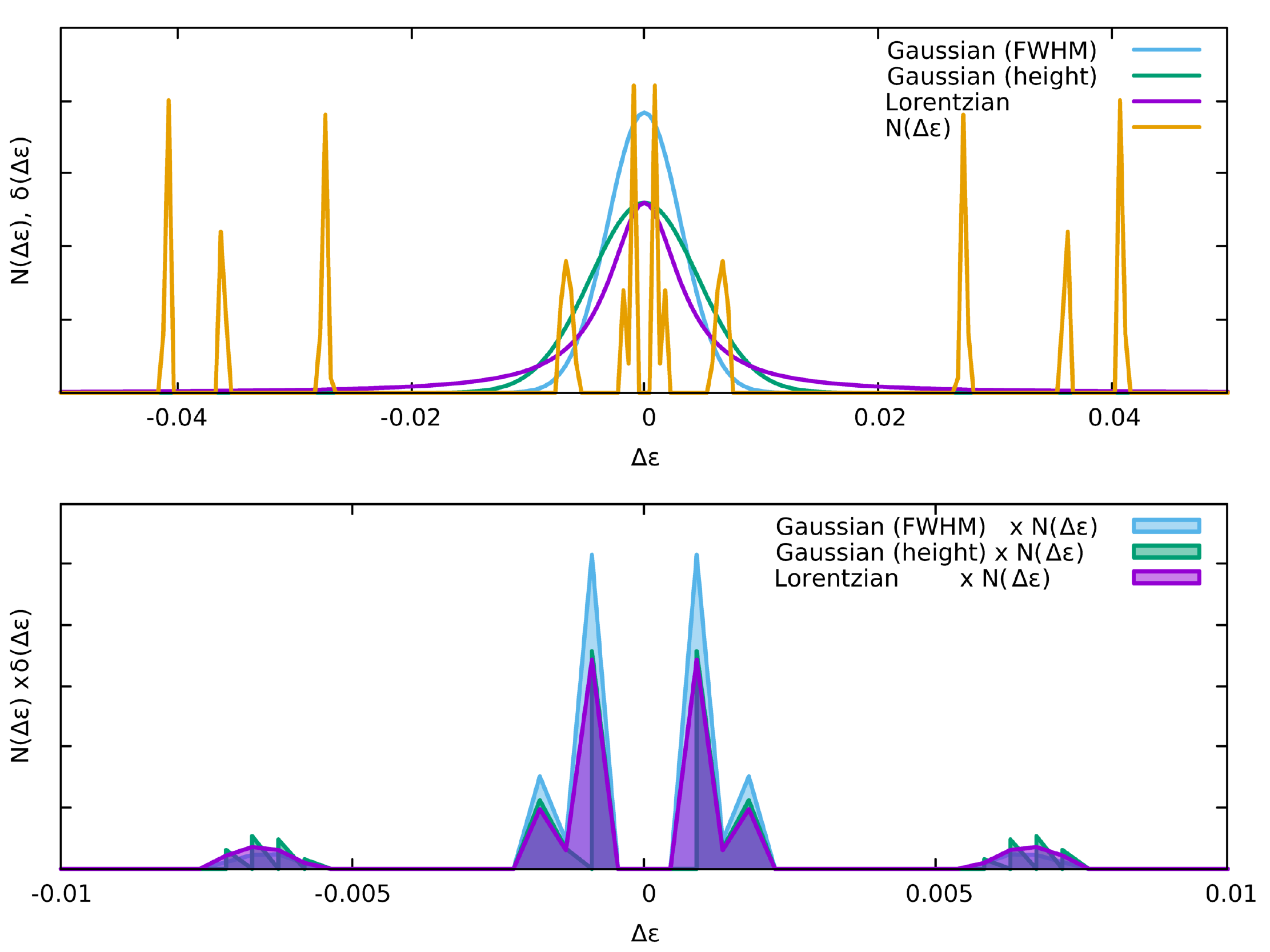}
\end{center}
\caption{\label{fig:effect-of-delta-rep-inter-bands-al4} 
Dirac delta-function representations and number of pairs of bands with 
the same inter-band energy difference (top), and their 
product (bottom) for Al fcc with density of $2.7$ g/cm$^3$ at T=$31.6$kK.  }
\end{figure}

Plots of the three approximate representations of the Dirac delta function
$\delta(\Delta\epsilon)$ and $N(\Delta\epsilon)$ for inter-band
contributions (top), as well as their product (bottom),
are given in
\figtworef{fig:effect-of-delta-rep-inter-bands-al4}{fig:effect-of-delta-rep-inter-bands-al16}
for the respective examples given in
\figtworef{fig:effect-of-delta-rep-al4}{fig:effect-of-delta-rep-al16}.

The sparsity of the $N(\Delta\epsilon)$ for fcc Al leads to functions
with different areas when multiplied by the different delta
function representations. In contrast, for the example from the MD step, the
disorder is reflected in an almost uniform $N(\Delta\epsilon)$ which
in turn yields distributions with almost the same area when 
multiplied with the approximate delta function representations.
Therefore, it is the sparsity of the distribution of differences in
energies $N(\Delta\epsilon)$ that seems to determine the success of the
different delta-function representations in the dc conductivity calculation.

\begin{figure}
\begin{center}
\includegraphics[width=\textwidth,viewport=1 1 710 495,clip]{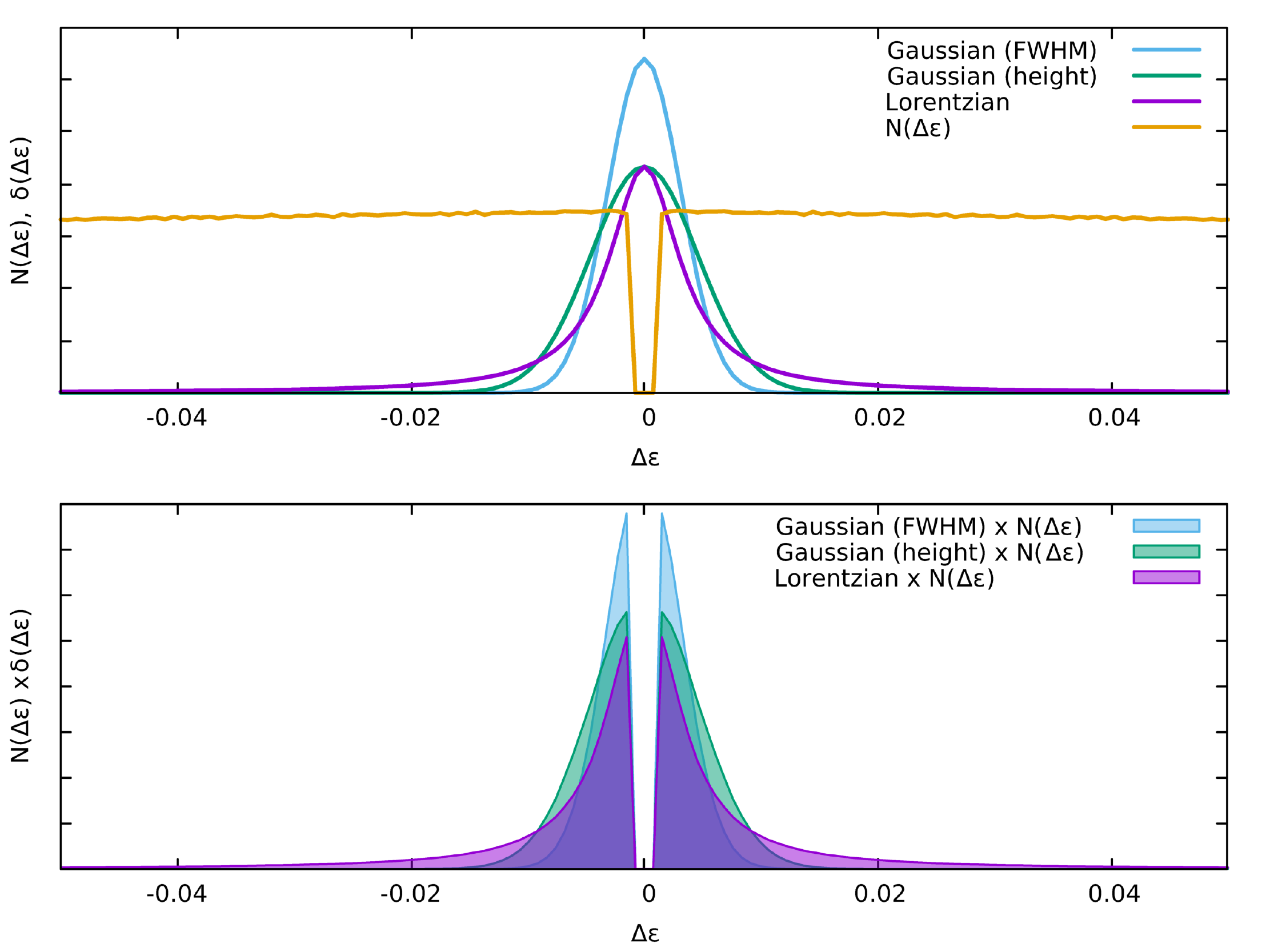}
\end{center}
\caption{\label{fig:effect-of-delta-rep-inter-bands-al16} 
Dirac delta-function representations and number of pair of bands with the 
same inter-band energy difference (top), and their product (bottom) 
for a molecular dynamics step of Al at $0.1$ g/cm$^3$ and T=$30$ kK.}
\end{figure}

\subsection{\label{sec:paw-quality}PAW quality}

\begin{figure}
\includegraphics[width=\textwidth]{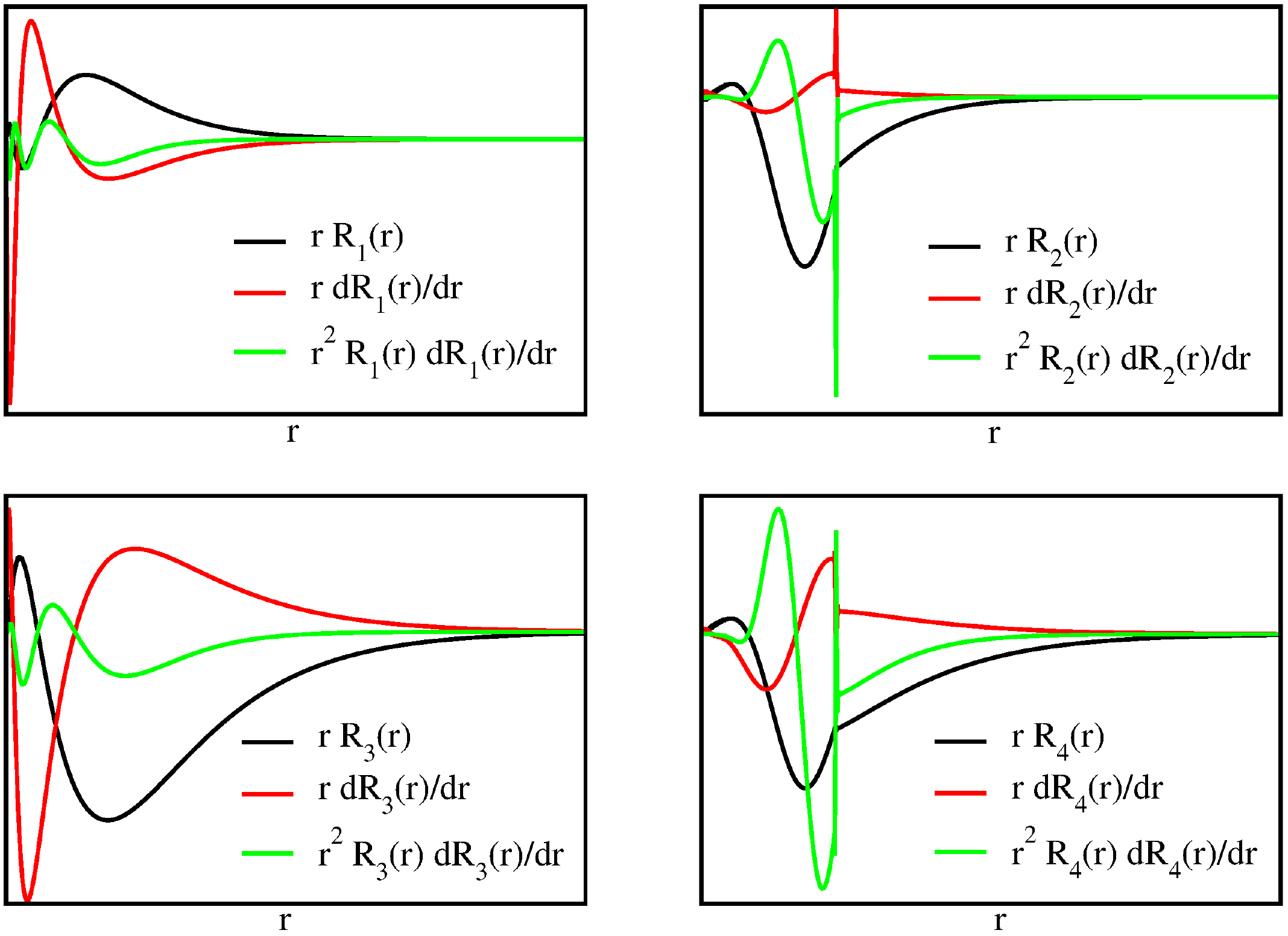}
\caption{\label{fig:paw-problems-aewfc} Atomic $3s$ ($R_1$ in the plot) and $3p$ ($R_3$ in the plot) wave functions of Al and their corresponding augmented waves ($R_2$ for $3s$ and $R_4$ for $3p$).}
\end{figure}

\begin{figure}
\includegraphics[width=\textwidth]{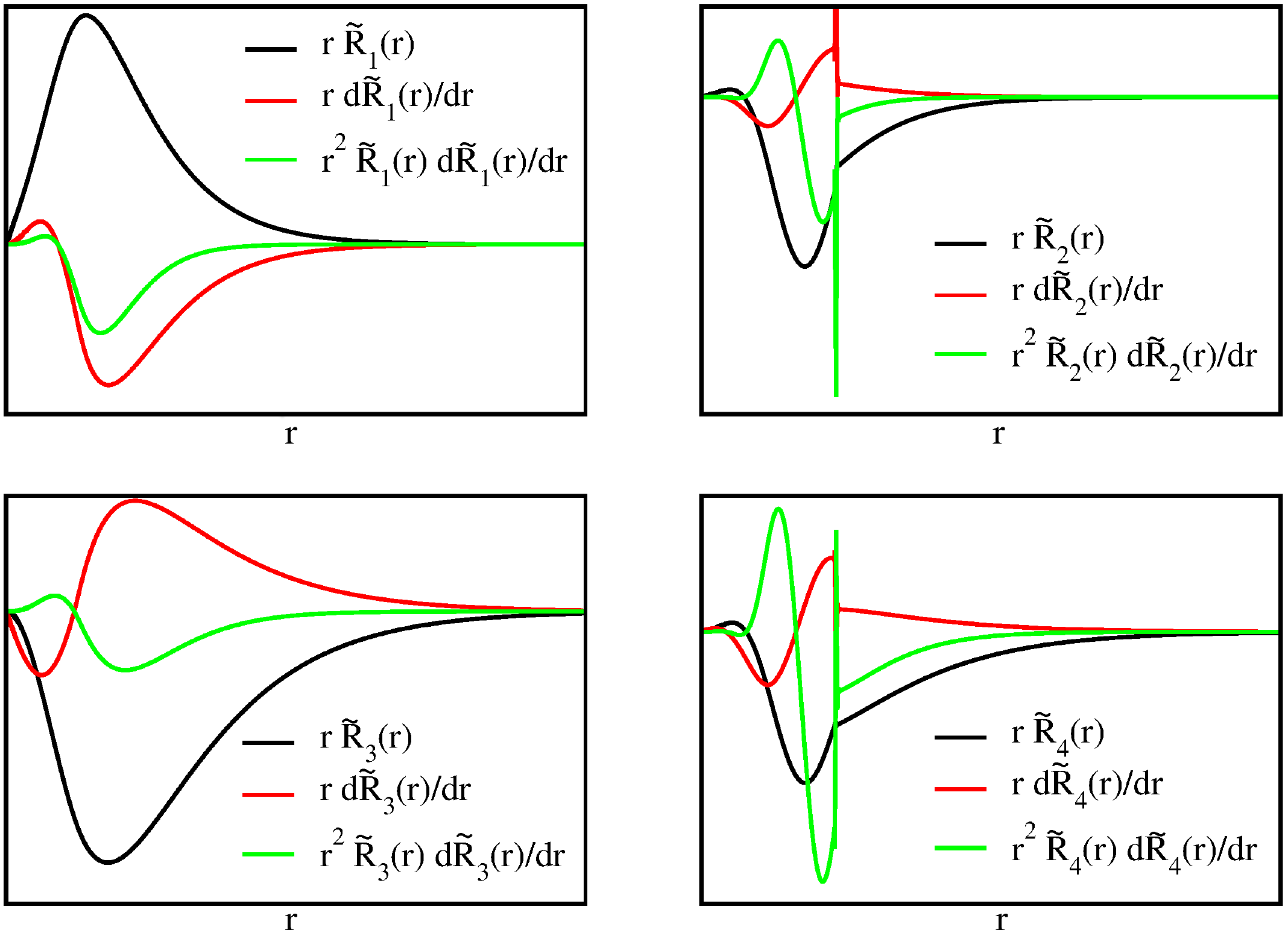}
\caption{\label{fig:paw-problems-pswfc} Atomic $3s$ ($\tilde{R}_1$ in the plot) and $3p$ ($\tilde{R}_3$ in the plot) pseudized wave functions of Al and their corresponding pseudized-augmented waves ($\tilde{R}_2$ for $3s$ and $\tilde{R}_4$ for $3p$).}
\end{figure}

During development of KGEC, we noticed some problems with the
numerical derivatives involved in the calculation of the gradient
matrix elements in the PAW approach.  Close inspection of the radial
atomic wave functions and pseudo-wave functions revealed that there
seems to be a systematic problem in the generation of the augmented
waves that is carried over to the corresponding pseudo-waves. The
situation is represented in \figref{fig:paw-problems-aewfc} for the
atomic wave functions and in \figref{fig:paw-problems-pswfc} for the
pseudo wave functions generated for Al with three valence electrons 
and four projectors.
Notice from \figref{fig:paw-problems-pswfc} that the pseudo waves
$\tilde{R}_1(r)$ and $\tilde{R}_3(r)$, generated from the $3s$ and
$3p$ natural atomic states respectively, are smooth but $\tilde{R}_2(r)$ and
$\tilde{R}_4(r)$, which were generated from the corresponding 
augmented-pseudized waves $\tilde{R}_1(r)$ and $\tilde{R}_3(r)$ 
respectively, are not.

The problem can be traced to the augmented waves themselves as
corroborated by \figref{fig:paw-problems-aewfc}.  We found the problem
in four different PAW data sets generated with ATOMPAW
\cite{HOLZWARTH2001329} and LD1 \cite{QE2009}.  This issue may need a
bit more investigation, but it seems that the cancellation of
errors that occurs in $\spr{R}{dR/dr}-\spr{\tilde{R}}{d\tilde{R}/dr}$
provides a way to get accurate results for properties calculated in
the PAW scheme.

Another cancellation that occurs is in the product of the projectors
and the pseudo-wave functions.  That is given by the projectors being
strictly zero starting at the radii where discontinuities in the first
derivative of the pseudo wave functions appear and extending all the
way to infinity. That should keep the dual orthogonality intact but
it is not guaranteed, as illustrated by some tests described next.

\subsection{PAW duality and wave functions orthogonality problems }
Another difficulty observed during code development is related to
the orthogonality required between pseudo-wavefunctions and
projectors.  Depending on the PAW dataset, there could be failures
with errors larger 
than $10^{-4}$.  That led to inclusion of a check for the dual
orthogonality condition in KGEC.  

There also are cases in which the reconstructed PAW all-electron orbitals
are not orthogonal. This problem can be related to the dual
orthogonality problem just described, but it can also arise from an
inadequate plane wave basis set. KGEC is able to check for that sort
of problem as well; it provides warnings and points the user to another
output file for more information.

\section{\label{sec:remarks}Remarks}

From the perspective of a new computational implementation, we have
reviewed the state of the art of electrical conductivity calculations
using the Kubo Greenwood (KG) approach and derived all the necessary
analytical expressions for its implementation using PAW data sets with
a plane wave basis set. The analysis and derivations were done for
both the original KG formula and it most popular version, which we
have found contains approximations that often do not lead to the same
results as the original one.

The derived formulae were used to design a user-friendly algorithm
with capabilities to face the challenges of simulations of matter
under extreme conditions. The algorithms have been coded in modular Fortran
90 as a post-processing tool for Quantum Espresso. Named KGEC, from 
the initials of ``Kubo-Greenwood Electrical Conductivity'', the 
code has the following special features:
\begin{itemize}
 \item  Calculates the full complex conductivity tensor, not just the average trace.
 \item  Uses either the original KG formula or the more popular, although approximated one in terms of a Dirac delta function.
 \item Performs a decomposition into intra- and inter-band contributions as well as degenerate state contributions.
 \item  Calculates the direct-current conductivity tensor directly.
 \item  Provides both Gaussian and Lorentzian representations of the Dirac delta function.
 \item  Provides MPI parallelization over k-points, bands and plane waves, with an option to recover the plane waves process for their use in bands parallelization as well.
 \item  Gives faster convergence with respect to k-point density than the implementation in the Abinit code.
\end{itemize}
KGEC is downloadable from http://www.qtp.ufl.edu/ofdft under GPL.

These features make the code versatile and innovative.  There also 
are several underlying advances. An example is that the
calculation of the direct-current tensor using the most popular KG
formula is based on the removal of the singularity at zero frequency,
an approach 
not reported before.  That leads to analytical formulae, with the result 
that no fitting of a Drude term by the user is needed. Another example is
the analysis which undergirds the systematic inclusion of both intra-band and degenerate
state contributions in KGEC.  A third example is the recovery of the plane waves MPI-processes on the fly, a procedure 
based on redefinition of the communicators and exploitation of MPI
Single-Program-Multi-Data (SPMD) characteristics.

The code should have wide, deep impact on the calculation of
electrical conductivities of materials ranging from small to large
systems in normal to extreme environments. On one hand the possibility
of doing full tensor calculations with no ballistic approximation
should make the code attractive for the simulation of electronic
materials. On the other, its parallel capabilities are very useful to
accelerate simulations in general, but especially for large systems at high
temperatures. For those,  the plane wave cutoff energies and number of bands
are very large, its parallel capabilities, including the recovery on
the fly of idle processes, should make of KGEC an essential tool.

In the near future, the next release of our group's Profess@QuantumEspresso \cite{KARASIEV20143240}  will
include our new finite-temperature generalized gradient 
approximation (GGA)functional \cite{arXiv1612.06266v2}.  QE compiled for the
Profess@QE suite is compatible
with KGEC, so full free-energy DFT electrical conductivity calculations at the
GGA level of refinement will be possible. 
  Farther out,  work within the context of electrical conductivity
 is likely to include  incorporation of spin
polarization, non-local corrections to the gradient matrix elements
for systematic use of conventional non-local pseudopotentials, and 
inclusion of spin-orbit corrections.  More broadly, we are considering
generalizing to calculation of the thermal conductivity via general
calculation of Onsager coefficients.


%
%
\setcounter{secnumdepth}{0}

\section{Acknowledgments}
We thank Xavier Gonze and Vanina Recoules for helpful 
conversations and Kai Luo, 
 Travis Sjostrom, and DeCarlos Taylor for beta testing.  We 
acknowledge the support of the U.S.\ Dept.\ of Energy via grant DE-SC0002139
and thank the University of Florida Research Computing organization 
for computational resources. 

\clearpage
\setcounter{secnumdepth}{1}

\appendix

\section{\label{app:csh}Spherical harmonic definitions}

The complex spherical harmonics are given by\cite{jackson_classical_1999}
\begin{align}
 Y_{lm}(\theta,\varphi)=\sqrt{\frac{2l+1}{4\pi}}\sqrt{\frac{(l-m)!}{(l+m)!}} P_{l}^{m}(\cos\theta) e^{im\varphi}
\end{align}
with the sign conventions and definitions of the associated Legendre polynomials $P_{l}^{m}(x)$.

In the context of the PAW method it also is useful to use the real spherical harmonics, defined as 
\begin{equation}
 S_{lm}(\theta,\varphi)= \sqrt{\frac{2l+1}{4\pi}}\sqrt{\frac{(l-|m|)!}{(l+|m|)!}} P_{l}^{|m|}(\cos\theta)
   \begin{cases}
     \sqrt{2}\sin |m|\theta,&  m < 0 \\
     {1}, & m=0 \\
     \sqrt{2}\cos m\theta,& {m > 0} 
  \end{cases}
\end{equation}

\section{$I$ integrals \label{app:i-integrals-csh} for complex spherical harmonics}
In the usual Cartesian coordinate system, the unit vectors relative to the spherical coordinates are 
\begin{align}
 \hat{\v{e}}_r      =& \sin(\theta) \cos(\varphi)\, \hat{\v{e}}_x  + \sin(\theta) \sin(\varphi)\, \hat{\v{e}}_y + \cos(\varphi)\, \hat{\v{e}}_z
\nonumber\\
 \hat{\v{e}}_\theta =& \cos(\theta) \cos(\varphi)\, \hat{\v{e}}_x  + \cos(\theta) \sin(\varphi)\, \hat{\v{e}}_y - \sin(\varphi)\, \hat{\v{e}}_z
\nonumber\\
 \hat{\v{e}}_\phi   =& -\sin(\varphi)\, \hat{\v{e}}_x  + \cos(\varphi)\, \hat{\v{e}}_y.
\end{align}

For $ \v{I}^{(r)}_{ll'mm'}$  we have 

\begin{align}
\v{I}^{(r)}_{ll'mm'}=
{\int_0^{\pi} \sin(\theta) d\theta \int_0^{2\pi} d\varphi Y^*_{lm}(\theta,\varphi)  Y_{l' m'}(\theta,\varphi )
\hat{ \v{e} }_r(\theta,\varphi),
} 
\end{align}
which yields
\begin{align}
I^{(r)}_{ll'mm', \hat x}=&
\int_0^{\pi} \sin(\theta) d\theta \int_0^{2\pi} d\varphi Y^*_{lm}(\theta,\varphi)  Y_{l' m'}(\theta,\varphi )
\sin(\theta) \cos(\varphi)
\nonumber
\\
=&
C_{lm} C_{l'm'} \int_0^{\pi} \sin^2(\theta)   
{P_{l}^{m}(\cos(\theta)}  P_{l'}^{m'}(\cos(\theta) )
d\theta
{\int_0^{2\pi}} \cos(\varphi) e^{i(m'-m)\varphi} d\varphi
\nonumber
\\
=&
\underbrace{2\pi C_{lm} C_{l'm'}
 \int_{-1}^{1} \sqrt{1-x^2}  
{P_{l}^{m}(x)}  P_{l'}^{m'}(x )
dx}
_{P^{(1)}_{lml'm'}}
\underbrace{\frac{1}{2\pi}
\int_0^{2\pi} \cos(\varphi) e^{i(m'-m)\varphi} d\varphi}
_{A^{(c)}_{mm'}}
\nonumber\\
=&
P^{(1)}_{lml'm'}A^{(c)}_{mm'},
\end{align}
where 
\begin{equation}
C_{lm}=\sqrt{\frac{2l+1}{4\pi}}\sqrt{\frac{(l-|m|)!}{(l+|m|)!}} \;.
\end{equation}
Then
\begin{align}
I^{(r)}_{ll'mm', \hat y}=&
{\int_0^{\pi} \sin(\theta) d\theta \int_0^{2\pi} d\varphi Y^*_{lm}(\theta,\varphi)  Y_{l' m'}(\theta,\varphi )
\sin(\theta) \sin(\varphi)
} 
\nonumber
\\
=&
\underbrace{{2\pi C_{lm} C_{l'm'}} \int_{-1}^{1} \sqrt{1-x^2}  
{P_{l}^{m}(x)}  P_{l'}^{m'}(x )
dx}
_{P^{(1)}_{lml'm'}}
\underbrace{\frac{1}{2\pi} \int_0^{2\pi} \sin(\varphi) e^{i(m'-m)\varphi} d\varphi}
_{A^{(s)}_{mm'}}
\nonumber\\
=&
P^{(1)}_{lml'm'}A^{(s)}_{mm'};
\end{align}
\begin{align}
I^{(r)}_{ll'mm', \hat z}=&
\int_0^{\pi} \sin(\theta) d\theta \int_0^{2\pi} d\varphi Y^*_{lm}(\theta,\varphi)  Y_{l' m'}(\theta,\varphi )
\cos(\theta)
\nonumber
\\
=&
2\pi C_{lm} C_{l'm'} \delta_{mm'}
\int_0^{\pi} \sin(\theta) {P_{l}^{m}(\cos(\theta))}  P_{l'}^{m'}(\cos(\theta) )
\cos(\theta)
d\theta
\nonumber
\\
=&
\underbrace{2\pi C_{lm} C_{l'm'} 
\int_{-1}^{1} 
x {P_{l}^{m}(x)}  P_{l'}^{m'}(x)
dx}
_{P^{(2)}_{lml'm'}} 
\delta_{mm'}
\nonumber\\
=&
{P^{(2)}_{lml'm'}} 
\delta_{mm'};
\end{align}

For $\v{I}^{(\theta)}_{ll'mm'}$  we have 
\begin{align}
\v{I}^{(\theta)}_{ll'mm', \hat x}
=&
\int_0^{\pi} \sin(\theta) d\theta \int_0^{2\pi} d\varphi Y^*_{lm}(\theta,\varphi ) 
\frac{\partial Y_{l'm'}( \theta, \varphi )}{\partial \theta} 
\hat{\v{e}}_\theta(\theta,\varphi),
\end{align}
which yields
\begin{align}
I^{(\theta)}_{ll'mm', \hat x}
=&
\int_0^{\pi} \sin(\theta) d\theta \int_0^{2\pi} d\varphi Y^*_{lm}(\theta,\varphi ) 
\frac{\partial Y_{l'm'}( \theta, \varphi )}{\partial \theta} 
\cos(\theta)\cos(\varphi) 
\nonumber\\
=&
2\pi C_{lm} C_{l'm'} \int_0^{\pi} \sin(\theta) {P_{l}^{m}(\cos(\theta) )} 
\frac{\partial P_{l'}( \cos(\theta))}{\partial \theta} 
\cos(\theta) 
d\theta
\frac{1}{2\pi}
\int_0^{2\pi} \cos(\varphi) e^{i(m'-m)\varphi} d\varphi
\nonumber\\
=&
\underbrace{-2\pi C_{lm} C_{l'}m' \int_{-1}^{1} x \sqrt{1-x^2} {P_{l}^{m}(x)} 
\frac{d P_{l'}^{m'}(x)}{dx} 
dx}
_{P^{(3)}_{lml'm'}}
\underbrace{\frac{1}{2\pi}
\int_0^{2\pi} \cos(\varphi) e^{i(m'-m)\varphi} d\varphi}
_{A^{(c)}_{mm'}}
\nonumber\\
=&
P^{(3)}_{lml'm'}
{A^{(c)}_{mm'}};
\end{align}

\begin{align}
I^{(\theta)}_{ll'mm', \hat y}
=&
\int_0^{\pi} \sin(\theta) d\theta \int_0^{2\pi} d\varphi Y^*_{lm}(\theta,\varphi ) 
\frac{\partial Y_{l'm'}( \theta, \varphi )}{\partial \theta} 
\cos(\theta)\sin(\varphi) 
\nonumber\\
=&
\underbrace{-2\pi C_{lm} C_{l'm'} \int_{-1}^{1} x \sqrt{1-x^2} {P_{l}^{m}(x)}
\frac{d P_{l'}^{m'}(x)}{dx} 
dx}
_{P^{(3)}_{lml'm'}}
\underbrace{\frac{1}{2\pi}
\int_0^{2\pi} \sin(\varphi) e^{i(m'-m)\varphi} d\varphi}
_{A^{(s)}_{mm'}}
\nonumber\\
=&
P^{(3)}_{lml'm'}
{A^{(s)}_{mm'}};
\end{align}
and
\begin{align}
I^{(\theta)}_{ll'mm', \hat z}
=&
\int_0^{\pi} \sin(\theta) d\theta \int_0^{2\pi} d\varphi Y^*_{lm}(\theta,\varphi ) 
\frac{\partial Y_{l'm'}( \theta, \varphi )}{\partial \theta} 
( -\sin(\theta) )
\nonumber\\
=&
2\pi C_{lm} C_{l'm'} \delta_{mm'} \int_0^{\pi} \sin(\theta) {P_{l}^{m}(\cos(\theta))} 
\frac{\partial P_{l'}^{m'}(\cos( \theta) )}{\partial \theta} 
( -\sin(\theta) ) d\theta
\nonumber\\
=&
\underbrace{
-2\pi C_{lm} C_{l'm'}  \int_{-1}^{1} (1-x^2) {P_{l}^{m}(x)} 
\frac{d P_{l'}^{m'}(x)}{dx} 
 dx}
_{P^{(4)}_{lml'm'}}\delta_{mm'}
\nonumber\\
=& {P^{(4)}_{lml'm'}}\delta_{mm'}.
\end{align}

For $\v{I}^{(\varphi)}_{ll'mm'}$  we have 
\begin{align}
\v{I}^{(\varphi)}_{ll'mm'}=&
\int_0^{\pi} d\theta \int_0^{2\pi} d\varphi Y^*_{lm}(\theta,\varphi ) \frac{\partial Y_{l'm'}( \theta, \varphi )}{\partial \varphi}
   \hat{\v{e}}_\varphi(\theta,\varphi) 
\end{align}
which yields
\begin{align}
I^{(\varphi)}_{ll'mm', \hat x}=&
\int_0^{\pi} d\theta \int_0^{2\pi} d\varphi Y^*_{lm}(\theta,\varphi ) \frac{\partial Y_{l'm'}( \theta, \varphi )}{\partial \varphi}
(-\sin(\varphi)) 
\nonumber
\\
=&
2\pi C_{lm} C_{l'm'}
\int_0^{\pi} {P_{l}^{m}(\cos(\theta))} P_{l'}^{m'}(\cos(\theta))
d\theta
\frac{i m'}{2\pi}
\int_0^{2\pi}(-\sin(\varphi) e^{i(m'-m)\varphi}) d\varphi 
\nonumber
\\
=&
\underbrace{2\pi C_{lm} C_{l'm'}
\int_{-1}^{1} \frac{1}{\sqrt{1-x^2}} {P_{l}^{m}(x)} P_{l'}^{m'}(x)
dx}
_{P^{(5)}_{lml'm'}}
\underbrace{\frac{i m'}{2\pi}
\int_0^{2\pi}(-\sin(\varphi) e^{i(m'-m)\varphi}) d\varphi}
_{-im'A^{(s)}_{mm'}}
\nonumber\\
=&
-im'
{P^{(5)}_{lml'm'}}
A^{(s)}_{mm'}
\end{align}
\begin{align}
I^{(\varphi)}_{ll'mm', \hat y}=&
\int_0^{\pi} d\theta \int_0^{2\pi} d\varphi Y^*_{lm}(\theta,\varphi ) \frac{\partial Y_{l'm'}( \theta, \varphi )}{\partial \varphi}
\cos(\varphi)
\nonumber
\\
=&
\underbrace{
2\pi C_{lm} C_{l'm'}
\int_{-1}^{1} \frac{1}{\sqrt{1-x^2}} {P_{l}^{m}(x)} P_{l'}^{m'}(x)
dx}
_{P^{(5)}_{lml'm'}}
\underbrace{\frac{i m'}{2\pi}
\int_0^{2\pi}\cos(\varphi) e^{i(m'-m)\varphi} d\varphi}
_{im'A^{(c)}_{mm'}}
\nonumber\\
=&
im'
{P^{(5)}_{lml'm'}}
A^{(c)}_{mm'};
\end{align}
and
\begin{align}
I^{(\varphi)}_{ll'mm', \hat z}=0.
\end{align}

\vspace*{0.25in}
\section{Calculation of $P$ integrals \label{app:p-integrals} }
The $P$ integral general form is 
\begin{align}
 P^{(1)}_{lml'm'}=2\pi C_{lm} C_{l'm'}
 \int_{-1}^{1} \sqrt{1-x^2}  
{P_{l}^{m}(x)}  P_{l'}^{m'}(x )
dx;
\end{align}
\begin{align}
 P^{(2)}_{lml'm'}=2\pi C_{lm} C_{l'm'}
 \int_{-1}^{1} x  
{P_{l}^{m}(x)}  P_{l'}^{m'}(x )
dx;
\end{align}
\begin{align}
 P^{(3)}_{lml'm'}=2\pi C_{lm} C_{l'm'}
 \int_{-1}^{1} x\sqrt{1-x^2}  
{P_{l}^{m}(x)}  \frac{P_{l'}^{m'}(x )}{dx}
dx;
\end{align}
\begin{align}
 P^{(4)}_{lml'm'}=-2\pi C_{lm} C_{l'm'}
 \int_{-1}^{1} (1-x^2)  
{P_{l}^{m}(x)}  \frac{dP_{l'}^{m'}(x )}{dx}
dx;
\end{align}
and
\begin{align}
 P^{(5)}_{lml'm'}=2\pi C_{lm} C_{l'm'}
 \int_{-1}^{1} \frac{1}{\sqrt{1-x^2}}
{P_{l}^{m}(x)}  P_{l'}^{m'}(x)
dx.
\end{align}

Each of these five matrices has $16$ by $16$ elements for $0\leq l \leq 3 $.
They were calculated symbolically using Maple.

\vspace*{0.25in}
\section{Calculation of $A$ integrals \label{app:a-integrals} }
The $A$ integrals for complex spherical harmonics are
\begin{align}
A^{(c)}_{mm'}=&
 {\frac{1}{2\pi}
\int_0^{2\pi} \cos(\varphi) e^{i(m'-m)\varphi} d\varphi}
\nonumber\\
=&
\frac{1}{2} ( \delta_{m+1,m'}+\delta_{m-1,m'}),
\end{align}
and
\begin{align}
A^{(s)}_{mm'}=&
 {\frac{1}{2\pi}
\int_0^{2\pi} \sin(\varphi) e^{i(m'-m)\varphi} d\varphi}
\nonumber\\
=&
\frac{i}{2}(\delta_{m+1,m'} - \delta_{m-1,m'}).
\end{align} 

For real spherical harmonics we have 
\begin{align}
A^{(c)}_{mm'}=&
 {\frac{1}{2\pi}
\int_0^{2\pi} \cos(\varphi) 
\Phi_m(\varphi) \Phi_{m'}(\varphi)
d\varphi};
\end{align}
\begin{align}
A^{(s)}_{mm'}=&
 {\frac{1}{2\pi}
\int_0^{2\pi} \sin(\varphi) 
\Phi_m(\varphi) \Phi_{m'}(\varphi)
d\varphi};
\end{align} 
\begin{align}
A^{(c,d)}_{mm'}=&
 {\frac{1}{2\pi}
\int_0^{2\pi} \cos(\varphi) 
\Phi_m(\varphi) \frac{\partial \Phi_{m'}(\varphi)}{\partial\varphi}
d\varphi};
\end{align}
and
\begin{align}
A^{(s,d)}_{mm'}=&
 {\frac{1}{2\pi}
\int_0^{2\pi} \sin(\varphi) 
\Phi_m(\varphi) \frac{\partial \Phi_{m'}(\varphi)}{\partial\varphi}
d\varphi}.
\end{align}

\section{$I$ integrals \label{app:i-integrals-rsh} for real spherical harmonics}
For $ \v{I}^{(r)}_{ll'mm'}$  we have
\begin{align}
\v{I}^{(r)}_{ll'mm'}=
{\int_0^{\pi} \sin(\theta) d\theta \int_0^{2\pi} d\varphi S_{lm}(\theta,\varphi)  S_{l' m'}(\theta,\varphi )
\hat{ \v{e} }_r(\theta,\varphi),
} 
\end{align}
which yields
\begin{align}
I^{(r)}_{ll'mm', \hat x}=&
\int_0^{\pi} \sin(\theta) d\theta \int_0^{2\pi} d\varphi S_{lm}(\theta,\varphi)  S_{l' m'}(\theta,\varphi )
\sin(\theta) \cos(\varphi)
\nonumber
\\
=&
C_{lm} C_{l'm'} \int_0^{\pi} \sin^2(\theta)   
{P_{l}^{m}(\cos(\theta))}  P_{l'}^{m'}(\cos(\theta) )
d\theta
{\int_0^{2\pi}} \cos(\varphi) 
\Phi_m(\varphi)\Phi_{m'}(\varphi)
d\varphi
\nonumber
\\
=&
\underbrace{2\pi C_{lm} C_{l'm'}
 \int_{-1}^{1} \sqrt{1-x^2}  
{P_{l}^{m}(x)}  P_{l'}^{m'}(x )
dx}
_{P^{(1)}_{lml'm'}}
\underbrace{\frac{1}{2\pi}
\int_0^{2\pi} \cos(\varphi) 
\Phi_m(\varphi)\Phi_{m'}(\varphi)
d\varphi}
_{A^{(c)}_{mm'}}
\nonumber\\
=&
P^{(1)}_{lml'm'}A^{(c)}_{mm'},
\end{align}
where we have  used 
\begin{equation}
\Phi_m(\varphi)=
   \begin{cases}
     \sqrt{2}\sin |m|\theta,&  m < 0 \\
     {1}, & m=0 \\
     \sqrt{2}\cos m\theta,& {m > 0}. 
  \end{cases}
\end{equation}

For the rest of the components of $\v{I}^{(r)}_{ll'mm'}$ we have
\begin{align}
I^{(r)}_{ll'mm', \hat y}=&
{\int_0^{\pi} \sin(\theta) d\theta \int_0^{2\pi} d\varphi S_{lm}(\theta,\varphi)  S_{l' m'}(\theta,\varphi )
\sin(\theta) \sin(\varphi)
} 
\nonumber
\\
=&
\underbrace{{2\pi C_{lm} C_{l'm'}} \int_{-1}^{1} \sqrt{1-x^2}  
{P_{l}^{m}(x)}  P_{l'}^{m'}(x )
dx}
_{P^{(1)}_{lml'm'}}
\underbrace{\frac{1}{2\pi} \int_0^{2\pi} \sin(\varphi) 
\Phi_m(\varphi)\Phi_{m'}(\varphi)
d\varphi}
_{A^{(s)}_{mm'}}
\nonumber\\
=&
P^{(1)}_{lml'm'}A^{(s)}_{mm'};
\end{align}

\begin{align}
I^{(r)}_{ll'mm', \hat z}=&
\int_0^{\pi} \sin(\theta) d\theta \int_0^{2\pi} d\varphi S_{lm}(\theta,\varphi)  S_{l' m'}(\theta,\varphi )
\cos(\theta)
\nonumber
\\
=&
2\pi C_{lm} C_{l'm'} \delta_{mm'}
\int_0^{\pi} \sin(\theta) {P_{l}^{m}(\cos(\theta))}  P_{l'}^{m'}(\cos(\theta) )
\cos(\theta)
d\theta
\nonumber
\\
=&
\underbrace{2\pi C_{lm} C_{l'm'} 
\int_{-1}^{1} 
x {P_{l}^{m}(x)}  P_{l'}^{m'}(x)
dx}
_{P^{(2)}_{lml'm'}} 
\delta_{mm'}
\nonumber\\
=&
{P^{(2)}_{lml'm'}} 
\delta_{mm'}.
\end{align}

For $\v{I}^{(\theta)}_{ll'mm'}$  we have 
\begin{align}
\v{I}^{(\theta)}_{ll'mm', \hat x}
=&
\int_0^{\pi} \sin(\theta) d\theta \int_0^{2\pi} d\varphi S_{lm}(\theta,\varphi ) 
\frac{\partial S_{l'm'}( \theta, \varphi )}{\partial \theta} 
\hat{\v{e}}_\theta(\theta,\varphi),
\end{align}
which yields
\begin{align}
I^{(\theta)}_{ll'mm', \hat x}
=&
\int_0^{\pi} \sin(\theta) d\theta \int_0^{2\pi} d\varphi S_{lm}(\theta,\varphi ) 
\frac{\partial S_{l'm'}( \theta, \varphi )}{\partial \theta} 
\cos(\theta)\cos(\varphi) 
\nonumber\\
=&
2\pi C_{lm} C_{l'm'} \int_0^{\pi} \sin(\theta) {P_{l}^{m}(\cos(\theta) )} 
\frac{\partial P_{l'}( \cos(\theta))}{\partial \theta} 
\cos(\theta) 
d\theta
\frac{1}{2\pi}
\int_0^{2\pi} \cos(\varphi) 
\Phi_m(\varphi)\Phi_{m'}(\varphi)
d\varphi
\nonumber\\
=&
\underbrace{-2\pi C_{lm} C_{l'}m' \int_{-1}^{1} x \sqrt{1-x^2} {P_{l}^{m}(x)} 
\frac{d P_{l'}^{m'}(x)}{dx} 
dx}
_{P^{(3)}_{lml'm'}}
\underbrace{\frac{1}{2\pi}
\int_0^{2\pi} \cos(\varphi)
\Phi_m(\varphi)\Phi_{m'}(\varphi)
d\varphi}
_{A^{(c)}_{mm'}}
\nonumber\\
=&
P^{(3)}_{lml'm'}
{A^{(c)}_{mm'}};
\end{align}
\begin{align}
I^{(\theta)}_{ll'mm', \hat y}
=&
\int_0^{\pi} \sin(\theta) d\theta \int_0^{2\pi} d\varphi S_{lm}(\theta,\varphi ) 
\frac{\partial S_{l'm'}( \theta, \varphi )}{\partial \theta} 
\cos(\theta)\sin(\varphi) 
\nonumber\\
=&
\underbrace{-2\pi C_{lm} C_{l'm'} \int_{-1}^{1} x \sqrt{1-x^2} {P_{l}^{m}(x)}
\frac{d P_{l'}^{m'}(x)}{dx} 
dx}
_{P^{(3)}_{lml'm'}}
\underbrace{\frac{1}{2\pi}
\int_0^{2\pi} \sin(\varphi)
\Phi_m(\varphi)\Phi_{m'}(\varphi)
d\varphi}
_{A^{(s)}_{mm'}}
\nonumber\\
=&
P^{(3)}_{lml'm'}
{A^{(s)}_{mm'}};
\end{align}
and
\begin{align}
I^{(\theta)}_{ll'mm', \hat z}
=&
\int_0^{\pi} \sin(\theta) d\theta \int_0^{2\pi} d\varphi S_{lm}(\theta,\varphi ) 
\frac{\partial S_{l'm'}( \theta, \varphi )}{\partial \theta} 
( -\sin(\theta) )
\nonumber\\
=&
2\pi C_{lm} C_{l'm'} \delta_{mm'} \int_0^{\pi} \sin(\theta) {P_{l}^{m}(\cos(\theta))} 
\frac{\partial P_{l'}^{m'}(\cos( \theta) )}{\partial \theta} 
( -\sin(\theta) ) d\theta
\nonumber\\
=&
\underbrace{
-2\pi C_{lm} C_{l'm'}  \int_{-1}^{1} (1-x^2) {P_{l}^{m}(x)} 
\frac{d P_{l'}^{m'}(x)}{dx} 
 dx}
_{P^{(4)}_{lml'm'}}\delta_{mm'}
\nonumber\\
=& {P^{(4)}_{lml'm'}}\delta_{mm'}.
\end{align}

For $\v{I}^{(\varphi)}_{ll'mm'}$  we have 
\begin{align}
\v{I}^{(\varphi)}_{ll'mm'}=&
\int_0^{\pi} d\theta \int_0^{2\pi} d\varphi S_{lm}(\theta,\varphi ) \frac{\partial S_{l'm'}( \theta, \varphi )}{\partial \varphi}
   \hat{\v{e}}_\varphi(\theta,\varphi) 
\end{align}
which yields
\begin{align}
I^{(\varphi)}_{ll'mm', \hat x}=&
\int_0^{\pi} d\theta \int_0^{2\pi} d\varphi S_{lm}(\theta,\varphi ) \frac{\partial S_{l'm'}( \theta, \varphi )}{\partial \varphi}
(-\sin(\varphi)) 
\nonumber
\\
=&
2\pi C_{lm} C_{l'm'}
\int_0^{\pi} {P_{l}^{m}(\cos(\theta))} P_{l'}^{m'}(\cos(\theta))
d\theta
\frac{1}{2\pi}
\int_0^{2\pi}(-\sin(\varphi) )
\Phi_m(\varphi)\frac{\partial \Phi_{m'}(\varphi)}{\partial \varphi}
d\varphi
\nonumber
\\
=&
\underbrace{2\pi C_{lm} C_{l'm'}
\int_{-1}^{1} \frac{1}{\sqrt{1-x^2}} {P_{l}^{m}(x)} P_{l'}^{m'}(x)
dx}
_{P^{(5)}_{lml'm'}}
\underbrace{\frac{1}{2\pi}
\int_0^{2\pi}(-\sin(\varphi))
\Phi_m(\varphi)\frac{\partial \Phi_{m'}(\varphi)}{\partial \varphi}
d\varphi
}
_{-A^{(s,d)}_{mm'}}
\nonumber\\
=&
-
{P^{(5)}_{lml'm'}}
A^{(s,d)}_{mm'}
\end{align}

\begin{align}
I^{(\varphi)}_{ll'mm', \hat y}=&
\int_0^{\pi} d\theta \int_0^{2\pi} d\varphi S_{lm}(\theta,\varphi ) \frac{\partial S_{l'm'}( \theta, \varphi )}{\partial \varphi}
\cos(\varphi)
\nonumber
\\
=&
\underbrace{
2\pi C_{lm} C_{l'm'}
\int_{-1}^{1} \frac{1}{\sqrt{1-x^2}} {P_{l}^{m}(x)} P_{l'}^{m'}(x)
dx}
_{P^{(5)}_{lml'm'}}
\underbrace{\frac{1}{2\pi}
\int_0^{2\pi}\cos(\varphi) 
\Phi_m(\varphi)\frac{\partial \Phi_{m'}(\varphi)}{\partial \varphi}
d\varphi
}
_{A^{(c,d)}_{mm'}}
\nonumber\\
=&
{P^{(5)}_{lml'm'}}
A^{(c,d)}_{mm'};
\end{align}
and
\begin{align}
I^{(\varphi)}_{ll'mm', \hat z}=0.
\end{align}


\vspace*{8pt}
\section{\label{app:kgec-l-g} Lorentzian and Gaussian}
The Lorentzian with a full width at half the maximum of $\delta$ \\
has the expression
\begin{equation}
 L(x)=\frac{1}{\pi}\frac{\delta /2}{x^2+{\delta^2}{/4}}
\end{equation}
which equals half of its maximum amplitude for $x=\pm \delta/2$.

The Gaussian with width $\sigma_g$ is defined as
\begin{equation}
 G(x)=\frac{1}{ \sigma_g \sqrt{\pi}} \exp{\left(-\frac{ x^2}{\sigma_g^2}\right)}.
\end{equation}

Both functions are normalized to one. For the two to have the same
height, the Gaussian width must be
\begin{equation}
 \sigma_g=\frac{\delta}{2 \sqrt{\pi}}\;,
\end{equation}
while for equal FWHMs
\begin{equation}
 \sigma_g=\frac{\delta}{2 \sqrt{\ln{2}}}.
\end{equation}

%
%
\setcounter{secnumdepth}{0}
\section{References}
\bibliography{bib}{}

\end{document}